\documentclass[aps,prd,twocolumn,superscriptaddress,nofootinbib]{revtex4-1}%
\usepackage{epsf,epsfig}
\usepackage[utf8]{inputenc}
\usepackage{amssymb,amsmath,amsfonts}
\usepackage{color}
\usepackage{slashed}
\usepackage{float}
\usepackage{tensor}
\usepackage{braket}
\usepackage[colorlinks=true]{hyperref}  
\hypersetup{
    bookmarks=true,         
    unicode=false,          
    pdftoolbar=true,        
    pdfmenubar=true,        
    pdffitwindow=false,     
    pdfstartview={FitH},    
    colorlinks=true,       
    linkcolor=magenta, 
    citecolor=blue,        
    filecolor=magenta,      
    urlcolor=cyan           
} 

\newcommand{\bea}{\begin{eqnarray}}
\newcommand{\eea}{\end{eqnarray}}
\newcommand{\ba}{\begin{eqnarray}}
\newcommand{\ea}{\end{eqnarray}}

\newcommand{\nn}{\nonumber \\}

\newcommand{\beq}{\begin{equation}}
\newcommand{\eeq}{\end{equation}}
\newcommand{\beqa}{\begin{eqnarray}}
\newcommand{\eeqa}{\end{eqnarray}}
\newcommand{\beqar}{\begin{eqnarray*}}
	\newcommand{\eeqar}{\end{eqnarray*}}

\usepackage{color}

\newcommand{\req}[1]{(\ref{#1})} 

\begin{document}

\title{Gravitational ringing of rotating black holes in higher-derivative gravity}

\author{Pablo A. Cano}
\email{pabloantonio.cano@kuleuven.be}
\affiliation{Institute for Theoretical Physics, KU Leuven. Celestijnenlaan 200D, B-3001 Leuven, Belgium}

\author{Kwinten Fransen}
\email{kwinten.fransen@kuleuven.be}
\affiliation{Institute for Theoretical Physics, KU Leuven. Celestijnenlaan 200D, B-3001 Leuven, Belgium}

\author{Thomas Hertog}
\email{thomas.hertog@kuleuven.be}
\affiliation{Institute for Theoretical Physics, KU Leuven. Celestijnenlaan 200D, B-3001 Leuven, Belgium}

\author{Simon Maenaut}
\email{simon.maenaut@kuleuven.be}
\affiliation{Institute for Theoretical Physics, KU Leuven. Celestijnenlaan 200D, B-3001 Leuven, Belgium}


\begin{abstract}

We study gravitational perturbations of slowly-rotating black holes in a general effective-field-theory extension of general relativity that includes up to eight-derivative terms. We show that two Schr\"odinger-like equations with spin-dependent effective potentials govern the odd - and even-parity master variables.
These equations are coupled for parity-violating corrections, and this coupling affects the quasinormal modes even at linear order in the higher-derivative corrections, due to their isospectrality in general relativity. 
We provide results for the shifts in the fundamental quasinormal mode frequencies at linear order in the spin, which we expect to be valuable for high-precision phenomenology through future gravitational wave observations.

\end{abstract}

\maketitle

\section{Introduction}

Gravitational wave (GW) observations probe the dynamics of highly curved regions of spacetime and are therefore
paving the way to test general relativity (GR) in this regime \cite{TheLIGOScientific:2016src,Yunes:2016jcc,Berti:2018cxi,Berti:2018vdi,Barack:2018yly,Abbott:2018lct,LIGOScientific:2019fpa,Cabero:2019zyt,LIGOScientific:2020tif}. In particular, GWs from black hole binaries test the Kerr hypothesis, \textit{i.e.} the GR prediction that the geometry of a rotating black hole is given by the Kerr one, or even whether black objects are black holes at all \cite{Cardoso:2016rao}. 
In the light of these developments it is of great interest to study and identify what are the possible signatures of corrections to GR in gravitational wave signals \cite{Okounkova:2019zjf,Sennett:2019bpc,Nair:2019iur,Carson:2020cqb,Carson:2020ter,Carson:2020iik,Okounkova:2020rqw,deRham:2020ejn,Perkins:2021mhb}. 

Such corrections are to be expected. Fundamental gravity theory and general effective-field-theory (EFT) arguments suggest that GR is modified by higher-derivative terms in the action \cite{Gross:1986iv,Gross:1986mw,Bergshoeff:1989de}. As it turns out, the Kerr geometry is typically not a solution of modified or `corrected' gravity theories.\footnote{See \textit{e.g.} \cite{Konno:2007ze,Yunes:2009hc,Pani:2009wy,Kleihaus:2011tg,Pani:2011gy,Yagi:2012ya,Ayzenberg:2014aka,Maselli:2015tta,Kleihaus:2015aje,Delsate:2018ome,Cardoso:2018ptl,Cano:2019ore,Adair:2020vso} for rotating black hole solutions in some of these models.} This may give rise to deviations from GR predictions that might be observable. Since higher-derivative terms modify gravity at short distances, the most sensitive probes are those that involve physics in the near-horizon region of black holes. In particular the ringdown signal, determined by the black holes's quasinormal modes (QNMs) \cite{Berti:2009kk,Konoplya:2011qq}, deserves special attention \cite{Berti:2018vdi,Barack:2018yly}.  

The corrections to the QNM frequencies of non-rotating black holes have been obtained in a number of theories, including dynamical Chern-Simons gravity \cite{Cardoso:2009pk,Molina:2010fb}, Einstein-dilaton-Gauss-Bonnet gravity \cite{Blazquez-Salcedo:2016enn,Blazquez-Salcedo:2017txk}, and theories with quartic  \cite{Cardoso:2018ptl,McManus:2019ulj} and cubic \cite{deRham:2020ejn} curvature terms, among other models \cite{Tattersall:2018nve,Konoplya:2020bxa,Moura:2021eln,Moura:2021nuh}. In addition, theory-independent approaches were studied in \cite{Tattersall:2017erk,Cardoso:2019mqo,McManus:2019ulj}, while eikonal QNMs beyond GR were investigated in \cite{Glampedakis:2019dqh,Silva:2019scu,Bryant:2021xdh}.

The calculation of QNMs of astrophysically more relevant, rotating black holes is much more involved. Only recently, some progress has been made. In the case of a test scalar field, ref.~\cite{Cano:2020cao} gave a way of dealing with the non-separability of the wave equation and obtained the QNMs for rotating black holes in a large class of higher-derivative theories were computed.\footnote{Another strategy consists in performing a general theory-independent parametrization of the departure from the GR prediction, as in ref.~\cite{Maselli:2019mjd}, but this requires the introduction of a large number of parameters.} However, when it comes to gravitational perturbations the situation is more precarious, given the difficulty to perform full-fledged perturbation theory for rotating black holes in modified gravity theories. Nevertheless, this problem can be alleviated by assuming that the spin is small \cite{Pani:2012bp,Pani:2013pma}. Working at linear order in the spin, the QNM frequencies of slowly-rotating black holes in Einstein-dilaton-Gauss-Bonnet gravity were recently obtained in \cite{Pierini:2021jxd}, and an analogous analysis for dynamical Chern-Simons theory was done in \cite{Wagle:2021tam,Srivastava:2021imr}. 

Here we consider a general effective-field-theory (EFT) extension of general relativity, without introducing additional degrees of freedom. The resulting action, eq. \req{eq:EFTofGR} below, represents the most general diffeomorphism-invariant purely metric deformation of GR and is thus of phenomenological interest. Despite its generality, the action only contains five parameters up to the eight-derivative level. The goal of this paper is to compute the QNM frequencies of slowly-rotating black holes at linear order in the spin in this class of theories. 
 
Taking an EFT viewpoint, we assume that the effect of the beyond-GR corrections is small and work perturbatively in the higher-order couplings. Still, given that a length scale $\ell \geq 1$ km is required for modifications of gravity  to be detectable through gravitational waves generated in processes involving astrophysical black holes, one may wonder what are the experimental bounds on the couplings in the theory \req{eq:EFTofGR}. In the weak field limit, the leading correction to the Newtonian potential is $\sim \ell^4 M^3/r^7$. This means that the relative correction $\Delta$ compared to GR predictions is $\Delta\sim\ell^4 M^2/r^6$. If, say, $\ell\sim 10$ km, then $\Delta_{\odot}\sim 10^{-31}$ and $\Delta_{\oplus}\sim 10^{-30}$ at the surfaces of the sun and the earth, respectively. Hence the corrections we consider have a completely negligible effect on solar system physics. On the other hand, for a black hole of 10 solar masses $\Delta_{\rm BH}\sim 10^{-2}-10^{-3}$, large enough to induce effects that are potentially observable through high-precision GW observations. 

Also, no constraints can be imposed on \req{eq:EFTofGR} from cosmological observations. The reason is that the corrections to BH geometries depend only on the Weyl curvature, and in fact, in the EFT \req{eq:EFTofGR} one can trade all the Riemann tensors for Weyl tensors. On the other hand, terms with three or more Weyl tensors are irrelevant to cosmology, since FRLW metrics are conformally flat, and hence this EFT cannot be constrained via cosmology. Therefore GWs from stellar-mass black hole binaries provide a uniquely powerful window to search for higher-derivative deviations of Einstein's theory of the kind we consider, justifying our interest in QNMs in these theories.

The outline of the paper is as follows: we start in Section \ref{sec:EFT} by reviewing the EFT under consideration as well as the slowly rotating black holes in this theory. We analyse gravitational perturbations of these slowly rotating black holes in Section \ref{sec:perturbationequations}, first describing our general approach and then applying this to derive master equations for respectively the parity-preserving corrections and the parity-violating corrections. We then proceed to use these master equations to compute quasinormal mode frequencies in \ref{sec:qnms} and briefly conclude in Section \ref{sec:conclusion}. 


\section{The EFT of gravity and its black hole solutions}
\label{sec:EFT}
We consider a general EFT extension of GR, including no fields but the metric. To eight derivatives one can expand the action as
\begin{equation}\label{eq:EFTofGR}
S=\frac{1}{16\pi G}\int d^4x\sqrt{|g|}\left\{R+\ell^4\mathcal{L}_{(6)}+\ell^6\mathcal{L}_{(8)}+\ldots \right\}\, ,
\end{equation}
where $\ell$ is a length scale and where $\mathcal{L}_{(n)}$ denotes a Lagrangian density with $n$ derivatives of the metric.
Note that the leading corrections appear at the six-derivative level, since all four-derivative theories have the same vacuum solutions as GR, and hence can be ignored from the EFT viewpoint we adopt.   
In order to derive the most general form of the Lagrangians $\mathcal{L}_{(6)}$ and $\mathcal{L}_{(8)}$, one has to take into account that terms with explicit Ricci curvature are either trivial - in the sense that they do not affect the vacuum solutions of GR - or that they can be removed via a redefinition of the metric, and hence neglected. Therefore, one is left with a set of pure Riemann invariants, where the number of independent operators can be further reduced using diverse identities\footnote{See \cite{Endlich:2017tqa,Cano:2019ore} for a more complete discussion on the construction of this effective theory.}.  The resulting six-derivative Lagrangian contains two cubic invariants, one preserving and the other breaking parity, 

\begin{equation}
	\mathcal{L}_{(6)} = \lambda_{\rm ev}\tensor{R}{_{\mu\nu }^{\rho\sigma}}\tensor{R}{_{\rho\sigma }^{\delta\gamma }}\tensor{R}{_{\delta\gamma }^{\mu\nu }}+\lambda_{\rm odd}\tensor{R}{_{\mu\nu }^{\rho\sigma}}\tensor{R}{_{\rho\sigma }^{\delta\gamma }} \tensor{\tilde R}{_{\delta\gamma }^{\mu\nu }} \, ,
\end{equation}
where
\begin{equation}
{\tilde R}^{\mu\nu\rho\sigma}=\frac{1}{2}\epsilon^{\mu\nu\alpha\beta}\tensor{R}{_{\alpha\beta}^{\rho\sigma}} \, .
\end{equation}

\noindent
On the other hand, the eight-derivative Lagrangian can be expressed as the sum of three quartic Riemann operators,
\begin{equation}\label{eq:quarticL}
	\mathcal{L}_{(8)}=\epsilon_1\mathcal{C}^2+\epsilon_2\tilde{\mathcal{C}}^2+\epsilon_3\mathcal{C}\tilde{\mathcal{C}}\, ,
\end{equation}

\noindent
where

\begin{equation}
	\mathcal{C}=R_{\mu\nu\rho\sigma} R^{\mu\nu\rho\sigma}\, ,\quad \tilde{\mathcal{C}}=R_{\mu\nu\rho\sigma} \tilde{R}^{\mu\nu\rho\sigma}\, .
\end{equation}
In this case, the $\epsilon_1$ and $\epsilon_2$ interactions preserve parity, while $\epsilon_3$ violates it. However, the $\epsilon_2$ term has the peculiarity of vanishing for spherically symmetric configurations, as it is the square of an odd-parity term. 
Thus, the corrections to GR are controlled by only five dimensionless couplings, $\lambda_{\rm ev, odd}$, and $\epsilon_{1,2,3}$ and the overall length scale $\ell$, which can be absorbed in these couplings making them dimensionful.  Let us mention that some constraints can be set on these coupling from imposing physical requirements such as causality \cite{Gruzinov:2006ie,Camanho:2014apa,Endlich:2017tqa} or unitarity \cite{Bern:2021ppb}. However, we will allow for the couplings to be arbitrary. 
The Einstein field equations of the theory \req{eq:EFTofGR} are given by

\begin{equation}
\mathcal{E}_{\mu\nu}=G_{\mu \nu} +\ell^4\mathcal{E}^{(6)}_{\mu\nu}+\ell^6\mathcal{E}^{(8)}_{\mu\nu}=0\, ,
\end{equation}
where 

\begin{equation}
\mathcal{E}^{(n)}_{\mu\nu}=\tensor{P}{^{(n)}_{(\mu}^{\rho \sigma \gamma}} \tensor{R}{_{\nu) \rho \sigma \gamma}} -\frac{1}{2}g_{\mu \nu} \mathcal{L}_{(n)}+2 \nabla^\sigma \nabla^\rho P^{(n)}_{(\mu| \sigma|\nu)\rho}\, ,
\end{equation}
and in each case the tensor $P^{(n)}_{\mu\nu\rho \sigma}$ is given by

\begin{align}
\notag
P^{(6)}_{\mu\nu\rho \sigma}&=3\lambda_{\rm ev} \tensor{R}{_{\mu\rho}^{\alpha \beta}}\tensor{R}{_{\alpha\beta\rho\sigma}}\\
&+\frac{3\lambda_{\rm odd}}{2}\left(\tensor{R}{_{\mu\rho}^{\alpha \beta}}\tensor{\tilde R}{_{\alpha\beta\rho\sigma}}+\tensor{R}{_{\mu\rho}^{\alpha \beta}}\tensor{\tilde R}{_{\rho\sigma\alpha\beta}}\right)\, ,\\\notag
P^{(8)}_{\mu\nu\rho \sigma}&=4\epsilon_1 \mathcal{C} R_{\mu\nu\rho \sigma}+2 \epsilon_2\mathcal{\tilde C}\left(\tilde R_{\mu\nu\rho \sigma}+\tilde R_{\rho \sigma\mu\nu}\right)\\
&+\epsilon_{3}\left[2\mathcal{\tilde C}R_{\mu\nu\rho \sigma}+\mathcal{C}\left(\tilde R_{\mu\nu\rho \sigma}+\tilde R_{\rho \sigma\mu\nu}\right)\right]\, .
\end{align}
The (slowly-)rotating black hole solutions of these theories have been studied in refs.~\cite{Cardoso:2018ptl,Cano:2019ore} at different orders in the spin expansion, but here we only work at first order in the spin, in which case the metric takes a relatively simple form, namely
\begin{equation}\label{eq:rotatingBH}
ds^2=-N^2 f dt^2+\frac{dr^2}{f}-2 a h\sin^2\theta dt d\phi+r^2(d\theta^2+\sin^2\theta d\phi^2)\, .
\end{equation}

Working perturbatively in the higher-order couplings and in the spin $a=M \chi$, we find the following solution for the functions $N$, $f$ and $h$

\begin{widetext}
\begin{align}\notag
f&=1-\frac{2M}{r}+24\lambda_{\rm ev}\ell^4M^2\left(\frac{9}{r^6}-\frac{49M}{3r^7}\right)
-\lambda_{\rm odd}\ell^4\chi \cos\theta\frac{27 (2 M-r)}{14 M^2 r^8} \big(704 M^5-80 M^4 r-32 M^3 r^2-12 M^2 r^3\\\notag
&-4 Mr^4-r^5\big)+1152\epsilon_{1}\ell^6M^3\left(\frac{4}{r^9}-\frac{67M}{9r^{10}}\right)-\epsilon_{3}\ell^6\chi \cos\theta\frac{3 (2 M-r) }{20 M^4 r^{11}}\big(139008 M^8-13312 M^7 r-5824 M^6 r^2\\\label{eq:fsol}
&-2496 M^5
   r^3-1040 M^4 r^4-416 M^3 r^5-156 M^2 r^6-52 M r^7-13
   r^8\big)+\mathcal{O}\left(\ell^8,\chi^2\right)\, .\\
\notag
N&=1-\frac{108\lambda_{\rm ev}\ell^4M^2}{r^6}+\lambda_{\rm odd}\ell^4\chi  \frac{9\cos\theta}{14 M^2 r^7} \left(-1152 M^5+80 M^4 r+32 M^3 r^2+12 M^2 r^3+4 M
   r^4+r^5\right)\\\notag
&-1792\epsilon_{1}\ell^6\frac{M^3}{r^9}+\epsilon_{3}\ell^6\chi \frac{\cos\theta}{20 M^4 r^{10}}\big(-177408 M^8+13312 M^7 r+5824 M^6 r^2+2496 M^5 r^3+1040 M^4
   r^4+416 M^3 r^5\\
   &+156 M^2 r^6+52 M r^7+13 r^8\big)+\mathcal{O}\left(\ell^8,\chi^2\right)\, ,\\
h&=\frac{M}{r}-20\frac{\lambda_{\rm ev}\ell^4M^3}{r^7}-\frac{64\epsilon_{1}\ell^6M^3}{11r^{10}}(108r-121M)-6912\epsilon_{2}\ell^6\frac{M^3}{11r^9}+\mathcal{O}\left(\ell^8,\chi\right)\, .
\end{align}
\end{widetext}
It is important to note that, due to the corrections, the position of the horizon is modified. It corresponds to the largest root of $f$, which yields
\begin{equation}
r_{+}=2M-\frac{5\lambda_{\rm ev}\ell^4}{8M^3}-\frac{5\epsilon_{1}\ell^6}{4M^5}+\mathcal{O}(\ell^8,\chi^2)\, .
\end{equation}
In this solution $M$ represents the black hole mass, while the total angular momentum is $J=\chi M^2$. Other properties of these black hole geometries were studied in Refs~\cite{Cardoso:2018ptl,Cano:2019ore}.

\section{Perturbations of slowly-rotating black holes}
\label{sec:perturbationequations}
Let us now consider a perturbation $g_{\mu\nu}=g^{\rm BH}_{\mu\nu}+h_{\mu\nu}$ over the black hole metric \req{eq:rotatingBH}. The goal of this section is to simplify the corresponding linearized equations into a reduced set of equations for some master variables. To do this, it is useful to review first the case of non-rotating black holes. 

\subsection{Spherically symmetric black holes}
In a spherically symmetric spacetime, the metric perturbations can be separated in terms of the tensor spherical harmonics. These come in even-parity and odd-parity types and are labeled by the angular momentum and magnetic numbers, $l$ and $m$ respectively. These are described in \cite{Martel:2005ir}, but we review them here for the sake of convenience. Let us denote by $a$, $b$, ... the indices $t, r$ and $A$, $B$, ... those of the 2-sphere. 

The even-parity (polar) sector has the following components
\begin{align}
h^{lm}_{{(+)}ab}&=e^{-i \omega t}H_{ab}^{lm}Y^{lm}\, ,\\
h^{lm}_{{(+)}aB}&=e^{-i \omega t}j_{a}^{lm}Y_{B}^{lm}\, ,\\
h^{lm}_{{(+)}AB}&=e^{-i \omega t}r^2\left(K^{lm}\Omega_{AB}Y^{lm}+G^{lm}Y_{AB}^{lm}\right)\, ,
\end{align}
where we are already separating the time dependence as $e^{-i \omega t}$.
Here $H_{ab}^{lm}$, $j_{a}^{lm}$, $K^{lm}$ and $G^{lm}$ are functions of $r$, $\Omega_{AB}$ is the metric of the 2-sphere and  $Y_{B}^{lm}$ and $Y_{AB}^{lm}$ are the even-parity vector and tensor spherical harmonics. These are defined in terms of the scalar spherical harmonics $Y^{lm}$ according to

\begin{align}
Y_{A}^{lm}&=D_{A}Y^{lm}\, ,\\
Y_{AB}^{lm}&=\left[D_{A}D_{B}+\frac{1}{2}l(l+1)\Omega_{AB}\right]Y^{lm}\, ,
\end{align}
where $D_A$ is the covariant derivative of the 2-sphere. 

On the other hand, the odd-parity (axial) perturbations have the following structure,
\begin{align}
h^{lm}_{{(-)}ab}&=0\, ,\\
h^{lm}_{{(-)}aB}&=e^{-i \omega t}h_{a}^{lm}X_{B}^{lm}\, ,\\
h^{lm}_{{(-)}AB}&=e^{-i \omega t}h_{2}^{lm}X_{AB}^{lm}\, ,
\end{align}
where again $h_{a}^{lm}$ and $h_{2}^{lm}$ are functions of $r$, and the odd-parity vector and tensor spherical harmonics read

\begin{align}
X_{A}^{lm}&=-\epsilon_{A}^{\,\,\,\, B}D_{B}Y^{lm}\, ,\\
X_{AB}^{lm}&=-\epsilon_{(A}^{\,\,\,\, C}D_{B)}D_{C}Y^{lm}\, .
\end{align}

It is important to note that the vector and tensor spherical harmonics of odd and even parity satisfy a number of orthogonality relationships that can be found in \cite{Martel:2005ir}.

A general metric perturbation can be written as a linear superposition of all the modes $h^{lm}_{{(\pm)}\mu\nu}$. By definition, the quasinormal modes only contain modes with a fixed value of $l$ and $m$. In addition, since GR preserves parity, modes of different parity are decoupled and hence the quasinormal modes are either given by $h_{\mu\nu}=h^{lm}_{{(+)}\mu\nu}$ or $h_{\mu\nu}=h^{lm}_{{(-)}\mu\nu}$. 
However, some of the higher-derivative corrections we consider break parity, and as a consequence odd-parity and even-parity modes mix. Thus, in that case the QNMs will necessarily have the form
\begin{equation}\label{eq:sphericpert}
h_{\mu\nu}^{lm}=h^{lm}_{{(+)}\mu\nu}+h^{lm}_{{(-)}\mu\nu}\, .
\end{equation}

\noindent
Finally, we have presented the perturbations $h^{lm}_{{(+)}\mu\nu}$  and $h^{lm}_{{(-)}\mu\nu}$ in a general gauge, but to simplify the computations one can choose the Regge-Wheeler gauge, in which the following components vanish,
\begin{equation}\label{eq:RWg}
j_{a}^{lm}=0,\quad G^{lm}=0\, ,\quad h^{lm}_{2}=0\, .
\end{equation}
The higher-derivative corrections do not change this fact and we can always work in this gauge. 
Then, evaluating the linearized equations of motion on \req{eq:sphericpert} one gets a system of equations for the remaining variables. These equations can be simplified to yield two equations for two master variables, which are coupled if there  are parity-breaking corrections --- see \cite{Cardoso:2018ptl} for the case of the quartic theories and \cite{deRham:2020ejn} for the parity-preserving cubic one.

\subsection{Slowly-rotating black holes: linear order in spin}
In the presence of rotation, the tensor harmonics $h^{lm}_{{(\pm)}\mu\nu}$ are not eigenfunctions of the linearized Einstein's equations anymore, but one can always expand the metric perturbation using these functions since they form a basis. Thus, in general we can write

\begin{equation}
h_{\mu\nu}=\sum_{l,m}h^{lm}_{{(+)}\mu\nu}+\sum_{l,m}h^{lm}_{{(-)}\mu\nu}\, .
\end{equation}

\noindent
Now, in the case of a slowly-rotating black hole, at linear order in the spin, one can see that the quasinormal modes will necessarily have the following structure

\begin{equation}\label{eq:hmunuslow}
h_{\mu\nu}=h^{lm}_{\mu\nu}+\chi h'_{\mu\nu}+\mathcal{O}(\chi^2)\, ,
\end{equation}
where $h'_{\mu\nu}$ contains the sum of modes with $l'\neq l$, 

\begin{equation}
h'_{\mu\nu}=\sum_{l'\neq l } \left[h^{l'm}_{{(+)}\mu\nu}+h^{l'm}_{{(-)}\mu\nu}\right]\, .
\end{equation}

In words, the perturbation is composed of a leading term corresponding to the result for the spherically symmetric case, and the additional terms will be of order $\chi$. At the same time, the linearized equations have the following structure,

\begin{equation}
\mathcal{E}_{\mu\nu}=\mathcal{E}_{\mu\nu}^{(0)}+\chi \mathcal{E}_{\mu\nu}^{(1)}+\mathcal{O}(\chi^2)=0\, . \label{eqn:Einsteinperturbativechi}
\end{equation}
The tensor harmonics are eigenfunctions of $\mathcal{E}_{\mu\nu}^{(0)}$ but not of $\mathcal{E}_{\mu\nu}^{(1)}$. Each of these equations contains a contribution from the Einstein's equations as well as contributions from the higher-derivative terms. 

When we evaluate the equations in the $h_{\mu\nu}$ defined in \req{eq:hmunuslow}, we find

\begin{align}\notag
&\mathcal{E}_{\mu\nu}^{(0)}(h^{lm})+\chi \left[\mathcal{E}_{\mu\nu}^{(1)}(h^{lm})+\mathcal{E}_{\mu\nu}^{(0)}(h')\right] +\mathcal{O}(\chi^2) \, .
\end{align}

Then, we can project on the tensor harmonics with index $l$ and this projection kills all the terms $\mathcal{E}_{\mu\nu}^{(0)}(h')$, since $h'$ only contains modes with $l'\neq l$.
Therefore we obtain equations involving only $h^{lm}$ and these can be written as

\begin{align}
P_a=\int d\Omega \bar{X}^{B,lm}\mathcal{E}_{aB}(h^{lm})=&0\, ,\\
P_2=\int d\Omega \bar{X}^{AB,lm}\mathcal{E}_{AB}(h^{lm})=&0\, ,
\end{align}

and 

\begin{align}
	Q_{ab}=\int d\Omega \bar{Y}^{lm}\mathcal{E}_{ab}(h^{lm})=&0\, , \\
	Q_{a}=\int d\Omega \bar{Y}^{B,lm}\mathcal{E}_{aB}(h^{lm})=&0\, ,\\
	Q_{2}=\int d\Omega \bar{Y}^{AB,lm}\mathcal{E}_{AB}(h^{lm})=&0\, ,\\
	Q_{3}=\int d\Omega \bar{Y}^{lm}\Omega^{AB}\mathcal{E}_{AB}(h^{lm})=&0\, .
\end{align}
This is the set of consistent equations from which we will obtain the quasinormal mode frequencies. 

In addition, we will make use of the Regge-Wheeler gauge \req{eq:RWg}, which can be applied to the slowly-rotating case as well. 
To show this, consider first a gauge transformation with parameter $\xi_{\mu}$ that would put the $h^{lm}_{\mu\nu}$ term in \req{eq:hmunuslow} in the RW form had we set $\chi=0$. However, in the presence of rotation, this transformation will fail to achieve the RW gauge by $\mathcal{O}(\chi)$ terms:
\begin{equation}
h_{\mu\nu}+2\nabla_{(\mu}\xi_{\nu)}=h^{lm,\rm RW}_{\mu\nu}+\chi h'_{\mu\nu}+\chi\delta h_{\mu\nu} +\mathcal{O}(\chi^2)\, .
\end{equation}
However, $\delta h_{\mu\nu}$ can be decomposed in spherical harmonics, so that 
\begin{equation}
h_{\mu\nu}+2\nabla_{(\mu}\xi_{\nu)}=\left(h^{lm,\rm RW}_{\mu\nu}+\chi\delta h^{lm}_{\mu\nu}\right)+\chi \left(h'_{\mu\nu}+\delta h'_{\mu\nu}\right)+\mathcal{O}(\chi^2)\, .
\end{equation}
Now, it is clear that a new gauge transformation of order $\chi$ can be made in order to put the term $\chi\delta h^{lm}_{\mu\nu}$ into the RW gauge up to $\mathcal{O}(\chi^2)$ terms. Likewise, $\mathcal{O}(\chi)$ transformations can be used to put all the terms in $h'_{\mu\nu}$ in the RW form, although this will not be relevant for our calculations. Thus, we can reduce the number of variables in the problem by imposing \req{eq:RWg} right away.

In order to make further progress, it is useful to distinguish between the higher-derivative corrections that preserve parity and those that do not.

\subsection{Parity-preserving corrections}
For the parity-preserving corrections, the operator $\mathcal{E}_{\mu\nu}^{(1)}(h^{lm})$ does not mix the odd and even part of the perturbation, $h^{lm}_{(+)}$ and $h^{lm}_{(-)}$. Therefore, these perturbations can be considered separately. 

\subsubsection{Decoupled equations for odd perturbations}

We start discussing the resolution of the odd-parity equations for parity-preserving corrections, which are simpler as they only involve two variables in the RW gauge, $h_0$ and $h_1$. We have three equations,\footnote{The even-parity sector of the equations vanishes identically in this case, $Q_{ab}=Q_a=Q_2=Q_3=0$} namely, $P_{t}=P_{r}=P_{2}=0$, but there is a relation among them implied by the Bianchi identity of the gravitational tensor, and hence it is enough to keep only $P_{r}$ and $P_{2}$, which are of first order in the case of Einstein gravity. 
Our goal is to solve these equations perturbatively in $\chi$ and in the higher order couplings, that we denote collectively by $\lambda$, keeping the terms of order $\lambda\chi$.
To achieve this, we first solve the equations with $\lambda=0$. This is easily accomplished since in that case both equations are of first order and they give us $h_{0}$ as a function of $h_{1}$ and $h_{1}'$, as well as a second-order equation for $h_{1}$. Then we consider the equations with $\lambda\neq0$, which contain higher derivatives of these variables. However, since we are only interested in the perturbative solution in $\lambda$, we can use the zeroth-order solution in all the terms proportional to $\lambda$, and this allows us to express those terms as functions of $h_1$ and $h_1'$ only. Then we can again solve the equations $P_{r}=P_{2}=0$, that give us a solution of the form
\begin{equation}
h_{0}=a(r) h_{1}+b(r)h_{1}'\, ,\quad h_{1}''+c(r) h_{1}+d(r) h_{1}'=0\, ,
\end{equation}
for certain functions $a(r)$, $b(r)$, $c(r)$, $d(r)$. Finally, we can introduce a rescaled master variable
\begin{equation}\label{eq:psiminusdef}
\Psi^{-}=\frac{A(r)r^2}{r-r_{+}}h_{1}\, ,
\end{equation}
where $A(r)$ is a factor that can be chosen so that $\Psi^{-}$ satisfies a Schr\"odinger-type equation in terms of a preferred radial variable. For instance, for the sake of numerical computations, we find it interesting to write the master equation in terms of the ``pseudo-tortoise coordinate'' $r_{*}$ defined as
\begin{equation}\label{eq:pseudot}
r_{*}=r+2M\log\left(\frac{r}{r_{+}}-1\right)\, .
\end{equation}
This has the property that $r_{*}\rightarrow -\infty$ at the event horizon. In the case of the Schwarzschild black hole, a Schr\"odinger-type equation for $\Psi^{-}$ in terms of $r_{*}$ is achieved for $A(r)=1$. Thus, in order to take into account the effect of the spin and of the higher-derivative corrections, we must choose $A(r)=1+\delta A(r)$. 
Then, by choosing $\delta A(r)$ appropriately  --- the choice is unique up to a constant term ---, we find the following equation for $\Psi^{-}$, 

\begin{equation}\label{eq:masterodd}
\frac{d^2\Psi^{-}}{dr_{*}^2}+\left(\omega^2-V^{-}_{lm}\right)\Psi^{-}=0\, .
\end{equation}
The potential $V^{-}_{lm}$ has a GR part plus a contribution from each of the higher-derivative terms,
\begin{align}\label{eq:potentialodd}
V^{-}_{lm}=V_{lm}^{-\rm GR}+\ell^4\lambda_{\rm ev} V_{lm}^{-\rm (ev)}+\ell^6\left(\epsilon_{\rm 1} V_{lm}^{-(1)}+\epsilon_{\rm 2} V_{lm}^{-(2)}\right)\, ,
\end{align}
 At the same, each of these contributions contains a term of zeroth order and another one of first order in the spin. For the GR part we have
\begin{align}\notag
&V_{lm}^{-\rm GR}=\left(1-\frac{2M}{r}\right)\left(\frac{l(l+1)}{r^2}-\frac{6M}{r^3}\right)\\&+m\chi\bigg(\frac{4M^2\omega}{r^3}+\frac{24M^2(3r-7M)(r-2M)}{l(l+1)r^{7}\omega}\bigg)\, ,
\label{eq:V-GR}
\end{align}
which corresponds to the usual Regge-Wheeler potential \cite{Regge:1957td} plus a spin correction. The expressions for the other potentials are lengthier and hence we show them in the appendix \ref{app:potentials}. 

\subsubsection{Decoupled equations for even perturbations}
The case of even perturbations is more intricate, as it involves four variables, $H_{0}$, $H_{1}$, $H_{2}$ and $K$, in the Regge-Wheeler gauge. The odd-parity sector of the equations is trivially satisfied, $P_{a}=P_2=0$. To solve the remaining equations we proceed iteratively, solving first the system with $\lambda=0$, then substituting that result in the terms proportional to $\lambda$ and solving the equations again.  
When $\lambda=0$, the equations $Q_2=Q_{rr}=0$ allow us to obtain $H_{1}$ and $H_{2}$ in terms of $H_{0}$, $K$ and their derivatives. Then, working at first order in $\chi$, the rest of the equations can be solved to obtain a relation of the form $H_{0}=\tilde a(r) K+\tilde b(r) K'$, as well as a second order differential equation for $K$. In this way, all the variables and their derivatives can be expressed as certain linear combinations of $K$ and $K'$. 
Now, coming back to $\lambda\neq 0$, since we are only interested in the linear corrections, we can substitute the solution of the zeroth order equations in the terms proportional to $\lambda$. By doing so, we rewrite all those terms as linear combinations of $K$ and $K'$. Then we can solve the system of equations in an analogous to the $\lambda=0$ case. Thus, we get a master second-order equation for $K$ and the rest of variables are linear combinations of $K$ and $K'$. 
Finally, instead of working with $K$, it is customary to introduce the Zerilli function $\Psi^{+}$ as
\begin{equation}\label{eq:psiplusdef}
\Psi^{+}=\frac{2r\tilde{A}(r)}{l(l+1)}\left[K+\frac{2 f}{\Lambda}\left(H_2-r K'\right)\right]\, ,
\end{equation}
where $\Lambda=(l-1)(l+2)+6M/r$, and $\tilde{A}(r)$ is a normalization factor. In the case of the Schwarzschild black hole in GR, this definition matches the original one by Zerilli when $\tilde A(r)=1$. However, that factor has been introduced so that $\Psi^{+}$ satisfies a Schr\"odinger-type equation once rotation and higher-derivative corrections are taken into account, so that in general we will have $\tilde A(r)=1+\delta \tilde A(r)$. Thus, by choosing $\delta\tilde A(r)$ suitably  --- again, the choice is unique up to a constant term ---, we find that $\Psi^{+}$ satisfies the equation

\begin{equation}\label{eq:mastereven}
\frac{d^2\Psi^{+}}{dr_{*}^2}+\left(\omega^2-V^{+}_{lm}\right)\Psi^{+}=0\, ,
\end{equation}
where  $r_*$ is the pseudo-tortoise coordinate defined in \req{eq:pseudot} and $V^{+}_{lm}$ is the effective potential for even-parity perturbations.  As before, the potential has contributions from the different terms in the gravitational action
\begin{align}\label{eq:potentialeven}
V^{+}_{lm}=V_{lm}^{+\rm GR}+\ell^4\lambda_{\rm ev} V_{lm}^{+\rm (ev)}+\ell^6\left(\epsilon_{\rm 1} V_{lm}^{+(1)}+\epsilon_{\rm 2} V_{lm}^{+(2)}\right)\, ,
\end{align}
and for instance, the GR part reads

\begin{widetext}
\begin{align}\notag
	&V_{lm}^{+\rm GR}=\frac{1}{\Lambda^2}\left(1-\frac{2M}{r}\right)\bigg\lbrack \mu^2\left(\frac{\mu+2}{r^2}+\frac{6M}{r^3}\right)+\frac{36M^2}{r^4}\left(\mu+\frac{2M}{r}\right) \bigg\rbrack \notag\\
	&+\frac{4 m M^2 \chi  \omega }{(\mu +2) r^6 \Lambda^3} \left(72 (\mu +6) M^3+36 \mu  (\mu +8) M^2 r+6 \mu ^2 (3 \mu +16) M r^2+\mu ^2 \left(\mu ^2+2 \mu -12\right) r^3\right)\, ,
	\label{eq:V+GR}
\end{align}
\end{widetext}
with $\mu = (l-1)(l+2)$. Note that for $\chi=0$ this reduces to the well-known Zerilli potential for the Schwarzschild black hole \cite{Zerilli:1970wzz,Zerilli:1970se}. The expressions for the other contributions are given in the appendix \ref{app:potentials}. 

\subsection{Parity-breaking corrections}
The higher-derivative terms controlled by $\lambda_{\rm odd}$ and $\epsilon_{3}$ violate parity, and hence their effect is somewhat different. Instead of modifying the equations of odd and even perturbations independently, they introduce a coupling between these modes. Proceeding along the same lines as in the previous subsection, we find the following coupled equations for the variables $\Psi^{\pm}$:\footnote{These are defined as in eqs.~\req{eq:psiminusdef} and \req{eq:psiplusdef}, but in the case of $\Psi^{+}$ we set $f=1-2M/r$ in \req{eq:psiplusdef} instead of using the actual value of the metric function given in \req{eq:fsol}.}

\begin{align}\label{eq:paritymix1}
\frac{d^2\Psi^{+}}{dr_{*}^2}+\left(\omega^2-V^{+\rm GR}_{lm}\right)\Psi^{+}&=U^{+}_{lm}\Psi^{-}+W^{+}_{lm}\frac{d\Psi^{-}}{dr_{*}}\, ,\\\label{eq:paritymix2}
\frac{d^2\Psi^{-}}{dr_{*}^2}+\left(\omega^2-V^{-\rm GR}_{lm}\right)\Psi^{-}&=U^{-}_{lm}\Psi^{+}+W^{-}_{lm}\frac{d\Psi^{+}}{dr_{*}}\, .
\end{align}
Here $V^{\pm\rm GR}_{lm}$ are the GR potentials we shown above, while the coefficients $U^{\pm}_{lm}$, $W^{\pm}_{lm}$ are controlled by the higher-derivative terms and hence they have two contributions, 
\begin{align}
U^{\pm}_{lm}&=\lambda_{\rm odd}\ell^4U^{\pm (\rm odd)}_{lm}+\epsilon_{3} \ell^6U^{\pm (3)}_{lm}\, ,\\
W^{\pm}_{lm}&=\lambda_{\rm odd}\ell^4W^{\pm (\rm odd)}_{lm}+\epsilon_{3} \ell^6W^{\pm (3)}_{lm}\, .
\end{align}
We show the explicit values of these quantities in the appendix \ref{app:potentials}. Also, notice that, in the presence of parity-breaking corrections, the radius of the horizon is not modified at leading order, $r_{+}=2M$. Hence, $r_{*}$ is the standard Schwarzschild tortoise coordinate, $r_{*}=r+2M\log\left(\frac{r}{2M}-1\right)$, so that 
\begin{equation}
\frac{d}{dr*}=\left(1-\frac{2M}{r}\right)\frac{d}{dr}\, .
\end{equation}

In the case of non-rotating black holes in the quartic theory, ref.~\cite{Cardoso:2018ptl} was able to find a change of variables for \req{eq:paritymix1}-\req{eq:paritymix2} that cancels the coefficients $W^{\pm}_{lm}$ and sets $U^{+}_{lm}=-U^{-}_{lm}$. However, this becomes less practical once the linear terms in spin are taken into account. 

Nevertheless, we do find interesting to perform a Chandrasekhar transformation of one of the variables, say $\Psi^{+}\rightarrow \tilde \Psi^{+}$, such that the zeroth-order equation $\frac{d^2\Psi^{+}}{dr_{*}^2}+\left(\omega^2-V^{+\rm GR}_{lm}\right)\Psi^{+}$ is transformed into $\frac{d^2\tilde\Psi^{+}}{dr_{*}^2}+\left(\omega^2-V^{-\rm GR}_{lm}\right)\tilde\Psi^{+}$. This is possible even at first order in the spin since $V^{+\rm GR}_{lm}$ and $V^{-\rm GR}_{lm}$ are isospectral at that order, and we show the explicit transformation in appendix~\ref{app:darboux}. Of course, the functions $U^{\pm}_{lm}$, $W^{\pm}_{lm}$ are also affected by this change of variables, but expressed in this form we can devise a way to study the system of equations by solving only uncoupled equations. The idea is to decompose the corrections in contributions that are symmetric and antisymmetric under the exchange of $\Psi^{\pm}$, since for each of these contributions one can decouple the equations. One can then add up appropriately the effect of each correction in order to obtain, \textit{e.g.}, the total shift in the QNM frequencies, that we study next. We explore this method in more detail in appendix~\ref{app:coupledbyuncoupled}.

\section{Quasinormal modes}
\label{sec:qnms}
Once we have set up the master equations that govern the perturbations of slowly-rotating black holes in our theory \req{eq:EFTofGR}, our goal is to compute the corrections to the quasinormal modes of the Kerr black hole. Recall that these correspond to solutions in which the variables $\Psi^{\pm}$ behave as outgoing waves both at infinity and at the horizon and that this is only possible for a discrete set of complex frequencies. We are mostly interested in the shifts in the Kerr QNM frequencies induced by the higher-derivative corrections:
\begin{equation}
\omega=\omega^{\rm Kerr}+\delta\omega\, .
\end{equation}

We recall that, in the case of Kerr, the spectrum of odd and even perturbations is identical. In order to show explicitly that, for our GR potentials to linear order in spin, this is still true, we present in appendix \ref{app:darboux} an extension to the Chandrasekhar transformation which relates the Regge-Wheeler and Zerilli potentials. Now, for the higher-derivative corrections that preserve parity, the shifts in the QNM frequencies of odd and even modes are typically different, and hence one loses isospectrality. Thus, we have two families of frequencies. 
\begin{equation}
\omega^{\pm}=\omega^{\rm Kerr}+\delta\omega^{\pm} \,.
\end{equation}
If several parity-preserving interactions appear in the Lagrangian at the same time, one may obtain the total shifts $\delta\omega^{\pm}$ by simply adding up each of the individual contributions. 

In the case of the corrections that violate parity, modes of even and odd parity get coupled and this actually produces a linear shift in the QNM frequencies \cite{McManus:2019ulj}. The reason for this is that both GR potentials $V^{\pm \rm GR}$ are isospectral, and in that case, the problem \req{eq:paritymix1}-\req{eq:paritymix2} is analogous to performing quantum-mechanical perturbation theory with a degenerate Hamiltonian. In this situation, contrarily to the result with non-degenerate Hamiltonians, off-diagonal components in the Hamiltonian perturbation do introduce a linear correction into the energy states --- the QNM frequencies in our case. 
In fact, one can see that, when there are only parity-breaking corrections, the shifts in the frequencies are opposite for the two families of modes

\begin{equation}
\omega^{\pm}=\omega^{\rm Kerr}\pm\delta\omega^{\rm break}\, , 
\end{equation}
for certain $\delta\omega^{\rm break}$.
Note that in this case the labels $\pm$ are unrelated to the parity of the perturbations. In addition to the frequency, since the QNMs in the presence parity-violating interactions contain both the even- and odd-parity sectors, it is also important to determine the relative normalization of both modes. Thus, we introduce the quantity,
\begin{equation}
\gamma=\lim_{r\rightarrow\infty}\frac{\Psi^{+}}{\Psi^{-}}\, ,
\end{equation}
which has to be determined as well (note that the two families of modes $\pm$ have relative normalizations $\pm\gamma$). \footnote{Let us remark that the variables $\Psi^{\pm}$ are defined according to \req{eq:psiminusdef} and \req{eq:psiplusdef}, and that their normalization factors satisfy $A=\tilde A=1$ at $r\rightarrow\infty$.}

Now, if there are several independent parity-violating operators in the action, the total value of $\delta\omega^{\rm break}$ cannot be easily obtained by adding up the contributions of each term. However, one can easily combine them if one knows separately the `symmetric' $\omega_s$ and `antisymmetric' $\omega_a$ contributions to the frequency with respect to interchange of $\Psi^+$ and $\Psi^-$. Roughly, by considering these contributions individually in equations \req{eq:paritymix1}-\req{eq:paritymix2}, they can be decoupled, and for these decoupled equations the different contributions add linearly as usual. There are two caveats, though. The first is that one cannot actually use \req{eq:paritymix1}-\req{eq:paritymix2} as one should make a transformation to ensure the uncorrected (isospectral) potentials are identical, \textit{i.e.} use the transformation given in appendix \ref{app:darboux}, as we mentioned earlier. Secondly, one must be able to recombine these `symmetric' and `antisymmetric' contributions. The procedure is analogous to what is used to combine the even-parity and odd-parity contributions discussed next but for convenience it is made explicit in appendix \ref{app:coupledbyuncoupled}. The final, full correction due to parity-breaking interactions can be written as $\delta \omega^{\rm break} = \sqrt{(\sum_i \omega^{(i)}_s)^2 + (\sum_i \omega^{(i)}_a)^2}$. Equivalently, in order not to work with these auxiliary frequencies, one can note that the ratio $\omega^{(i)}_a/\omega^{(i)}_s$ will characterize the mixing so in principle one can combine the different odd-parity corrections if in addition to $\delta\omega^{(i) \rm break}$ one is given the mixing coefficients $\gamma^{(i)}$.

Once the contributions to the QNM frequencies from parity-preserving and parity-violating terms have been independently determined, we would like to obtain the total shift in these frequencies in the case in which all the higher-derivative operators appear simultaneously in the action. To obtain that result, one can note that the problem of obtaining the corrected QNMs is formally equivalent to the problem of obtaining the eigenvectors and eigenvalues of a perturbed Hamiltonian. Thus, the usual formulas from quantum-mechanical perturbation theory can be applied --- see appendix \ref{app:pert} and references \cite{Leaver:1986gd,Zimmerman:2014aha,Mark:2014aja,McManus:2019ulj}. Therefore, one can derive the total shift from the eigenvalue equation
\begin{equation} \label{eq:dettotal}
   \begin{vmatrix}
   \delta\omega^{+}-\delta\omega &\delta\omega^{\rm break}\\
    \delta\omega^{\rm break} &\delta\omega^{-}-\delta\omega
   \end{vmatrix}=0\, ,
\end{equation}
coming from the perturbation of a degenerate $2\times 2$ Hamiltonian. 
This yields the result

\begin{equation}\label{eq:totalshift}
\delta\omega^{(\pm)}_{\rm tot}=\frac{1}{2}\left(\delta\omega^{+}+\delta\omega^{-}\right)\pm\sqrt{\frac{1}{4}\left(\delta\omega^{+}-\delta\omega^{-}\right)^2+\left(\delta\omega^{\rm break}\right)^2}\, .
\end{equation}
Note that we keep the $\pm$ labels, but one should bear in mind that these no longer represent defined-parity modes whenever $\delta\omega^{\rm break}\neq 0$.  
Finally, the relative normalization between the even- and odd-parity master variables can be determined by applying again essentially formulas from a $2\times 2$ matrix eigenvalue problem. We get in that case

\begin{equation}
\gamma^{\pm}_{\rm total}= \gamma^{\rm break} \frac{\delta\omega^{\pm}-\delta\omega^{+}}{\delta\omega^{\rm break}}\, .
\label{eq:defsgammafull}
\end{equation}

Let us now compute the coefficients $\delta\omega^{\pm\rm, break}$ and $\gamma^{\rm break}$ for each of the terms in \req{eq:EFTofGR}. 

\subsection{Cubic corrections}

\begin{figure}[t!]
	\begin{center}
		\includegraphics[width=0.49\textwidth]{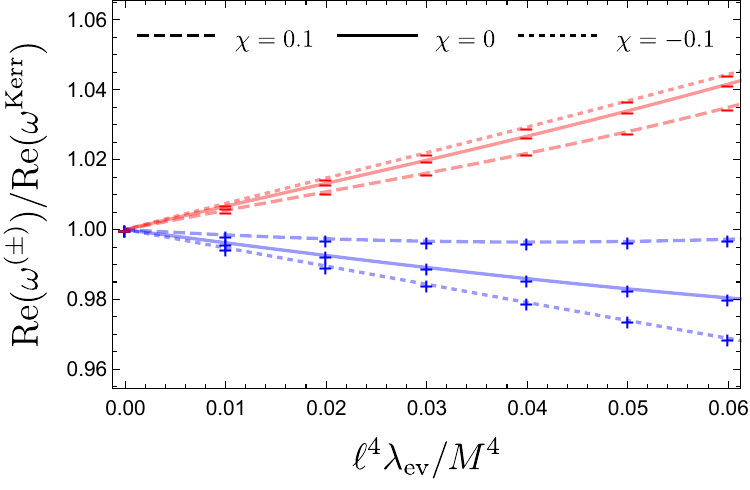}  \includegraphics[width=0.49\textwidth]{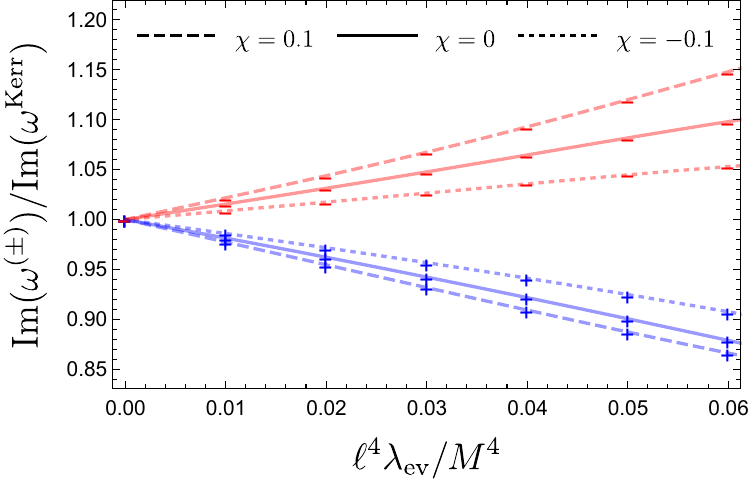} 
		\caption{Fundamental QNM frequencies $l=m=2$ relative to the Kerr values as we increase the cubic coupling $\lambda_{\rm ev}$. Top: Ratio between the real parts of the corrected QNM frequency and of the Kerr one. Bottom: ratio of the imaginary parts. The lines represent a quadratic fit, while the points are numerical data. We observe that $``+"$ and $``-"$ modes are no longer isospectral and that the effect of the spin on the corrections is not negligible.  }
		\label{fig:evomega}
	\end{center}
\end{figure}

In the cubic Lagrangian there is one parity-preserving correction and one parity-violating one. Thus, we have
\begin{align}
\delta\omega^{\pm}&=\lambda_{\rm ev}\frac{\ell^4}{M^5}\delta\omega^{\pm}_{\rm ev}\, ,\\
\delta\omega^{\rm break}&=\lambda_{\rm odd}\frac{\ell^4}{M^5}\delta\omega^{\rm break}_{\rm odd}\, ,
\end{align}
where the factors of $M$ have been introduced so that the coefficients $\delta\omega_{i}$ in the right-hand side are dimensionless. In addition, at first order in the spin one can see that the dependence of these modes on the magnetic number $m$ is always through the combination $m\cdot\chi$. Therefore, we can write
\begin{align}
\delta\omega^{\pm}_{\rm ev}&=\delta\omega^{\pm}_{\rm ev, 0}+m\chi \,\delta\omega^{\pm}_{\rm ev, 1}+\mathcal{O}(\chi^2)\, ,\\
\delta\omega^{\rm break}_{\rm odd}&=\delta\omega^{\rm break}_{\rm odd, 0}+m\chi\, \delta\omega^{\rm break}_{\rm odd, 1}+\mathcal{O}(\chi^2)\, ,
\end{align}
and we only have to determine the $\delta\omega_{0,1}$ coefficients numerically for several values of $l$. As a check, we will also use our numeric results to obtain the Kerr QNM frequencies, that we express as 
\begin{equation}
\omega^{\rm Kerr}=\omega_0+m\chi \,\omega_1+\mathcal{O}(\chi^2)\, .
\end{equation}

On the other hand, the relative normalization factor $\gamma_{\rm odd}$ for the parity-breaking corrections is mildly dependent on the value of $\lambda_{\rm odd}$. However, as a first approximation, we only need to consider its value in the limit $\lambda_{\rm odd}\rightarrow 0$, as the running of this parameter is sensitive to the presence of additional higher-derivative corrections. This value is a function of the spin and therefore we also expand it as

\begin{equation}
\gamma_{\rm odd}=\gamma_{\rm odd, 0}+m\chi\, \gamma_{\rm odd,1}+\mathcal{O}(\chi^2)\, .
\end{equation}

\begin{figure}[t!]
	\begin{center}
		\includegraphics[width=0.49\textwidth]{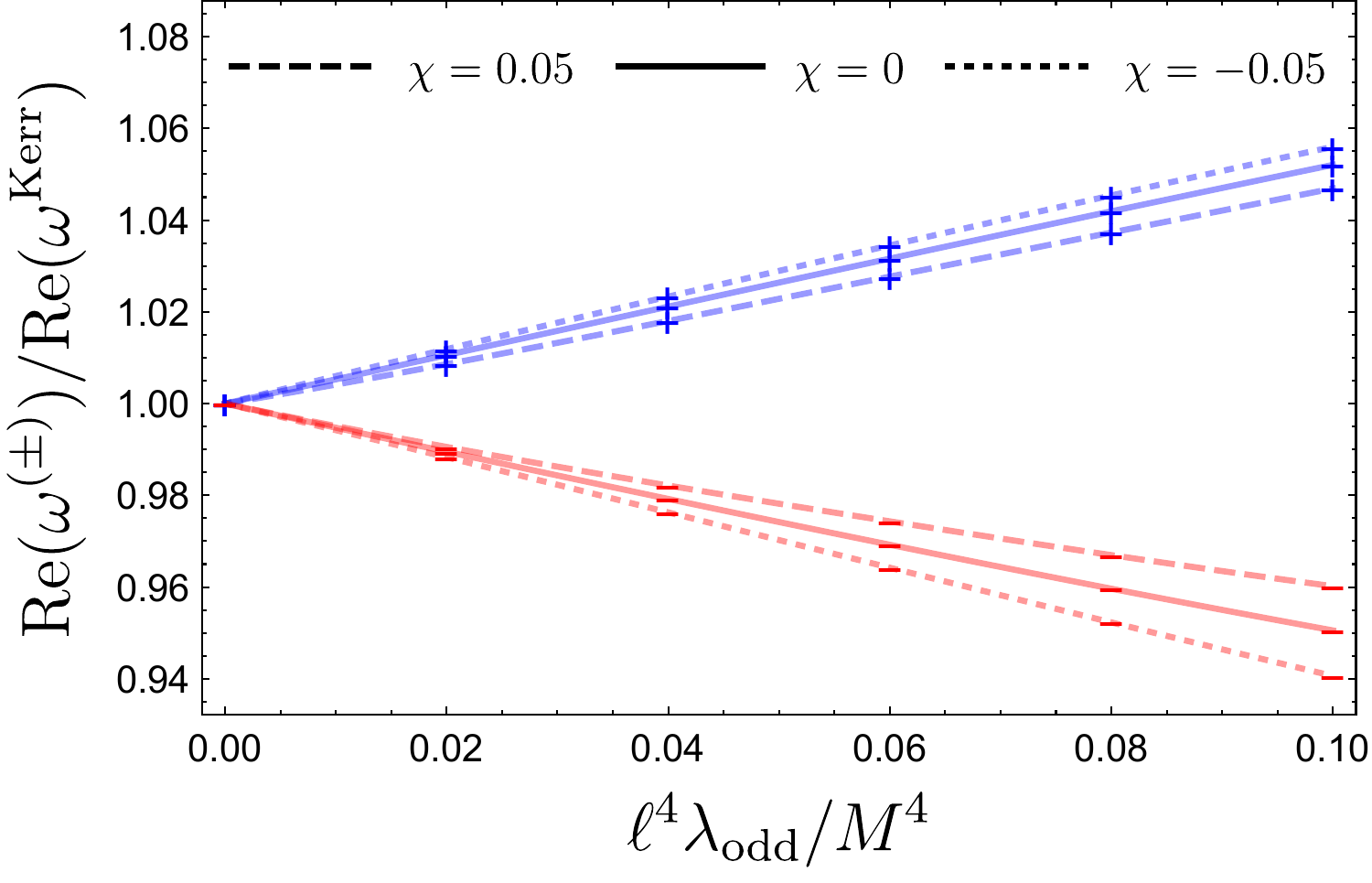}  \includegraphics[width=0.49\textwidth]{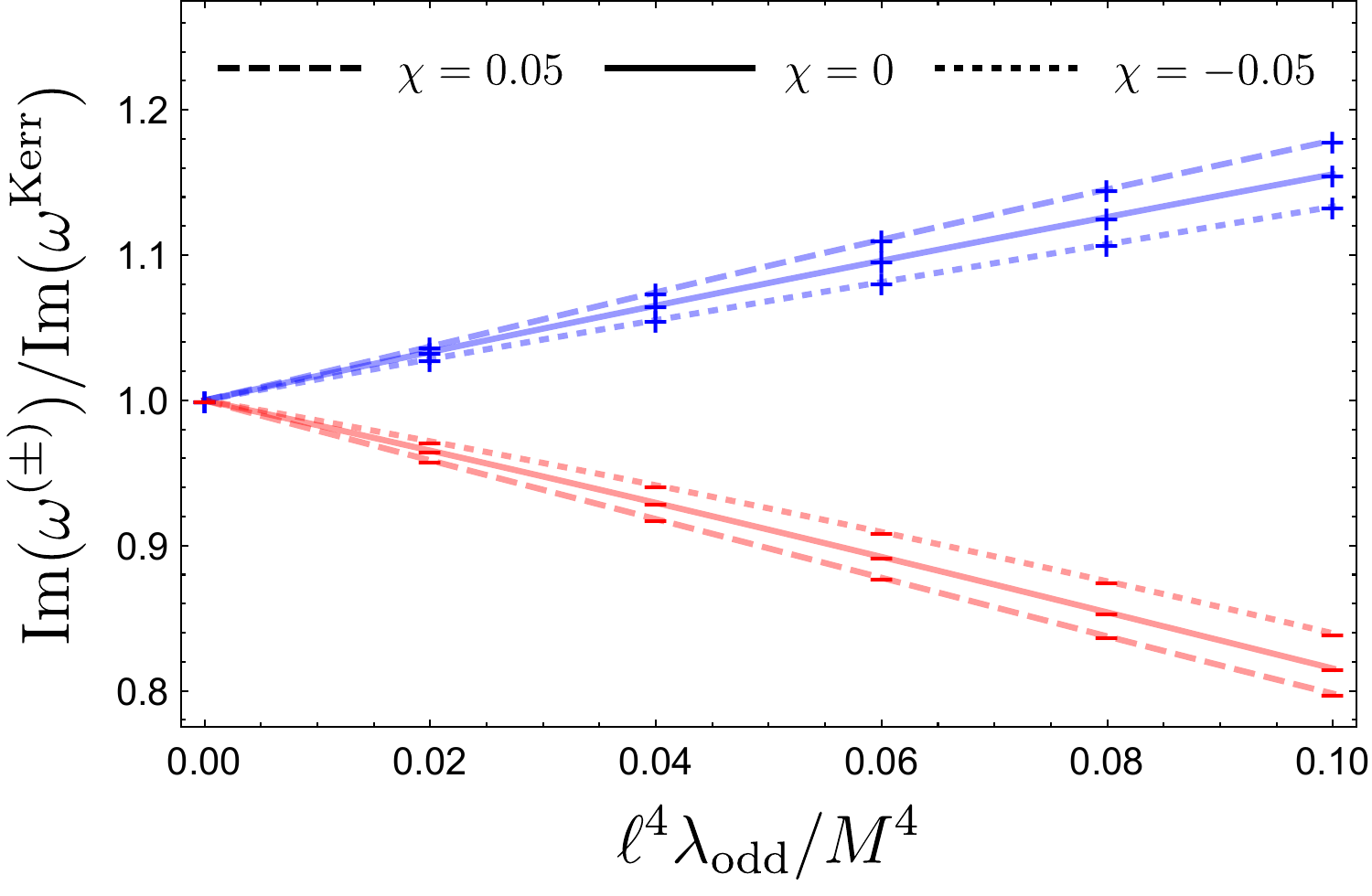} 
		\caption{Fundamental QNM frequencies $l=m=2$ relative to the Kerr values as we increase the cubic coupling $\lambda_{\rm odd}$. Top: Ratio between the real parts of the corrected QNM frequency and of the Kerr one. Bottom: ratio of the imaginary parts. The lines represent a quadratic fit. We use $``+"$ and $``-"$ to label the different branches but these no longer represent defined-parity modes. The rotation $\chi$ is explicitly kept small with respect to the parity-violating term for these numerical results to avoid interference from the loss of isospectrality due to higher order in $\chi$ contributions.}
		\label{fig:oddomega}
	\end{center}
\end{figure}

In order to compute all of these coefficients, we have performed a numerical integration of Eqs.~\req{eq:mastereven}, \req{eq:masterodd} and \req{eq:paritymix1}-\req{eq:paritymix2}. This allows us to obtain the fundamental QNM frequencies for a range of values of $\chi$ and the corresponding coupling, $\lambda_{\rm ev}$ or $\lambda_{\rm odd}$. By fitting the resulting values to a quadratic polynomial in $\chi$ and $\lambda_{\rm ev, odd}$, we are able to extract the Schwarzschild value together with the linear shifts in the spin and in the coupling, as well as the coefficient of $\lambda_{\rm ev, odd}\cdot \chi$. 
In table~\ref{tableev} (in appendix \ref{app:numeric}) we show the best-fit coefficients associated to the $\lambda_{\rm ev}$ correction, while in table~\ref{tableodd} we show the corresponding values for the $\lambda_{\rm odd}$ one, including in that case the value for $\gamma_{\rm odd}$.  Also, in Fig.~\ref{fig:evomega} we show the data points for the corrected QNM frequencies relative to the Kerr ones together with a quadratic fit in the case of $\lambda_{\rm ev}$ corrections, and in  Fig.~\ref{fig:oddomega} we show an analogous plot for the parity-breaking corrections $\lambda_{\rm odd}$.

As a test of our results, we observe that the Kerr QNM frequencies obtained from these fits are in good agreement with the known values.\footnote{There is only some discrepancy in the imaginary part of $\omega_1$ as it is relatively small and therefore harder to capture.} In addition, we have compared our results for non-rotating black holes with those in \cite{deRham:2020ejn} for the $\lambda_{\rm ev}$ correction, and we have found that they agree within $\sim 1\%-5\%$ accuracy.\footnote{For the comparison we take into account that the $\lambda_{\rm ev}$ correction is related to the paramater $d_9$ in \cite{deRham:2020ejn} according to $d_9=\frac{M^2\lambda_{\rm ev}\ell^4}{16\pi G}$, where $M$ is an energy scale used in that reference.} 

We note that the potentials $V^{\pm (\rm ev)}_{lm}$ both scale with $l^2$ when we take $m=l\rightarrow \infty$ and $\omega \propto l$. As a consequence, the eikonal limit is well-defined even when the corrections are present, and thus, the real part of the QNMs scales with $l$ while the imaginary part tends to a constant for $l\rightarrow\infty$. This trend is already captured by the values in table \ref{tableev}. However, these eikonal QNM frequencies will not in general correspond to the result obtained from unstable photon orbits \cite{Cardoso:2008bp}.

Finally, using the results in both tables together with \req{eq:totalshift} we can derive the total shift in the frequencies. For instance, for the $l=2$ modes we get, explicitly,

\begin{widetext}
\begin{align}\notag
\delta\omega^{\pm}=&\frac{\ell^4}{M^5}\bigg[\lambda_{\rm ev}\left(0.0533+0.0255 i+m\chi(0.0812-0.0489 i)\right)\\
&\pm\Big(\left(0.0152-0.0556 i-m\chi (0.155+0.0249 i)\right)\lambda_{\rm ev}^2+\left(0.0139-0.0580 i-m\chi (0.152+0.0086 i)\right)\lambda_{\rm odd}^2\Big)^{1/2}\bigg]\, .
\end{align}
\end{widetext}

\subsection{Quartic corrections}
At quartic order, we have two parity-preserving terms and one parity-breaking one, and the shifts in the QNM frequencies accordingly take the form
\begin{align}
\delta\omega^{\pm}&=\frac{\ell^6}{M^7}\left(\epsilon_{1}\delta\omega^{\pm}_{1}+\epsilon_{2}\delta\omega^{\pm}_{2}\right)\, ,\\
\delta\omega^{\rm break}&=\epsilon_{3}\frac{\ell^6}{M^7}\delta\omega^{\rm break}_{3}\, ,
\end{align}
where  the coefficients $\delta\omega$ in the right-hand-side are dimensionless. Again, at linear order in the spin, these coefficients can be expanded as
\begin{align}
\delta\omega^{\pm}_{1}&=\delta\omega^{\pm}_{1, 0}+m\chi \,\delta\omega^{\pm}_{1, 1}+\mathcal{O}(\chi^2)\, ,\\
\delta\omega^{\pm}_{2}&=\delta\omega^{\pm}_{2, 0}+m\chi \,\delta\omega^{\pm}_{2, 1}+\mathcal{O}(\chi^2)\, ,\\
\delta\omega^{\rm break}_{3}&=\delta\omega^{\rm break}_{3, 0}+m\chi\, \delta\omega^{\rm break}_{3, 1}+\mathcal{O}(\chi^2)\, .
\end{align}
In the same fashion, we express the relative normalization between the $\Psi^{+}$ and $\Psi^{-}$ variables in the presence of $\epsilon_3$ corrections as

\begin{equation}
\gamma_{3}=\gamma_{3, 0}+m\chi\, \gamma_{3,1}+\mathcal{O}(\chi^2)\, .
\end{equation}

The values of these coefficients obtained by a quadratic fit of numerical data are shown in tables \ref{table1}, \ref{table2} and \ref{table3} (in appendix \ref{app:numeric}) for the $\epsilon_1$, $\epsilon_2$ and $\epsilon_3$ corrections, respectively. As a first consistency check, we observe that we recover the Kerr QNM frequencies quite accurately from these fits. In addition, we have checked that our results for the corrections to the QNMs of non-rotating black holes are in good agreement with those in Refs \cite{Cardoso:2018ptl} and \cite{McManus:2019ulj} for the $\epsilon_{1,2}$ and $\epsilon_{3}$ corrections, respectively --- the discrepancies we get are of the order of $1\%-6\%$ depending on the case.\footnote{For the comparison, we have to take into account that the $\epsilon_i$ parameters of reference \cite{Cardoso:2018ptl} compare to our own $\epsilon_i$ according to $(\epsilon_1,\epsilon_2,\epsilon_3)\big|_{\rm there}=-\ell^6/M^6(\epsilon_1,\epsilon_2/4,\epsilon_3/2)\big|_{\rm here}$}
As already noted in the first of those references, the $\epsilon_1$ corrections to the even-parity QNM frequencies and the $\epsilon_2$ corrections to the odd-parity ones scale as $l^2$, so the eikonal limit is ill-defined in those cases (the corrections become dominant with respect to GR). Here, we observe that the same applies to the linear dependence in the spin, which grows fast with $l$. On the other hand, a new feature with respect to the static case is that now the even-parity QNMs do get a correction from $\epsilon_2$. In this case, the $l\rightarrow \infty$ limit seems to be well-behaved. 
Finally, a puzzling observation is that, in the case of $\epsilon_3$ corrections the value for the relative normalization factor $\gamma_{3}$ is essentially coincident with the one for the cubic $\lambda_{\rm odd}$ corrections. Clearly, this cannot be a coincidence and indicates that there is more to be understood on the structure of QNMs with parity-breaking interactions. This will be explored elsewhere.

\section{Discussion}
\label{sec:conclusion}
We have presented a complete calculation of the QNMs of black holes in a general effective theory extension of GR, taking into account the rotation at leading order. The ultimate goal of this program is to predict the corrections to the QNM spectrum of fully rotating black holes, so that one can experimentally search for these deviations from GR. 

A key question that immediately arises is whether the linear approximation is good enough for this purpose. While it is difficult to provide a definitive answer to this, we can see that, in the case of the Kerr QNM frequencies (without corrections), the linear dependence in the spin provides a reasonably good approximation even for moderate values of the rotation $\chi\sim 0.5$. Thus, it would seem plausible to assume that the linear-in-spin approximation for the shifts in the QNM frequencies which we compute, will prove sufficient for initial comparison with future detailed GW ringdown observations from black hole binaries. 

Higher-derivative corrections to GR yield a rich phenomenology for the modified Kerr QNM spectrum. First, we observe that the isospectrality of the even- and odd-parity modes is broken. Furthermore, in the presence of parity-breaking corrections, even - and odd modes get coupled, so that QNMs are composed by a certain linear combination of both types of perturbations. Interestingly, the parity-breaking corrections also introduce a linear shift in the QNM frequencies due to the isospectrality of the Kerr QNMs. When all types of higher-derivative corrections (parity-breaking and parity-preserving) are taken into account, the shift in the QNM frequencies is actually a non-linear function of the higher-order couplings, as evidenced in eq. \req{eq:totalshift}. In addition to these properties, which have already been observed for non-rotating black holes \cite{Cardoso:2018ptl,McManus:2019ulj}, we note the well-known splitting of the different $m$ modes for rotating black holes. The dependence on $m$, both for the Kerr QNM frequencies and for the corrections, always appears through the product $m\cdot \chi$ at linear order in the spin. 

Regarding future directions, it would be interesting to obtain the corrections to the overtone frequencies, as they seem to be relevant for the ringdown signal \cite{Sago:2021gbq,Oshita:2021iyn}. Also, even though the QNMs at leading order in the spin have already been computed for Einstein-dilaton-Gauss-Bonnet theory \cite{Pierini:2021jxd} and for dynamical Chern-Simons gravity  \cite{Wagle:2021tam,Srivastava:2021imr}, it remains important to generalize those results to the more general scalar-tensor theory presented in \cite{Cano:2019ore} so that one has a complete catalogue of QNMs in EFT extensions of GR. The next major goal in this program should be to compute the QNMs for fully-rotating corrected Kerr black holes, but it is unclear at present what is the best strategy towards that objective. Decoupling the perturbation equations should be easier in the Newman-Penrose formalism, but this should be developed for higher-order gravities first. In any case, if decoupling of the equations is achieved, then one should be able to effectively separate the master equation by using the same method as in \cite{Cano:2020cao}.

\textit{Acknowledgements.}  This work is supported by the
C16/16/005 grant of the KU Leuven, the COST Ac-
tion GWverse CA16104, and by the FWO Grant No.
G092617N. The work of PAC is supported by a post-
doctoral fellowship from the FWO (12ZH121N). KF is
Aspirant FWO-Vlaanderen (ZKD4846-ASP/18).

\onecolumngrid
\appendix
\section{Numerical results}\label{app:numeric}
In the tables below we show the Kerr QNM frequencies as well as the shifts associated to each type of higher-derivative correction obtained from fits to numerical data. By applying different numerical approaches, we estimate that the error in the shifts is of the order of $1\%-5\%$ depending on the case, which is also in line with the comparison of these numbers with those in refs.~\cite{deRham:2020ejn,Cardoso:2018ptl,McManus:2019ulj} for non-rotating black holes. Thus, one should probably  keep only  the first two significant digits for the $\delta\omega_{i}^{\pm\rm break}$ coefficients.

\bgroup
\def\arraystretch{1.24}
\setlength{\tabcolsep}{4pt}
\begin{table}[H]
	\centering
	\begin{tabular}{|c||c|c|c|c|c|c|c|c|}
		\hline
		$l$&2&3&4&5\\
		\hline\hline
		$\omega^{+}_{0}$			&  ${\color{black} 0.37367-0.08896 i} $ & ${\color{black}0.59943-0.09270} i$ &${\color{black}0.80917-0.09417 }i$&${\color{black}1.0122}8-{\color{black}0.09487} i$\\ \hline
		$\omega^{+}_{1}$ 			&${\color{black} 0.063}5+{\color{black}0.001}2 i $& ${\color{black}0.067}8+{\color{black}0.0008}9 i$ & ${\color{black}0.07}01+{\color{black}0.000}54 i$& ${\color{black}0.071}5+{\color{black}0.0003}6 i$\\ \hline
		$\delta\omega_{\rm ev, 0}^{+}$ &  ${\color{black}-0.13}7+{\color{black}0.1}61 i$ &${\color{black}-0.25}8+{\color{black}0.151} i$ &${\color{black}-0.35}7+{\color{black}0.149} i$ & ${\color{black}-0.45}2+{\color{black}0.148} i$\\ \hline
		$\delta\omega_{\rm ev, 1}^{+}$ &${\color{black}0.30}6+{\color{black}0.18}8 i$ & ${\color{black}0.19}9+{\color{black}0.14}2 i$  & ${\color{black}0.17}8+{\color{black}0.103} i$& ${\color{black}0.17}2+{\color{black}0.080} i$ \\ \hline\hline
		$\omega^{-}_{0}$&  ${\color{black} 0.37367-0.08896 i}$& ${\color{black} 0.59944-0.09270 }i$ &${\color{black}0.80917-0.09416 i}$&${\color{black}1.0122}9-{\color{black}0.09487} i$\\ \hline
		$\omega^{-}_{1}$ &${\color{black}0.063}0+{\color{black}0.001}1 i$&${\color{black}0.067}7+{\color{black}0.0007}8 i$&${\color{black}0.0702}+{\color{black}0.0004}8 i$&${\color{black}0.071}5+{\color{black}0.0003}3 i$\\ \hline
		$\delta\omega_{\rm ev, 0}^{-}$&  ${\color{black}0.24}4-{\color{black}0.13}0i$ &${\color{black}0.34}0-{\color{black}0.13}2 i$ &${\color{black}0.44}0-{\color{black}0.13}2 i$&${\color{black}0.54}0-{\color{black}0.133} i$\\ \hline
		$\delta\omega_{\rm ev, 1}^{-}$&${\color{black}-0.144}-{\color{black}0.28}6 i$ & ${\color{black}-0.123}-{\color{black}0.14}9 i$ &${\color{black}-0.12}3-{\color{black}0.09}92 i$&${\color{black}-0.12}5-{\color{black}0.074}6 i$ \\ \hline
	\end{tabular}
	\caption{Best-fit coefficients for the fundamental quasinormal frequencies with $\lambda_{\rm ev}$ corrections. In the upper half of the table we show the results for the Kerr QNM frequencies and for the corresponding shifts in the case of even-parity perturbations, and in the lower half we show the results for odd-parity ones.}
	\label{tableev}
\end{table}
\egroup

\bgroup
\def\arraystretch{1.24}
\setlength{\tabcolsep}{4pt}
\begin{table}[H]
	\centering
	\begin{tabular}{|c||c|c|c|c|c|c|c|c|}
		\hline
		$l$&2&3\\
		\hline\hline
		$\omega^{+}_{0}$&  ${\color{black}0.3736}6-{\color{black}0.0889}3 i$ & ${\color{black}0.59944-0.09270} i$ \\ \hline
		$\omega^{+}_{1}$ &${\color{black}0.063}1+{\color{black}0.001}1 i$&${\color{black}0.067}3+{\color{black}0.0006}7 i$\\ \hline
		$\delta\omega_{\rm odd, 0}^{\rm break}$&  ${\color{black}0.19}2-{\color{black}0.15}1 i$ &$0.304-0.144 i$ \\ \hline
		$\delta\omega_{\rm odd, 1}^{\rm break}$&$-0.233-0.207 i$ & $-0.164-0.146 i$  \\ \hline
		$\gamma_{\rm odd, 0}$&  ${\color{black}2.2}1-{\color{black}4.9}7 i$ &$0.705-3.253 i$ \\ \hline
		$\gamma_{\rm odd, 1}$&$-0.15+0.77 i$ & $-0.075+0.345 i$   \\ \hline
	\end{tabular}
	\caption{Best-fit coefficients for the fundamental quasinormal frequencies as well as for the relative normalization factor $\gamma$ with $\lambda_{\rm odd}$ corrections. }
	\label{tableodd}
\end{table}
\egroup

\bgroup
\def\arraystretch{1.24}
\setlength{\tabcolsep}{4pt}
\begin{table}[H]
	\centering
	\begin{tabular}{|c||c|c|c|c|c|c|c|c|}
		\hline
		$l$&2&3&4&5\\
		\hline\hline
		$\omega^{+}_{0}$			&  ${\color{black}0.37367-0.08896 i}$ & ${\color{black}0.59944-0.09270 i}$ &${\color{black}0.80918-0.09416 i}$&${\color{black}1.0122}9-{\color{black}0.09487} i$\\ \hline
		$\omega^{+}_{1}$ 			&${\color{black}0.063}2+{\color{black}0.001}2 i$& ${\color{black}0.067}7+{\color{black}0.0007}4 i$ & ${\color{black}0.070}2+{\color{black}0.0003}9 i$& ${\color{black}0.071}6+{\color{black}0.0002}2 i$\\ \hline
		$\delta\omega_{1, 0}^{+}$ &  ${\color{black}-0.17}2-{\color{black}0.24}5 i$ &${\color{black}-0.65}3-{\color{black}0.59}4i$ &${\color{black}-1.4}5-{\color{black}1.07} i$ & ${\color{black}-2}.71-{\color{black}1.6}5 i$\\ \hline
		$\delta\omega_{1, 1}^{+}$ &${\color{black}-0.35}8-{\color{black}0.2}05 i$ & ${\color{black}-0.60}8-{\color{black}0.2}05 i$  & ${\color{black}-0.9}47-{\color{black}0.2}68 i$& ${\color{black}-1}.35-{\color{black}0.3}32 i$ \\ \hline\hline
		$\omega^{-}_{0}$&  ${\color{black}0.37367-0.08896 i}$& ${\color{black}0.59944-0.09270 i}$ &${\color{black}0.80917-0.09417} i$&${\color{black}1.0122}9-{\color{black}0.09487} i$\\ \hline
		$\omega^{-}_{1}$ &${\color{black}0.063}0+{\color{black}0.001}3 i$&${\color{black}0.067}6+{\color{black}0.0007}8 i$&${\color{black}0.070}1+{\color{black}0.0004}3 i$&${\color{black}0.071}5+{\color{black}0.0002}6 i$\\ \hline
		$\delta\omega_{1, 0}^{-}$&  ${\color{black}-0.07}9-{\color{black}0.05}4i$ &${\color{black}0.0059}-{\color{black}0.03}8 i$ &${\color{black}0.04}0-{\color{black}0.017} i$&${\color{black}0.057}-{\color{black}0.001}8 i$\\ \hline
		$\delta\omega_{1, 1}^{-}$&$-0.096+0.171 i$ & ${\color{black}-0.04}1+{\color{black}0.04}8 i$ &${\color{black}-0.008}2+{\color{black}0.02}0 i$&${\color{black}0.007}7+{\color{black}0.01}3i$ \\ \hline
	\end{tabular}
	\caption{Best-fit coefficients for the fundamental quasinormal frequencies with $\epsilon_{1}$ corrections. In the upper half of the table we show the results for the Kerr QNM frequencies and for the corresponding shifts in the case of even-parity perturbations, and in the lower half we show the results for odd-parity ones.}
	\label{table1}
\end{table}
\egroup

\bgroup
\def\arraystretch{1.24}
\setlength{\tabcolsep}{4pt}
\begin{table}[H]
	\centering
	\begin{tabular}{|c||c|c|c|c|c|c|c|c|}
		\hline
		$l$&2&3&4&5\\
		\hline\hline
		$\omega^{+}_{0}$			&  ${\color{black}0.37367-0.08896 i}$ & ${\color{black}0.59944}-{\color{black}0.09270} i$ &${\color{black}0.80918-0.09416} i$&${\color{black}1.0122}9-{\color{black}0.09487} i$\\ \hline
		$\omega^{+}_{1}$ 			&${\color{black}0.0629}+{\color{black}0.0010} i$& ${\color{black}0.067}4+{\color{black}0.00065} i$ & ${\color{black}0.0698+0.00033} i$& ${\color{black}0.0712}+{\color{black}0.00019} i$\\ \hline
		$\delta\omega_{2, 0}^{+}$ &  ${\color{black}0}$ &${\color{black}0}$ &${\color{black}0}$ & ${\color{black}0}$\\ \hline
		$\delta\omega_{2, 1}^{+}$ &$-0.040+0.205 i$ & ${\color{black}-0.07}9+{\color{black}0.059} i$  & ${\color{black}-0.057+0.018} i$& ${\color{black}-0.041}+{\color{black}0.004}7  i$ \\ \hline\hline
		$\omega^{-}_{0}$&  ${\color{black}0.37367-0.08896 i}$& ${\color{black}0.59944-0.09270 i}$ &${\color{black}0.80917-0.09416} i$&${\color{black}1.0122}9-{\color{black}0.09486} i$\\ \hline
		$\omega^{-}_{1}$ &${\color{black}0.063}0+{\color{black}0.001}2 i$&${\color{black}0.067}7+{\color{black}0.0007}4 i$&${\color{black}0.070}2+{\color{black}0.0003}9 i$&${\color{black}0.071}7+{\color{black}0.0002}1 i$\\ \hline
		$\delta\omega_{2, 0}^{-}$&  ${\color{black}-0.20}7-{\color{black}0.35}4 i$ &${\color{black}-0.6}48-{\color{black}0.70}7 i$ &${\color{black}-1}.45-{\color{black}1.1}6 i$&$-{\color{black}2}.72-{\color{black}1}.73 i$\\ \hline
		$\delta\omega_{2, 1}^{-}$&$-{\color{black}0.27}3-{\color{black}0.4}16 i$ & ${\color{black}-0.5}86-{\color{black}0.3}13 i$ &${\color{black}-0.}943-{\color{black}0.3}32 i$&${\color{black}-1}.37-{\color{black}0.3}68 i$ \\ \hline
	\end{tabular}
	\caption{Best-fit coefficients for the fundamental quasinormal frequencies with $\epsilon_{2}$ corrections. In the upper half of the table we show the results for the Kerr QNM frequencies and for the corresponding shifts in the case of even-parity perturbations, and in the lower half we show the results for odd-parity ones.}
	\label{table2}
\end{table}
\egroup

\bgroup
\def\arraystretch{1.24}
\setlength{\tabcolsep}{4pt}
\begin{table}[H]
	\centering
	\begin{tabular}{|c||c|c|c|c|c|c|c|c|}
		\hline
		$l$&2&3\\
		\hline\hline
		$\omega^{+}_{0}$&  ${\color{black}0.3736}8-{\color{black}0.08896 i}$ & ${\color{black}0.5994}5-{\color{black}0.09269} i$ \\ \hline
		$\omega^{+}_{1}$ &${\color{black}0.063}0+{\color{black}0.001}0 i$&$0.0676+0.00071 i$\\ \hline
		$\delta\omega_{3, 0}^{\rm break}$&  ${\color{black}0.07}55+{\color{black}0.13}2 i$ &$0.332-0.304 i$ \\ \hline
		$\delta\omega_{3, 1}^{\rm break}$&$0.117+0.235 i$ & $0.255+0.130 i$  \\ \hline
		$\gamma_{3, 0}$&  ${\color{black}2.2}1-{\color{black}4.9}7 i$ &$0.705-3.25 i$ \\ \hline
		$\gamma_{3, 1}$&$-0.15+0.76 i$ & $-0.035+0.336 i$   \\ \hline
	\end{tabular}
	\caption{Best-fit coefficients for the fundamental quasinormal frequencies as well as for the relative normalization factor $\gamma$ with $\epsilon_{3}$ corrections. }
	\label{table3}
\end{table}
\egroup

\section{Integrals of associated Legendre functions}
\label{app:integrals}

\noindent
In projecting the Einstein equations onto (tensor) spherical harmonics, the following integrals are useful:

\begin{align}
	\int_{-1}^{1} dy \,  P_l^m(y) P_l^m(y) &= \frac{(l+m)!}{(l+1/2)(l-m)!} \, , \label{eqn:normalization}\\
	\int_{-1}^{1} dy \,  \frac{P_l^m(y) P_l^m(y)}{1-y^2} &= \frac{(l+m)!}{m(l-m)!} \, , \quad m \neq 0 \, , \label{eqn:int2} \\
	\int_{-1}^{1} dy \,  \frac{P_l^m(y) P_l^m(y)}{(1-y^2)^2} &= \frac{(-1+l+l^2+m^2)(l+m)!}{2m(m^2-1)(l-m)!} \, , \quad m \neq -1, 0, 1 \, , \\
	\int_{-1}^{1} dy \,  y P_l^m{}'(y) P_l^m(y) &= - \frac{(l+m)!}{(2l+1)(l-m)!} \, ,  \\
	\int_{-1}^{1} dy \,  \frac{y P_l^m{}'(y) P_l^m(y)}{1-y^2} &= -\frac{l(l+1)(l+m)!}{2 m (m^2-1)(l-m)!} \, , \quad m \neq -1, 0, 1 \, ,  \\
	\int_{-1}^{1} dy \,  P_l^m{}'(y) P_l^m{}'(y) (1-y^2) &=  \frac{(2l(1+l-m)-m)(l+m)!}{(2l+1)(l-m)!} \, ,  \\
	\int_{-1}^{1} dy \,  P_l^m{}'(y) P_l^m{}'(y)  &=  \frac{m(1+l+l^2-m^2)(l+m)!}{2(m^2-1)(l-m)!} \, , \quad m \neq -1, 1 \,  .	
\end{align}

Some of them possess exceptions for small values of $m$, but we find that in those cases the corresponding integral appears multiplied by a quantity that vanishes and therefore one can ignore these exceptions and  apply these integrals as if $m\neq -1, 0, 1$.

\section{Effective potentials}\label{app:potentials}
\allowdisplaybreaks

Here we show the higher-derivative contributions to the effective potentials in equations \req{eq:potentialodd} and \req{eq:potentialeven}, as well as the coupling terms in \req{eq:paritymix1}-\req{eq:paritymix2}.

\subsection{Potentials for odd perturbations}
Introducing $L=l(l+1)$, we have,

\begin{align}\notag
V_{lm}^{-(\rm ev)}=&\frac{102720 M^4}{r^{10}}+\frac{4 (-34609+1754 L)
	M^3}{r^9}-\frac{2 (-30469+3204 L)
	M^2}{r^8}+\frac{5 (-1729+288 L)
	M}{r^7}-\frac{5}{2 r^6}-\frac{25}{2 M
	r^5}\\\notag
&+\frac{5 (-3+2 L)}{4 M^2 r^4}+\left(\frac{752
	M^2}{r^6}-\frac{308
	M}{r^5}-\frac{10}{r^4}-\frac{5}{M r^3}-\frac{5}{2
	M^2 r^2}\right) \omega ^2+m\chi 
\Bigg[\frac{1}{\omega
}\Big(\frac{8640 (-1255+78 L)  M^6}{Lr^{13}}\\\notag
&+\frac{192 \left(91873-9327 L+102
		L^2\right)  M^5}{L r^{12}}-\frac{288
		\left(36403-5576 L+141 L^2\right) M^4}{L
		r^{11}}\\\notag
	&+\frac{144 \left(18552-4172 L+177
		L^2\right)  M^3}{L r^{10}}-\frac{5040
		\left(48-16 L+L^2\right) M^2}{L r^9}-\frac{420
		 M}{L r^8}+\frac{180}{L r^7}\Big)\\
	 &+\left(\frac{16 (1008-373 L)  M^4}{L
	r^9}+\frac{16 (-1584+167 L) M^3}{L
	r^8}+\frac{40 (216+L) M^2}{L r^7}+\frac{20
	M}{r^6}+\frac{10}{r^5}\right) \omega \Bigg]\, ,
\end{align}

\begin{align}\notag
V_{lm}^{-(1)}=&-\frac{7502976 M^5}{r^{13}}-\frac{64 (-162809+1090
	L) M^4}{r^{12}}+\frac{32 (-150773+2232 L)
	M^3}{r^{11}}-\frac{16 (-46651+1152 L)
	M^2}{r^{10}}-\frac{40
	M}{r^9}\\\notag
&-\frac{20}{r^8}-\frac{10}{M
	r^7}-\frac{5}{M^2 r^6}-\frac{25}{M^3 r^5}+\frac{5
	(-3+2 L)}{2 M^4 r^4}+\Bigg(-\frac{27264
	M^3}{r^9}+\frac{15808 M^2}{r^8}-\frac{160
	M}{r^7}-\frac{80}{r^6}-\frac{40}{M
	r^5}-\frac{20}{M^2 r^4}\\\notag
&-\frac{10}{M^3
	r^3}-\frac{5}{M^4 r^2}\Bigg) \omega ^2+m \chi 
\Bigg(\frac{-\frac{11688960 M^7}{L
		r^{16}}+\frac{16229376 M^6}{L
		r^{15}}-\frac{7511040 M^5}{L
		r^{14}}+\frac{1161216 M^4}{L r^{13}}-\frac{840}{L
		M r^8}+\frac{360}{L M^2 r^7}}{\omega}\\\notag
&+\bigg(\frac{256 (-216+545 L) M^5}{L
	r^{12}}-\frac{256 (-1188+3419 L) M^4}{11 L
	r^{11}}+\frac{640 M^3}{r^{10}}+\frac{320
	M^2}{r^9}+\frac{160
	M}{r^8}+\frac{80}{r^7}+\frac{40}{M
	r^6}\\
&+\frac{20}{M^2 r^5}\bigg) \omega \Bigg)\, ,
\end{align}

\begin{align}\notag
V_{lm}^{-(2)}=&\frac{2304 (-2+L) L M^3}{r^{11}}-\frac{1152 (-2+L) L
	M^2}{r^{10}}+m \chi  \Bigg(\frac{1}{\omega }\bigg(-\frac{594542592
		M^7}{L r^{16}}-\frac{110592 (-9219+88 L) M^6}{L
		r^{15}}\\\notag
	&+\frac{165888 (-3899+76 L) M^5}{L
		r^{14}}-\frac{18432 (-9729+290 L) M^4}{L
		r^{13}}+\frac{184320 (-99+4 L) M^3}{L
		r^{12}}\bigg)\\
	&+\left(\frac{4644864 M^5}{L
	r^{12}}-\frac{9216 \left(5544-19 L+11 L^2\right)
	M^4}{11 L r^{11}}+\frac{4608 \left(252-2
	L+L^2\right) M^3}{L r^{10}}\right) \omega \Bigg)\, .
\end{align}

\subsection{Potentials for even perturbations}

Writing

\begin{equation}
	V_{lm}^{+(\rm ev)}= V_{0,lm}^{+(\rm ev)} + m \chi V_{1,lm}^{+(\rm ev)} \, ,
\end{equation}

\begin{equation}
	V_{lm}^{+(1)}= V_{0,lm}^{+(1)} + m \chi V_{1,lm}^{+(1)} \, ,
\end{equation}
and $\Lambda=(l-1)(l+2)+6M/r$,

we have

\begin{align}\notag
		V_{0,lm}^{+(\rm ev)}=&\frac{1}{4 L M^2 r^{15} \Lambda^4}\bigg(8364238848 M^{11}+248832 (-92541+19150 L)
			M^{10} r+20736 \left(1308448-548805 L\right.\\\notag
			&\left.+42964
			L^2\right) M^9 r^2+3456
			\left(-5132084+3258299 L-516962 L^2+13188
			L^3\right) M^8 r^3-1728
			\left(-4019168\right.\\\notag
			&\left.+3414047 L-812084 L^2+37754
			L^3+2108 L^4\right) M^7 r^4-96
			\left(16952832-17894441 L+5520092 L^2\right.\\\notag
			&\left.-278082
			L^3-73184 L^4+4736 L^5\right) M^6 r^5-16
			\left(-13181184+16225522 L-5626732 L^2-92978
			L^3\right.\\\notag
			&\left.+321100 L^4-37831 L^5+766 L^6\right) M^5
			r^6+8 \left(-1451520+1816690 L-360884
			L^2-390702 L^3+210364 L^4\right.\\\notag
			&\left.-33683 L^5+1476
			L^6\right) M^4 r^7-20 L \left(-17072+32584
			L-25860 L^2+10372 L^3-2015 L^4+144
			L^5\right) M^3 r^8\\\notag
			&-10 (-2+L)^2 L
			\left(592-172 L+L^2\right) M^2 r^9+10
			(-2+L)^3 L (-50+19 L) M r^{10}+5 (-2+L)^4 L
			(-3+2 L) r^{11}\bigg)\\\notag
			&+\frac{\omega ^2}{2 L M^2 r^9
				\Lambda^2}	\left(1693440 M^7+1152
			(-1899+341 L) M^6 r+96 \left(9864-3811 L+188
			L^2\right) M^5 r^2\right.\\\notag
			&\left.+16 \left(-8640+5575 L-838
			L^2+94 L^3\right) M^4 r^3-8 L \left(293-278
			L+77 L^2\right) M^3 r^4-20 L (1+L)^2 M^2
			r^5\right.\\
			&\left.-10 L \left(-8+2 L+L^2\right) M r^6-5
			(-2+L)^2 L r^7\right)\, ,
\end{align}

\begin{align}\notag
	V_{1,lm}^{+(\rm ev)}=& \frac{2 \omega }{L^2
		r^{16} \Lambda^5} \left(-12546358272 M^{11}-124416
		(-294577+70217 L) M^{10} r-62208
		\left(737286-340861 L\right.\right.\\\notag
		&\left.\left.+32442 L^2\right) M^9
		r^2-20736 \left(-1533921+1019412 L-181163
		L^2+8146 L^3\right) M^8 r^3\right.\\\notag
		&\left.+3456
		\left(-3808590+3153553 L-718088 L^2+34364
		L^3+119 L^4\right) M^7 r^4+288
		\left(11234736-10128768 L\right.\right.\\\notag
		&\left.\left.+1741658 L^2+487218
		L^3-123315 L^4+4142 L^5\right) M^6 r^5+48
		\left(-9046944+6195408 L+3521758 L^2\right.\right.\\\notag
		&\left.\left.-4029866
		L^3+1125606 L^4-103781 L^5+1915 L^6\right)
		M^5 r^6+8 \left(3027456+3521826 L-13061557
		L^2\right.\right.\\\notag
		&\left.\left.+10777762 L^3-3822484 L^4+602090
		L^5-34424 L^6+383 L^7\right) M^4 r^7-4 L
		\left(2064888-4682494 L\right.\right.\\\notag
		&\left.\left.+4215145 L^2-1895140
		L^3+435860 L^4-45914 L^5+1502 L^6\right) M^3
		r^8+10 (-2+L)^2 L \left(9864\right.\right.\\\notag
		&\left.\left.-17311 L+10452
		L^2-2574 L^3+218 L^4\right) M^2 r^9+5
		(-2+L)^3 L \left(-156-19 L+22 L^2+2
		L^3\right) M r^{10}.\right.\\\notag
		&\left.+5 (-2+L)^4 L \left(-9-2
		L+L^2\right) r^{11}\right)\\
		&-\frac{3456 M^4 (2
		M-r) \omega ^3}{L^2 r^10 \Lambda^3} \left(1470 M^2+(-1262+343 L) M r-144
		(-2+L) r^2\right)\, ,
\end{align}

\begin{align}\notag
		V_{0,lm}^{+(1)}&=\frac{1}{2 L M^4 r^{18} \Lambda^4}\big(355663872000 M^{14}+331776
			(-3019888+597283 L) M^{13} r+55296
			\left(21835984-8823125 L\right.\\\notag
			&\left.+685122 L^2\right)
			M^{12} r^2+9216 \left(-87570496+54127539
			L-8660268 L^2+296918 L^3\right) M^{11}
			r^3+4608 \left(70120960\right.\\\notag
			&\left.-58846013 L+14509004
			L^2-1028218 L^3+9732 L^4\right) M^{10}
			r^4-768 \left(100874496-107609577 L+36206620
			L^2\right.\\\notag
			&\left.-3932118 L^3+56604 L^4+2179 L^5\right)
			M^9 r^5-128 \left(-80416512+104530213
			L-44699404 L^2+6417646 L^3\right.\\\notag
			&\left.-23804 L^4-27265
			L^5+1042 L^6\right) M^8 r^6+64
			\left(-9123840+13984507 L-7132820
			L^2+1105890 L^3+112684 L^4\right.\\\notag
			&\left.-34901 L^5+2232
			L^6+72 L^7\right) M^7 r^7-32 L
			\left(-27643+83732 L-97122 L^2+54164
			L^3-14203 L^4+1200 L^5\right.\\\notag
			&\left.+72 L^6\right) M^6
			r^8-80 L \left(-566+4 L+6 L^2+4
			L^3+L^4\right) M^5 r^9-40 L \left(2134-860
			L+6 L^2+4 L^3+L^4\right) M^4 r^{10}\\\notag
			&-20 L
			\left(-3248+2632 L-516 L^2+4 L^3+L^4\right)
			M^3 r^{11}-10 (-2+L)^2 L \left(592-172
			L+L^2\right) M^2 r^{12}\\\notag
			&+10 (-2+L)^3 L
			(-50+19 L) M r^{13}+5 (-2+L)^4 L (-3+2 L)
			r^{14}\big)-\frac{\omega ^2}{L M^4 r^{12}\Lambda^2}\big(-39518208 M^{10}\\\notag
			&-23040
			(-2240+279 L) M^9 r+2304 \left(-9728+2389
			L+190 L^2\right) M^8 r^2+384 \left(8448-2425
			L-1146 L^2\right.\\\notag
			&\left.+127 L^3\right) M^7 r^3-320 L
			\left(335-338 L+83 L^2\right) M^6 r^4+160 L
			(1+L)^2 M^5 r^5+80 L (1+L)^2 M^4 r^6\\
			&+40 L
			(1+L)^2 M^3 r^7+20 L (1+L)^2 M^2 r^8+10 L
			\left(-8+2 L+L^2\right) M r^9+5 (-2+L)^2 L
			r^{10}\big)\, ,
\end{align}

\begin{align}\notag
		V_{1,lm}^{+(1)}=&-\frac{4\omega }{11 L^2
			M^2 r^{19} \Lambda^5} \big(5868453888000 M^{14}+29196288
			(-592694+127903 L) M^{13} r+3649536
			\left(5992680\right.\\\notag
			&\left.-2584481 L+216607 L^2\right)
			M^{12} r^2+165888 \left(-92491168+59683272
			L-10024141 L^2+355036 L^3\right) M^{11}
			r^3\\\notag
			&-4608 \left(-1398467136+1196999584
			L-298889214 L^2+19870074 L^3+97405
			L^4\right) M^{10} r^4-2304
			\left(701931648\right.\\\notag
			&\left.-742219456 L+238883894
			L^2-19356366 L^3-1443058 L^4+153857
			L^5\right) M^9 r^5-384
			\left(-583159104\right.\\\notag
			&\left.+714095360 L-254062790
			L^2+4525850 L^3+11488868 L^4-1663586
			L^5+69707 L^6\right) M^8 r^6-64
			\left(205324416\right.\\\notag
			&\left.-242618112 L+25243342
			L^2+72914198 L^3-36477068 L^4+6504172
			L^5-455692 L^6+3091 L^7\right) M^7 r^7\\\notag
			&+64 L
			\left(9989760-22242871 L+19551757
			L^2-8465638 L^3+1803986 L^4-152627 L^5-673
			L^6+396 L^7\right) M^6 r^8\\\notag
			&-352 L
			\left(96768-245755 L+250393 L^2-127630
			L^3+32210 L^4-2759 L^5-235 L^6+36 L^7\right)
			M^5 r^9\\\notag
			&-440 L \left(-486-241 L+10 L^2+20
			L^3+20 L^4+10 L^5+2 L^6\right) M^4
			r^{10}-220 L \left(1512-538 L-341 L^2\right.\\\notag
			&\left.+20
			L^3+20 L^4+10 L^5+2 L^6\right) M^3
			r^{11}-110 (-2+L)^2 L \left(-504-31 L+84
			L^2+18 L^3+2 L^4\right) M^2 r^{12}\\\notag
			&-55
			(-2+L)^3 L \left(-156-19 L+22 L^2+2
			L^3\right) M r^{13}-55 (-2+L)^4 L \left(-9-2
			L+L^2\right) r^{14}\big)\\\notag
			& +\frac{3072 M^3
			(2 M-r)\omega ^3}{L^2 r^{13} \Lambda^3} \left(-77184 M^4-48 (-1363+335 L)
			M^3 r+216 \left(-66+29 L+2 L^2\right) M^2
			r^2\right.\\
			&\left.+126 (-2+L)^2 L M r^3+7 (-2+L)^3 L
			r^4\right)  \, ,
\end{align}

\begin{align}\notag
		V_{lm}^{+(2)}&=\frac{27648 \chi m \omega M^3}{11 L r^{15} \Lambda^3} \big(487872 M^5+72
			(-13860+3853 L) M^4 r+12 \left(68200-38804
			L+4431 L^2\right) M^3 r^2\\
			&+6
			\left(-56144+48896 L-11562 L^2+575
			L^3\right) M^2 r^3+(-2+L)^2 \left(17424-3170
			L+L^2\right) M r^4+726 (-2+L)^3 r^5\big)\, .	
\end{align}

\subsection{Coupled terms for parity-breaking interactions}

Again with

\begin{equation}
U_{lm}^{+(\rm odd)}= U_{0,lm}^{+(\rm odd)} + m \chi U_{1,lm}^{+(\rm odd)} \, ,
\end{equation}

\begin{equation}
W_{lm}^{+(\rm odd)}= W_{0,lm}^{+(\rm odd)} + m \chi W_{1,lm}^{+(\rm odd)} \, ,
\end{equation}

and similarly for the $U_{lm}^{-(\rm odd)}$, $W_{lm}^{-(\rm odd)}$  and the other higher-derivative couplings.

\begin{align}\notag
&U_{0,lm}^{+(\rm odd)}= \frac{i}{7 L M^2 \omega^2 r^{16}\Lambda^3} \Bigg(  \big(-4032 L^4+24192  L^3-48384 L^2+32256 L\big) M^3\omega^3 r^{11} +  \big(8064  L^4  -120960  L^3 +387072  L^2 \\\notag & -354816  L \big) M^4\omega^3 r^{10}  + \big(-10080 
L^5-30240  L^4 +(145152 \omega ^2 M^2  + 423360  
) L^3 +\left(-1016064  \omega ^2 M^2-1008000\right)
L^2 \\\notag& +\left(1451520  \omega ^2 M^2+725760 \right) L \big) M^3 	\omega r^9  + 
\big(44352  L^5 +258048  L^4 -3737664  L^3 +\left(870912  \omega
^2 M^2+10434816 \right) L^2 \\\notag&+\left(-2612736  \omega ^2 M^2-8692992 \right) L \big)	M^4 \omega r^8+ \big(-48384 L^5 -677376  L^4 
+12289536  L^3  -42094080  L^2 \\\notag&+L \left(1741824 \omega ^2
M^2+41223168 \right)\big) \omega  M^5  r^7 +  \big(564480  L^4-17902080 
L^3  +81527040 L^2 -95090688  L \big)
\omega M^6	r^6 \\&+ \big(9773568 L^3 -74221056 L^2+100638720  L  \big) \omega  M^7  r^5 +  \big(24385536  L^2  -24385536  L \big)
M^8 \omega	r^4 -20901888 L M^9 \omega  r^3 \Bigg) \, ,
\end{align}

\begin{align}\notag
&U_{1,lm}^{+(\rm odd)}=  \frac{i}{7 L M^2 \omega^2 r^{16}\Lambda^3} \Bigg(  \left(72  L^3-216  L^2+288   \right) \omega ^2 r^{13} + \left(288   L^3-648  
L^2 +1080   L-1872  \right)M \omega ^2  r^{12}
\\\notag&+(864  L^3 -432   L^2 -3456   L+5616  ) M^2 \omega^2
r^{11}+ \big(2304  L^3  \omega ^2 M^2+864 L^2  \omega ^2 M^2-5184 L  \omega ^2
M^2-3744   \omega ^2 M^2\big)M  r^{10}\\\notag&+\left(60480 L^4  \omega ^2 M^2-296640  L^3
\omega ^2 M^2+368064  L^2  \omega ^2 M^2+235008 L  \omega ^2 M^2-490176  
\omega ^2 M^2\right)  M^2 r^9\\\notag&+ \big(-169344  L^4  \omega ^2 M^2+1235520  L^3  \omega ^2
M^2-611712  L^2 \omega ^2 M^2-6732288 L  \omega ^2 M^2+8892288   \omega ^2
M^2\big) M^3  r^8 \\\notag&+\big(96768 L^4  \omega ^2 M^2-1040256 L^3  \omega ^2 M^2-9991296 
L^2 \omega ^2 M^2+49932288 L  \omega ^2 M^2-52690176  \omega ^2 M^2\big)
M^4 r^7 \\\notag&+  \big(143672832  \omega ^2 M^2-11612160 +L^3 \left(483840 -580608 
M^2 \omega ^2\right)+L^2 \left(37794816  M^2 \omega ^2-4838400 \right) +L
(13547520 \\\notag&-159349248  M^2 \omega ^2)\big) M^5 r^6 +
\big(-188227584 
\omega ^2 M^2-3290112  L^3+155409408 +L^2 \left(41852160  
-36481536   M^2 \omega^2\right) \\\notag&+L \left(238118400 i  M^8 \omega ^2-148248576 \right)\big) M^6 r^5+ \big(96546816 \omega ^2 M^2+7354368  L^3 -136055808
L^2 -866073600 \\\notag&+L \left(649603584 -137604096 M^2 \omega
^2\right)\big)  M^7 r^4 + \big(-5419008  L^3  +196826112  L^2 -1423070208 L 
+2572480512 \big)  M^8 r^3 \\&+  \big(-106831872 L^2 +1557190656 L 
-4293015552 \big)M^9 r^2 +\big(3813433344   -680472576  L 
\big) M^{10}  r-1407393792   M^{11} \Bigg) \, ,
\end{align}

\begin{align}\notag
&W_{0,lm}^{+(\rm odd)}= \frac{i}{7 L M^2 \omega r^{14}\Lambda^3} \bigg( -10450944 L r^3  M^8+  M^3  r^8 \left(-50400   L^4+302400  
L^3-604800  L^2+403200  L\right) \\\notag&+  M^4 r^7 \left(237888 L^4 -1995840  L^3 +5128704  L^2-4177152  L\right)\\\notag &+  M^5 r^6 \left(-274176  L^4 +4451328 L^3 -16257024  L^2+16902144  L\right)\\&+ 
M^6 r^5 \left(-3338496  L^3+22788864  L^2 -33094656  L \right)+M^7  r^4 \left(30772224  L
-11902464  L^2 \right) \bigg) \, ,
\end{align}

\begin{align}\notag
&W_{1,lm}^{+(\rm odd)}= \frac{i}{7 L M^2 \omega^2 r^{14}\Lambda^3} \Bigg(  \big(72 ^2 L^3-432
L^2+864  L-576  \big) \omega ^2 r^{12}+\big(288  L^3-432  L^2-1728  L+2880 \big)  M \omega ^2
r^{11} \\\notag&+\big(864 L^3-2592  L-1728  M^2 \big)  M^2 \omega ^2 r^{10}+\big(2304 L^3 +1728 L^2 -3456  L 
-2880  \big) \omega ^2 M^3 r^9 \\\notag&+\big(146880 L^3 -839808  L^2  +1689984  L -1133568 \big)   \omega ^2
M^4 r^8 \\\notag &+\big(-461952  L^3 +5367168  L^2 -15676416  L +13766400  \big) \omega ^2 M^5 r^7\\\notag&+\big(290304 
L^3 -10540800  L^2  +49171968  L  -58005504
\big)  \omega ^2 M^6 r^6 \\\notag&+\bigg(1680  L^3-  \left(-1451520  \omega ^2
M^2-94080 \right)-  L^2 \left(-88704  \omega ^2 M^2-16800  \right)-  L
\left(834048  \omega ^2 M^2+87360  \right)\bigg)72  M^5 r^5 \\\notag&+\bigg(-532224  L^3 
-8225280  L^2 -72 L \left(-306432  \omega ^2 M^2-827904 \right)-72 
\left(958464  \omega ^2 M^2+1139712  \right)\bigg) M^6 r^4\\\notag&+\left(580608  L^3 
+18289152  L^2 -212212224  L  +398877696   \right) M^7
r^3 \\\notag&+\left(-13353984  L^2 +335010816  L -971937792  \right)
M^8	r^2 \\&+\left(1185601536  -197406720 L \right)  M^9 r-578285568   M^{10} \Bigg) \, ,
\end{align}

\begin{align}\notag
&U_{0,lm}^{-(\rm odd)}= \frac{i}{14 L M^2 r^{16}\Lambda^3} \bigg( \big(-2016   L^4+12096   L^3-24192   L^2+16128 
L\big) M^3 \omega ^3 r^{11} \\\notag&+\big(4032  L^4 -60480 L^3 +193536  L^2 -177408  L \big) \omega ^3 M^4 r^{10} \\\notag&+\big(1120   L^4-10080    L^3+  \left(8064 \omega ^2 M^2+33600  
\right) L^2+ \left(-56448 \omega ^2 M^2-49280 \right) L+
\left(80640 \omega ^2 M^2+26880 \right) \big) 9 L M^3
\omega r^9 \\\notag&+\big(-4816   L^4+72352   L^3-333312   L^2+ \left(48384 \omega ^2
M^2+619136 \right) L+ \left(-145152 \omega ^2 M^2-406784 \right) \big) 9 L M^4
\omega  r^8 \\\notag&+\big(46368  L^5 -1538208  L^4 +11164608
L^3 -28119168 L^2 +9  L \left(96768 \omega ^2 M^2+2571520
\right)\big) \omega  M^5 r^7 \\\notag&+\big(1197504 L^4 -18398016  L^3 +71197056  L^2 -78382080  L \big) \omega  M^6 r^6 \\&+\big(11297664  L^3
-90260352  L^2+150135552  L \big) \omega  M^7
r^5+\big(45722880  L^2  -154151424  L  \big) M^8 \omega r^4+66189312  L
M^9 \omega  r^3 \bigg) \, ,
\end{align}

\begin{align}\notag
&U_{1,lm}^{-(\rm odd)}= \frac{i}{14 L M^2 r^{16}\Lambda^3} \bigg( \big(54   L^3-324  L^2+648 
L-432 \big) \omega ^2 r^{13}+\big(216  L^3-108  L^2-2160  L+3024  \big) M \omega ^2 r^{12} \\\notag&+\big(-27 
L^4+9  \left(72  M^2 \omega ^2+18 \right) L^3+9  \left(48  M^2 \omega ^2-36
\right) L^2+9  \left(24 -120  M^2 \omega ^2\right) L-4752 M^2 \omega ^2\big)
r^{11} \\\notag&+\big(-27  L^4+9  \left(192  M^2 \omega ^2-6  \right) L^3+9  \left(240
\omega ^2 M^2+96  \right) L^2+9\left(-96 \omega ^2 M^2-216  \right) L+9 
\left(144  -144  M^2 \omega ^2\right)\big) M r^{10} \\\notag&+\big(-6  L^4+ 
\left(13920  \omega ^2 M^2+6  \right) L^3+ \left(-79872  \omega ^2 M^2-12 
\right) L^2+  \left(161376  \omega ^2 M^2+312 \right) L \\\notag&+ \left(-107712 
\omega ^2 M^2-528  \right)\big) 9 M^2 r^9+\big(  \left(2016  M^2 \omega ^2-12 
\right) L^4+  \left(12  -69184  M^2 \omega ^2\right) L^3+  \left(641856 
\omega ^2 M^2+12  \right) L^2 \\\notag&+  \left(-1778688  \omega ^2 M^2-168  \right)
L+\left(1530176  \omega ^2 M^2+384  \right)\big) 9 M^3 r^8+\big(-3920 
L^5+ \left(23496  -4032  M^2 \omega ^2\right) L^4 \\\notag&+  \left(115136  M^2 \omega
^2-47016  \right) L^3+  \left(31384 -1674624  M^2 \omega ^2\right) L^2+ 
\left(7211904  M^2 \omega ^2-120 \right) L \\\notag&+  \left(-8533504  \omega ^2 M^2-96 
\right)\big)  M^4 r^7+\big(19824  L^5  - 196720 L^4 +  L \left(-12882432
\omega ^2 M^2-1163248  \right) \\\notag&+  L^3 \left(524208  -72576  M^2 \omega
^2\right)+  L^2 \left(1665408  M^2 \omega ^2-104784  \right)+  \left(23188224 
\omega ^2 M^2+1064256  \right)\big) 9 M^5 r^6 \\\notag&+\big(-31584  L^5  +619264  L^4 
-2557536  L^3 +  \left(-30618624  \omega ^2 M^2-15934848  \right)+ 
L^2 \left(-489216  \omega ^2 M^2-174624  \right) \\\notag&+  L \left(9313536  \omega ^2
M^2+14097184  \right)\big)9 M^6 r^5+\big(15232 L^5 -860416 L^4
+6557952 L^3 -1055168 L^2 \\\notag&+L \left(-1451520 \omega ^2
M^2-67066048 \right)+  \left(15731712  \omega ^2 M^2+99164928 
\right)\big) 9 M^7 r^4 \\\notag&+\big(3991680 L^4  -77379840  L^3 +106507008 L^2 
+1391078016  L -2957893632 \big) M^8 r^3 \\\notag&+\big(40787712 L^3 
-240786432 L^2 -1485008640 L +5504454144  \big)
M^9 r^2 \\&+\big(168666624  L^2 +425710080 L -5447181312  \big)
M^{10} r+228178944  L  M^{11}+2236502016  M^{11}\bigg) \, ,
\end{align}

\begin{align}\notag
&W_{0,lm}^{-(\rm odd)}= \frac{i}{7 L M^2 r^{13}\Lambda^2} \bigg( \left(1233792  L  -217728  L^2\right)\omega M^6 r^4  + M^5
\omega r^5 \left(-16128  L^3  +306432  L^2  -766080  L \right) \\&+M^4
\omega r^6 \left(18144  L^3  -139104  L^2 +205632  L  \right)+ M^3
\omega r^7 \left(-5040  L^3  +20160  L^2  -20160  L  \right)-725760  L M^7 r^3 \omega \bigg) \, ,
\end{align}

\begin{align}\notag
&W_{1,lm}^{-(\rm odd)}= \frac{i}{7 L M^2 r^{13}\Lambda^2} \bigg( -7257600  L  M^9+102332160 	M^9+  M^8 r
\big(-2080512  L^2  +61081344 L -196483968  \big) \\\notag&+M r^8 \big(-9 
L^2 \left(64  M^2 \omega ^2-2  \right)-9  L \left(32  M^2 \omega ^2+14  \right)-9
\left(-32  M^2 \omega ^2-20  \right)\big)\\\notag&+r^9 \big(-9  L^2 \left(24 M^2
\omega ^2+3 \right)+108  L -9  \left(12 -24  M^2 \omega ^2\right)\big) \\\notag&+ M^3 r^6
\big(-9  L^2 \left(-3200 M^2 \omega ^2-24 \right)-9 L \left(29504 M^2
\omega ^2-48 \right)-9  \left(-43904  M^2 \omega ^2-24  \right)\big) \\\notag&+ M^2 r^7
\big(-9  L^2 \left(1280  M^2 \omega ^2-8  \right)-9  L \left(-4352  M^2 \omega
^2-16  \right)-9 \left(4448  M^2 \omega ^2-8  \right)\big) \\\notag&+ M \omega ^2 r^{10} \big(-72
L^2  +72  L  +144  \big)+ \omega ^2 r^{11} \big(-18 
L^2 +72  L  -72  \big) \\\notag&+ M^7 r^2 \big(-145152  L^3 
+9982656  L^2  -85048704  L +150488064  \big) \\\notag&+  M^6 r^3 \big(459648 
L^3  -10280160  L^2 +47552256 L-57629952 \big) \\\notag&+  M^5 r^4
\big(-334656 L^3 +3890304 L^2 -9  L \left(5376  M^2 \omega ^2+1328896 
\right)-9  \left(-125952  M^2 \omega ^2-1227680 \right)\big) \\&+ M^4 r^5 \big(70560
L^3  -493344  L^2  -9  L \left(-43776  M^2 \omega ^2-125568  \right)-9
\left(133632  M^2 \omega ^2+94016 \right)\big)\bigg) \, , 
\end{align}

\begin{align}\notag
&U_{0,lm}^{+(3)}= \frac{i}{5 L M^4 \omega ^2  r^{19}\Lambda^3} \bigg(209018880  L r^3 \omega  M^{12}+  \omega  M^9 r^6 \big(13962240  L^4 -486328320  L^3
+2383395840  L^2 -2950594560  L\big) \\\notag&+ \omega  M^{10} r^5
\big(267356160  L^3 -2344273920 L^2 +3741050880 L
\big) \\\notag&+M^{11} \omega r^4 \big(892477440  L^2  -2063646720 L  \big)+ \omega ^3 M^6 r^{11} \big(134400 L^4 -806400 L^3 +1612800 L^2 -1075200 L \big) \\\notag&+ \omega ^3 M^7 r^{10} \big(-268800
L^4 +3663360 L^3-11427840 L^2 +10352640 L
\big) \\\notag&+  \omega  M^8 r^7 \big(-1059840 L^5-17464320 L^4 +332444160  L^3 -1181952000 L^2 -276480 L 
\left(126 M^2 \omega ^2-4307\right) \big) \\\notag&+  3840 \omega M^7 r^8 \big(3 L^6+266 L^5+1821 L^4-26376 L^3+74900
L^2-63216 L+\omega ^2 M^2\left(15336 L -5400 L^2 \right)\big) \\\notag&+  1920\omega  M^6 r^9 \big(-3 L^6-128 L^5-456 L^4+6048 L^3-14384 L^2+10368 L +\omega ^2 M^2 \left(-2136 L^3 +13944 L^2 -19344 L
\right)\big)\bigg) \, ,
\end{align}

\begin{align}\notag
&U_{1,lm}^{+(3)}= \frac{i}{5 L M^4 \omega ^2  r^{19}\Lambda^3} \bigg( \left(52  
L^3-156   L^2+208  \right) \omega ^2 r^{16}+\big(208 
L^3-468   L^2+780   L-1352\big)  M \omega ^2
r^{15} \\\notag&+\big(624  L^3-312  L^2-2496  
L+4056  \big)  M^2 \omega ^2 r^{14}+\big(1664  L^3 +624  L^2 
-3744  L -2704 \big) \omega ^2 M^3 r^{13} \\\notag&+\big(4160 
L^3  +3744 L^2  -4992  L -4576  \big) \omega ^2 M^4 r^{12}+ \big( \omega ^2 M^2 \left(-560 L^3 +3360 L^2 -6720 L +4480 \right) \\\notag&+ \left(52 L^3+65
L^2-26 L-39\right)\big) 192 \omega ^2  M^5 r^{11}+\big( 23296 L^3+34944 L^2-11648 \\\notag&+\omega ^2 M^2 \left(215040 L^3 -3225600 L^2 +10321920 L -9461760
\right) \big) \omega ^2 M^6
r^{10} \\\notag&+\big(2160 \left(7 L^2-40 L+52\right) M^2 \omega ^2+ \left(90 L^5+840 L^4+18928 L^3-139329
L^2+300078 L-207425\right) \big)  256 \omega ^2 M^7 r^9 \\\notag&+\big(-30 L^5-1005 L^4-15192 L^3+198173 L^2-574268 L+495707 +8640\omega ^2 M^2 \left( L -2
\right) \big) 1536 \omega ^2 M^8 r^8\\\notag&+\big(2227200 L^4 +21657600
L^3 -955763712  L^2  +3874799616 L -4130767872  \big) \omega ^2 M^9 r^7 \\\notag&+\big(65 L^4+290 L^3-12980 L^2+46360 L-44160 +4\omega ^2 M^2 \left(530 L^3 +69467 L^2
-437212 L +590201 \right) \big) 4608 M^8 r^6 \\\notag&+\big(-1198080 L^4-8064000 L^3+474531840 L^2-2249625600 L+2684805120 \\\notag&+\omega ^2 M^2\left(-603832320 L^2 +7800053760 L -14067191808
\right) \big) M^9 r^5 \\\notag&+\big(1198080 L^4+16220160 L^3-1410140160 L^2+9419304960 L-14660812800 \\\notag&+\omega ^2 M^2 \left(7184941056 -2703974400 L
\right) \big)  M^{10}
r^4+\big(-4248 L^3 +726624  L^2 -7654392 L
+1643453  \big) 2560 M^{11} r^3 \\&+\big(-831  L^2 +18308  L
-61942  \big) 1105920 M^{12} r^2+\big(58556252160  -8309882880  L 
\big)  M^{13} r-20676280320  M^{14} \bigg) \, ,
\end{align}

\begin{align}\notag
&W_{0,lm}^{+(3)}=  \frac{i}{5 L M^4 \omega^2  r^{17}\Lambda^3} \bigg( -104509440 L r^3 \omega  M^{11}+\omega  M^6 r^8 \big(30720  L^5 +399360  L^4 -3133440  L^3 +6758400 L^2 -4669440 L\big) \\\notag&+ \omega  M^7 r^7 \big(-61440 L^5 -1616640 L^4 \omega  M^7+17832960 L^3
-49966080 L^2 +42516480 L \big) \\\notag&+ \omega  M^8 r^6 \big(1635840 L^4 -33661440 L^3 +128839680 L^2 -136120320 L
\big) \\&+ \omega  M^9 r^5 \big(21058560 L^3 -130268160 L^2 +158883840 L\big)+ \omega  M^{10} r^4 \big(36771840 L^2 +13547520
L\big) \bigg) \, ,
\end{align}

\begin{align}\notag
&W_{1,lm}^{+(3)}=  \frac{i}{5 L M^4 \omega^2  r^{17}\Lambda^3} \bigg(\big(-52  L^3+312  
L^2-624   L+416  \big)  \omega ^2 r^{15}+\big(-208  
L^3+312   L^2+1248   L-2080 \big) M \omega ^2
r^{14} \\\notag&+\big(-624   L^3+1872  L+1248  
^2\big) M^2 \omega ^2 r^{13}+\big(-1664  L^3  -1248  L^2  +2496 L +2080 \big) \omega ^2 M^3 r^{12} \\\notag&+\big(-4160 L^3 -4992  L^2 +2496  L  +3328  \big)
\omega ^2 M^4	r^{11}+\big(-9984  L^3 -14976  L^2  +4992  \big) \omega ^2 M^5 r^{10} \\\notag&+\big(-23296  L^3  -39936 L^2 -9984 
L  +6656  \big) \omega ^2 M^6 r^9+\big(-120  L^4 +856  L^3  -3075  L^2 +3762 L  -1907  \big) 512 \omega ^2
M^7 r^8 \\\notag&+\big(122880 L^4 +878592 L^3
-10146816 L^2 +23596032 L -17756160 
\big) \omega ^2 M^8 r^7 \\\notag&+\big(-2764800  L^3  +70004736 L^2 -197572608  L  +132212736  \big) \omega ^2 M^9 r^6 \\&+4608 \big(-15 L^4-750 L^3-4820 L^2+28760 L-32000+24 \omega ^2 M^2 \left(-775 L^2 +4078 L -2640
\right) \big)  M^8 r^5 \\\notag&+\big(138240  L^4  +14008320 L^3 +140912640 
L^2 -368640 L \left(903 M^2 \omega ^2+3317\right) +73728  \left(2874 M^2
\omega ^2+23975\right)\big)  M^9 r^4 \\\notag&+ \big(-14192640  L^3 -296386560 L^2
+4228485120 L -8471715840  \big) M^{10} r^3+ 1105920 \big(187  L^2 
-5862 L +18347 \big) M^{11} r^2 \\&+\big(3713679360  L
-24271626240  \big) M^{12} r+11592253440 M^{13}\bigg) \, ,
\end{align}

\begin{align}\notag
&U_{0,lm}^{-(3)}=  \frac{i}{20 L M^4 r^{19}\Lambda^3} \bigg( -975421440  L r^3 \omega  M^{12}+  \omega  M^9 r^6 \big(-17971200 L^4+324725760 L^3
-1370649600 L^2 +1586165760 L \big)\\\notag&+ \omega  M^{10} r^5
\big(-191877120 L^3 +1628467200 L^2-2715863040 L
\big)+M^{11} \omega r^4 \big(2503249920 L  -781332480 L^2  \big) \\\notag&+ \omega ^3 M^6 r^{11} \big(80640 L^4 -483840 L^3 +967680 L^2-645120 L \big)+ 161280 \omega ^3
M^7 r^{10} \big(- L^4 +15 L^3-48 L^2 +44 L \big) \\\notag&+5760 \omega  M^6 r^9 \big( L^6-24 L^5+264 L^4-1088 L^3+1872 L^2-1152 L+504\omega ^2 M^2 \left(- L^3
+7 L^2 -10 L \right) \big) \\\notag&+ \omega  M^8 r^7 \big(-599040 L^5 +23938560 L^4 -206184960 L^3 +582773760 L^2 -184320 L   \left(189
M^2 \omega ^2+2836\right) \big)\\&+ 11520 \omega  M^7 r^8 \big( 1512 \omega ^2 M^2 \left(3 L - L^2\right)-10852 L^2+7952 L- L^6+50 L^5-913 L^4+5072 L^3\big)\bigg) \, ,
\end{align}

\begin{align}\notag
&U_{1,lm}^{-(3)}=  \frac{i}{20 L M^4 r^{19}\Lambda^3} \bigg(\big(78  L^3-468  
L^2+936  L-624  \big) \omega ^2 r^{16}+\big(312  
L^3-156   L^2-3120   L+4368 \big)
M \omega ^2 r^{15} \\\notag&+\big(-39 L^4+234 L^3-468 L^2+\omega ^2 M^2 \left(936 L^3 +624 L^2 -1560 L -6864 \right)+312
L\big) r^{14} \\\notag&+\big(-39 L^4-78 L^3+1248 L^2-2808 L+1872+\omega ^2 M^2 \left(2496 L^3 +3120 L^2 -1248 L -1872
\right)\big) r^{13} \\\notag&+\big(-78 L^4+78 L^3-156 L^2+4056 L-6864+\omega ^2 M^2 \left(6240 L^3 +9984 L^2 +1248 L -2496
\right)\big)
M^2	r^{12} \\\notag&+\big(-156 L^4+156 L^3+156 L^2-2184 L+4992+\omega ^2 M^2 \left(14976 L^3 +27456 L^2 +9984 L -2496\right)\big) M^3 r^{11} \\\notag&+\big(-312 L^4+312 L^3+312 L^2-1560
L-1248+\omega ^2 M^2\left(34944 L^3 +69888 L^2 +34944 L \right)\big)
M^4 r^{10} \\\notag&+\big( 16 \omega ^2 M^2 \left(-30 L^5 +240 L^4 -15736 L^3
+91901 L^2 -181790 L +120973\right)-65 L-52-13 L^4+13 L^3 +13 L^2\big) \\\notag &\times 48  M^5 r^9 +\big(16 \omega ^2 M^2 \left(30 L^5 -765 L^4 +32307 L^3
-322450 L^2 +926809 L -808654 \right) -13 L^4+13 L^3 +13 L^2 \\\notag&-65 L-52  \big) 96 M^6 r^8 +\big(3300 L^5+7187 L^4-122387 L^3+297613 L^2-216065 L-52 +16\omega ^2 M^2 (525 L^4 -20040 L^3
\\\notag&+379529 L^2 -1746962 L +2130419 )\big) 192 M^7
r^7 +\big(-10860 L^5+27707 L^4+322573 L^3-1530227 L^2+2231935 L\\\notag&-1019572+48 \omega ^2 M^2 \left(1035 L^3
-53838 L^2 +488617 L -920348 \right)\big) 384  M^8 r^6 +\big(11850 L^5-129263 L^4-5687 L^3 \\\notag&+2809573 L^2-7981905 L +6459788+144 \omega ^2 M^2 \left(1345 L^2 -54196 L
+195729 \right)\big) 768  M^9 r^5 \\\notag&+\big(-4290 L^5+147663 L^4-716253 L^3-1724215 L^2+13842007 L-17282692+1296 \omega ^2 M^2\left(295 L
-5471 \right)
\big) 1536 M^{10} r^4\\\notag&+\big(-163399680  L^4  +2362825728  L^3  -2646567936 
L^2  -37794106368 L +77002149888  \big) M^{11}
r^3 \\\notag&+\big(-27510 L^3  +169041 L^2  +527701  L 
-2311072  \big) 55296 M^{12} r^2+\left(-18990 L^2 +2831  L
+346436 \right) 331776  M^{13} r \\&-9794027520  L  M^{14}-43674992640 
M^{14} \bigg) \, ,
\end{align}

\begin{align}\notag
&W_{0,lm}^{-(3)}= \frac{i}{10 L M^4 r^{16}\Lambda^2} \bigg(55157760  L r^3 \omega  M^{10}+ \omega  M^6 r^7 \left(322560  L^3 -1290240  L^2+1290240  L \right) \\\notag&+ \omega  M^7 r^6 \left(-1330560  L^3 +9434880 L^2
-13547520  L\right)+ \omega  M^8 r^5 \left(1370880 L^3 -22417920
L^2 +52416000  L \right) \\&+r^4 M^9 \omega \left(17418240  L^2 
-88542720 L \right) \bigg) \, ,
\end{align}

\begin{align}\notag
&W_{1,lm}^{-(3)}= \frac{i}{10 L M^4 r^{16}\Lambda^2} \bigg(\big(-  L^2 +4  L 
-4  \big) 26 \omega ^2 r^{14}+104 \big(-  L^2  +  L  + 2  \big) M \omega ^2 r^{13} +39 \bigg( 8 \omega ^2 M^2 \left(1 -L^2 \right)- L^2+4 L-4\bigg) r^{12} \\\notag&+\bigg(16 \omega ^2 M^2 \left(-2 L^2 - L + 1\right)+ L^2-7 L+10\bigg) 26 M r^{11} +\big( 4 \omega ^2 M^2 \left(-5 L^2 -4 L + \right)+ L^2+2 L+1\big) 104 M^2 r^{10} \\\notag&+\big(16 \omega ^2 M^2\left(- L^2 - L \right)+ L^2+2 L+1\big) 312 M^3 r^9 +\big(  \omega ^2 M^2 \left(-14 L^2 -16 L -2 \right)+L^2+2 L+1\big) 832 M^4 r^8 \\\notag&+\big(16 \omega ^2 M^2 \left(1208 L^2 -5105 L +5027 \right)+65 L^2+130 L+65 \big) 32 M^5 r^7 \\\notag&+\big( 4 \omega ^2 M^2 \left(-684 L^2 +7883 L-13303 \right)+13 L^2+26 L+13 \big) 384 M^6 r^6\\\notag&+\big(-450 L^4-4500 L^3+88291 L^2-284218
L+259291+144\omega ^2 M^2 \left(2786 -789 L \right)\big) 128 M^7 r^5\\\notag&+\big(945 L^4+5130 L^3-303916 L^2+1275688 L-1390936+216\omega ^2 M^2 \left(35 L -754
\right)\big) 256 M^8 r^4\\\notag&+1536\big(-165  L^4
+1320 L^3 +110254  L^2 -732502  L +1016419 
\big) M^9 r^3 \\\notag&+18432\left(-255  L^3  -5622  L^2  +93908  L 
-190673 \right) M^{10} r^2 \\&+55296\left(-525 L^2  -17419  L 
+73484  \right) M^{11} r-59719680  L  M^{12}-1924300800  M^{12} \bigg) \, .
\end{align}

\section{Chandrasekhar transformation linear in spin}
\label{app:darboux}

The Regge-Wheeler and Zerilli equations are isospectral and this can be made explicit by performing a Darboux transformation, known in this context as the Chandrasekhar transformation, that relates them \cite{Glampedakis:2017rar}. We extend this transformation here to linear order in spin in order to connect $V^{- \text{GR}}_{lm}$ and  $V^{+ \text{GR}}_{lm}$ in \req{eq:V-GR} and \req{eq:V+GR}, respectively. Explicitly, we wish to transform

\begin{equation}
	\frac{d^2 \tilde{\Psi}^{+}}{d x^2} - V^{+ \text{GR}}_{lm} \tilde{\Psi}^{+}  = 0 \, ,
\end{equation}

into

\begin{equation}
	\frac{d^2 \Psi^{+}}{d x^2} - V^{- \text{GR}}_{lm} \Psi^{+}  = 0 \, .
\end{equation}

For a static black hole this can be done in a canonical way by 

\begin{equation}
	\Psi^{+} = \frac{d \tilde{\Psi}^{+}}{d x} - f_* \tilde{\Psi}^{+} \, ,
\end{equation}

with

\begin{equation}
	f_* = i \omega_* + \frac{3 M (r-2M)}{r^2(\mu r/2 + 3 M)} \, , \quad \omega_* = -i \frac{\mu (\mu/2+1)}{6 M} \, .
\end{equation}

In order to invert this transformation, note that

\begin{equation}
	V^{- \rm GR} + \frac{d f_*}{d x} - f_*^2 = (-\omega^2+\omega_*^2) \, ,
\end{equation}

so

\begin{equation}
\tilde{\Psi}^{+} = \frac{1}{(-\omega^2+\omega_*^2)} (\frac{d \Psi^{+}}{d x} + f_* \Psi^{+}) \, .
\end{equation}

To include spin, a slightly more general type of transformation is needed making the procedure more arbitrary \cite{Glampedakis:2017rar}. Nevertheless, an example that works to linear order in spin is given by 

\begin{equation}
	\Psi^{+} = (1+m \chi \beta)\frac{d \tilde{\Psi}^{+}}{d x} - (f_* + m \chi f^{(1)}_*) \tilde{\Psi}^{+} \, ,
	\label{eqn:darboux}
\end{equation}

with

\begin{equation}
	\beta = M\frac{24 M^2+6(l+2)(l-1)M r + l(-2-l+2l^2+l^3)r^2}{l(l+1)r^3(6 M + (l+2)(l-1)r)\omega} \, ,
\end{equation}

and

\begin{align}
	f^{(1)}_* &= -\frac{-864M^5-216(-4+l+l^2)rM^4-l(1+l)(l+2)^3(l-1)^3r^7\omega^2+36(l+2)(l-1)M^3r^2(-3+2r^2\omega^2)}{3 l (1+l)r^5(6+(l+2)(l-1)r)^2\omega} \nn \\
	&- \frac{3l(l+1)(l+2)^2(l-1)^2Mr^4(1+4r^2\omega^2)-6(l+2)(l-1)M^2r^3(2l^3+l^4+6r^2\omega^2-2l(1+3r^2\omega^2)-l^2(1+6r^2\omega^2))}{3 l (1+l)r^5(6+(l+2)(l-1)r)^2\omega}\, .
\end{align}

\section{QM-like perturbation theory}\label{app:pert}

We will define a symmetric bilinear form with respect to which the zeroth order wave equations are self-adjoint in the following sense

\begin{equation}
	\left \langle g |H_0 f \right\rangle = 	\left \langle H_0 g | f \right\rangle \, .
	\label{eq:adjoint}
\end{equation}

Up to boundary terms, this would hold for 

\begin{equation}
	\left \langle g | \psi \right\rangle = \int_C d r_*  g(r) \psi(r)  = \int_C d r (1+\frac{2 M}{r-r_+}) g(r) \psi(r) \, .
\end{equation}

This needs to be made well-defined on functions satisfying the QNM boundary conditions $\sim e^{-i \omega r_*}$ as $r \to r_H$ and $\sim e^{i \omega r_*}$ as $r \to \infty$ . In particular, given that the imaginary part of $\omega$ is negative, this means they diverge  for both $r_* \to \pm \infty$ which prohibits the otherwise natural choice $r \in ]r_+, \infty[$. Instead, one can take the contour $C$ from $r \to r_++\epsilon + i \infty$ down around the branchcut at $r_h$ and back up along $r \to r_+-\epsilon + i \infty$ \cite{Leaver:1986gd,Mark:2014aja}. Given two coupled equations, we can instead choose 

\begin{equation}
	\left \langle g | \psi \right\rangle  = \int_C d r (1+\frac{2 M}{r-r_+}) \begin{pmatrix}
		g^+(r) & g^-(r)
	\end{pmatrix} \begin{pmatrix}
		\psi^+(r) \\ \psi^-(r) \end{pmatrix} = \int_C d r (1+\frac{2 M}{r-r_+})
	(g^+(r) \psi^+(r)+ g^-(r) \psi^-(r) ) \, ,
\end{equation}

with respect to which the zeroth order uncoupled $H_0$ clearly satisfies \eqref{eq:adjoint}. Now consider

\begin{equation}
	\begin{pmatrix}
		\frac{d^2 }{d r_*^2} - (\omega^2-V^{+ \rm GR} -\delta V^{+}) &  \delta U^+ + \delta W^+ \frac{d }{d r_*} \\ \delta U^- + \delta W^- \frac{d }{d r_*}  & \frac{d^2 }{d r_*^2} - (\omega^2-V^{- \rm GR}-\delta V^{-}) \end{pmatrix}  \begin{pmatrix}
		\Psi^+(r) \\ \Psi^-(r) \end{pmatrix}  \equiv (H_0 + \delta H_+ + \delta H_-+ \delta H_{\text{break}}) \left| \psi \right\rangle = 0 \, .
\end{equation} 

Expand both $\left| \psi \right\rangle = \left| \psi^{(0)} \right\rangle + \left| \psi^{(1)} \right\rangle $ and $\omega = \omega^{(0)} + \omega^{(1)} $as a zeroth order piece and a small correction 

\begin{equation}
	H_0 \left| \psi^{(0)} \right\rangle = 0 \, , \quad  \left| \psi^{(0)} \right\rangle = c_+ \left| + \right\rangle + c_- \left| - \right\rangle \, ,
\end{equation}
with

\begin{equation}
	\left| + \right\rangle = 	\begin{pmatrix}
		\Psi^{+ \rm GR}(r) \\ 0 \end{pmatrix} \, , \quad \left| - \right\rangle = 	\begin{pmatrix}
		0 \\ \Psi^{- \rm GR}(r) \end{pmatrix} \, .
\end{equation}
The first order equation is given by 

\begin{equation}
	H_0 \left| \psi^{(1)} \right\rangle -2\omega^{(0)} \omega^{(1)} \left| \psi^{(0)}  \right\rangle  + (\delta H_+ + \delta H_-+ \delta H_{\text{break}}) \left| \psi^{(0)}  \right\rangle = 0 \, ,
\end{equation}
which, when acted on by respectively $\left \langle + \right| $, $\left \langle - \right|$  implies

\begin{align}
	-2\omega^{(0)} \omega^{(1)} c_+ \left \langle + | +\right \rangle + c_+ \left \langle + | \delta H_+ +\right \rangle + c_- \left \langle + | \delta H_{\rm break} -\right \rangle    &=0 \, , \\
	-2\omega^{(0)} \omega^{(1)} c_- \left \langle - | -\right \rangle + c_- \left \langle - | \delta H_- -\right \rangle + c_+ \left \langle - | \delta H_{\rm break} +\right \rangle    &=0 \, .
\end{align}
In order for this system to have a nontrivial solution, the following must hold

\begin{equation}
	\begin{vmatrix}
		\frac{\left \langle + | \delta H_+ +\right \rangle }{2\omega^{(0)}  \left \langle + | +\right \rangle }	- \omega^{(1)} & \frac{\left \langle + | \delta H_{\rm break} -\right \rangle}{2\omega^{(0)}  \left \langle + | +\right \rangle}\\ 
		\frac{\left \langle - | \delta H_{\rm break} +\right \rangle }{2\omega^{(0)}  \left \langle - | -\right \rangle }	& 	\frac{\left \langle - | \delta H_- -\right \rangle }{2\omega^{(0)}  \left \langle - | -\right \rangle }	- \omega^{(1)} 
	\end{vmatrix} = 0 \, .
\end{equation}
From the individual corrections we identify

\begin{equation}
	\delta \omega^+ = 	\frac{\left \langle + | \delta H_+ +\right \rangle }{2\omega^{(0)}  \left \langle + | +\right \rangle } \, , \quad  \delta \omega^- = \frac{\left \langle - | \delta H_- -\right \rangle }{2\omega^{(0)}  \left \langle - | -\right \rangle } \, , \quad (\delta \omega^{\rm break})^2 =  \frac{\left \langle - | \delta H_{\rm break} +\right \rangle }{2\omega^{(0)}  \left \langle - | -\right \rangle } \frac{\left \langle + | \delta H_{\rm break} -\right \rangle}{2\omega^{(0)}  \left \langle + | +\right \rangle} \, .
	\label{eq:defsdeltaomega}
\end{equation}
To conclude, we find \eqref{eq:dettotal} 

\begin{equation}
	(\delta \omega^+-\omega^{(1)} )(\delta \omega^--\omega^{(1)} )	- (\delta \omega^{\rm break})^2  = 0  \, .
\end{equation}
In deriving this results, we have assumed the frequency dependence in $H_0$  comes only from $\omega^2$. This is incorrect when including spin to linear order. However, this is easily corrected by simply replacing $2\omega^{(0)}$ by $(\frac{\partial (\omega^2-V^{\pm \rm GR})}{\partial \omega})_{\omega = \omega^{(0)}}$ in the normalizations of \eqref{eq:defsdeltaomega}. The associated eigenvectors can similarly be determined and, using

\begin{equation}
	(\gamma^{\rm break})^2 = \frac{\left \langle - | \delta H_{\rm break} +\right \rangle }{\left \langle + | \delta H_{\rm break} -\right \rangle}
\end{equation} 
they are found to lead to \req{eq:defsgammafull}

\begin{equation}
\gamma^{\pm}_{\rm total}= \gamma^{\rm break} \frac{\delta\omega^{\pm}-\delta\omega^{+}}{\delta\omega^{\rm break}}\, .
\end{equation}
when we write the eigenstate as

\begin{equation}
 c_+ \left| + \right\rangle +  c_- \left| - \right\rangle \propto	\gamma^{\pm}_{\rm total}  \left| + \right\rangle + \left| - \right\rangle \, ,
\end{equation}
which corresponds to the definition in the main text if asymptotically $\Psi^{(+) \rm GR} \sim \Psi^{(-) \rm GR}$.

\section{Solving the coupled equations using uncoupled equations}
\label{app:coupledbyuncoupled}

Consider the following set of coupled differential equations

\begin{align}
	\frac{d^2 \Psi^{-}}{d x^2} - V^- \Psi^{-} + \lambda(C_{-+}\Psi^{+} + D_{-+} \frac{d \Psi^{+}}{d x}) 	&= 0 \, , \\
	\frac{d^2 \Psi^{+}}{d x^2} - V^- \Psi^{+} + \lambda ( C_{+-}\Psi^{-} + D_{+-} \frac{d \Psi^{-}}{d x}) 	&= 0  \, , \label{eqn:pluscoupled}
\end{align}
with $\lambda$ a small parameter. Note the appearance of $V^-$ in the second equation: we can be bring the equations \req{eq:paritymix1}-\req{eq:paritymix2} into this form using the transformation described in appendix \ref{app:darboux} (we are including the term $\omega^2$ in $V^{-}$ to shorten the formulas). Suppose we can independently consider the symmetric and antisymmetric parts of the perturbation, this is

\begin{align}
	\frac{d^2 \Psi^{-}}{d x^2} - V^- \Psi^{-} + \lambda_s (\bar{C}\Psi^{+} + \bar{D} \frac{d \Psi^{+}}{d x})+ \lambda_a (\delta{C}\Psi^{+} + \delta{D} \frac{d \Psi^{+}}{d x}) 	&= 0 \, , \label{eqn:coupledas1} \\
	\frac{d^2 \Psi^{+}}{d x^2} - V^- \Psi^{+}  + \lambda_s (\bar{C}\Psi^{-} + \bar{D} \frac{d \Psi^{-}}{d x})- \lambda_a (\delta{C}\Psi^{-} + \delta{D} \frac{d \Psi^{-}}{d x}) 	&= 0 \, , \label{eqn:coupledas2}
\end{align}
where for any quantity $X$ we are defining

\begin{equation}
	\bar{X} = \frac{1}{2}(X_{-+} + X_{+-}) \, , \quad \delta X = \frac{1}{2}(X_{-+} - X_{+-}) \, .
\end{equation}
Here $\lambda_{s}$ and $\lambda_{a}$ are bookkeeping parameters, and we will eventually set them to $1$, hence recovering the original equation. Now we can study the effect of each term in these equations. 
The $\lambda_s$ coupled equations can be separated by $\Psi^{s} = \Psi^{-} \pm \Psi^{+}$ into an equation of the type

\begin{equation}
	\frac{d^2 \Psi^{s}}{d x^2} - V^- \Psi^{s} \pm \lambda_s (\bar{C}\Psi^{s} + \bar{D} \frac{d \Psi^{s}}{d x})	= 0 \, .
	\label{eqn:symmaster}
\end{equation}
On the other hand, the antisymmetric part can be separated by $\Psi^{a} =  \Psi^{-} \pm i \Psi^{+}$  into an equation of the type

\begin{equation}
	\frac{d^2 \Psi^{a}}{d x^2} - V^- \Psi^{a}  \mp i \lambda_a (\delta{C}\Psi^{a} + \delta{D} \frac{d \Psi^{a}}{d x}) \, .
	\label{eqn:antisymmaster}
\end{equation}

Keeping both, one can observe that diagonalizing either the symmetric or the antisymmetric part always makes the other completely off-diagonal and it is therefore generically not possible to diagnolize both at the same time. One possible resolution to nevertheless consider both together, as is the original problem, is to define projections onto the leading radial wavefunctions and to use standard perturbation theory from quantum mechanics \cite{McManus:2019ulj}. However, this is somewhat subtle as the QNMs are not square integrable \cite{Mark:2014aja} -- see appendix \ref{app:pert}. We will not take this approach but simply use it to argue for the form of the final equation in terms of solutions to \eqref{eqn:symmaster} and \eqref{eqn:antisymmaster}. In particular, we can write \eqref{eqn:coupledas1} and \eqref{eqn:coupledas2} suggestively as

\begin{equation}
	(H_0 + \lambda H') \equiv (H_0 + \lambda H_s + \lambda H_a)\left| \Psi \right\rangle \, .
\end{equation}
We know from degenerate perturbation theory, as a special case of \eqref{eq:dettotal}, that the correction due to an off-diagonal $H'$ is given by

\begin{equation}
	\omega_1 = \pm \sqrt{\left\langle \Psi^- \left| H' \right| \Psi^+ \right\rangle \left\langle \Psi^+ \left| H' \right| \Psi^- \right\rangle} = \pm \sqrt{ \omega_s^2 + \omega_a^2} \, , \label{eqn:uncoupledsol}
\end{equation} 
where in the second equality we wrote it in terms of the results from $H_s$ and $H_a$ separately. The value of $\omega_s$ can now be determined by the first order correction in \eqref{eqn:symmaster}, $\omega \approx \omega_0 + \lambda_s \omega_s$, while $\omega_a$ is similarly found from \eqref{eqn:antisymmaster} and $\omega \approx \omega_0 + \lambda_a \omega_a$. The cross-terms had to cancel due to the symmetry under exchange $\Psi^+ \leftrightarrow \Psi^-$. \\

To find the normalization coefficient, we can similarly use an analysis analogous to the one leading up to \req{eq:defsgammafull} to conclude that

\begin{equation}
 (\gamma^{\rm break})^2 = \frac{\left\langle \Psi^- \left| H' \right| \Psi^+ \right\rangle}{\left\langle \Psi^+ \left| H' \right| \Psi^- \right\rangle} = \frac{1 +(\frac{\omega_a}{\omega_s})^2}{(1-i \frac{\omega_a}{\omega_s})^2} \, .
\end{equation}
This again assumes asymptotically $\Psi^{(+) \rm GR} \sim \Psi^{(-) \rm GR}$. However, to compare with the results from the main text we actually need to translate to the original master variables $\tilde{\Psi}^{(+) \rm GR}$,  $\tilde{\Psi}^{(-) \rm GR}$. Asymptotically, we can read of from the transformation \eqref{eqn:darboux} 

\begin{equation}
	\frac{\tilde{\Psi}^{(+) \rm GR}}{\tilde{\Psi}^{(-) \rm GR}} = \frac{1}{i(\omega^{\rm GR}-\omega_*)}\frac{\Psi^{(+) \rm GR}}{\Psi^{(-) \rm GR}}  \, ,
\end{equation}
so finally, for the original master variables 

\begin{equation}
(\gamma^{\rm break}) = \pm \frac{1}{i(\omega^{\rm GR}-\omega_*)} \sqrt{\frac{1 +(\frac{\omega_a}{\omega_s})^2}{(1-i \frac{\omega_a}{\omega_s})^2}} \, .
\end{equation}

\bibliographystyle{apsrev4-1} 
\vspace{1cm}
\bibliography{Gravities} 

 \newcommand{\noop}[1]{}
\begin{thebibliography}{73}%
\makeatletter
\providecommand \@ifxundefined [1]{%
 \@ifx{#1\undefined}
}%
\providecommand \@ifnum [1]{%
 \ifnum #1\expandafter \@firstoftwo
 \else \expandafter \@secondoftwo
 \fi
}%
\providecommand \@ifx [1]{%
 \ifx #1\expandafter \@firstoftwo
 \else \expandafter \@secondoftwo
 \fi
}%
\providecommand \natexlab [1]{#1}%
\providecommand \enquote  [1]{``#1''}%
\providecommand \bibnamefont  [1]{#1}%
\providecommand \bibfnamefont [1]{#1}%
\providecommand \citenamefont [1]{#1}%
\providecommand \href@noop [0]{\@secondoftwo}%
\providecommand \href [0]{\begingroup \@sanitize@url \@href}%
\providecommand \@href[1]{\@@startlink{#1}\@@href}%
\providecommand \@@href[1]{\endgroup#1\@@endlink}%
\providecommand \@sanitize@url [0]{\catcode `\\12\catcode `\$12\catcode
  `\&12\catcode `\#12\catcode `\^12\catcode `\_12\catcode `\%12\relax}%
\providecommand \@@startlink[1]{}%
\providecommand \@@endlink[0]{}%
\providecommand \url  [0]{\begingroup\@sanitize@url \@url }%
\providecommand \@url [1]{\endgroup\@href {#1}{\urlprefix }}%
\providecommand \urlprefix  [0]{URL }%
\providecommand \Eprint [0]{\href }%
\providecommand \doibase [0]{http://dx.doi.org/}%
\providecommand \selectlanguage [0]{\@gobble}%
\providecommand \bibinfo  [0]{\@secondoftwo}%
\providecommand \bibfield  [0]{\@secondoftwo}%
\providecommand \translation [1]{[#1]}%
\providecommand \BibitemOpen [0]{}%
\providecommand \bibitemStop [0]{}%
\providecommand \bibitemNoStop [0]{.\EOS\space}%
\providecommand \EOS [0]{\spacefactor3000\relax}%
\providecommand \BibitemShut  [1]{\csname bibitem#1\endcsname}%
\let\auto@bib@innerbib\@empty
\bibitem [{\citenamefont {Abbott}\ \emph {et~al.}(2016)\citenamefont {Abbott}
  \emph {et~al.}}]{TheLIGOScientific:2016src}%
  \BibitemOpen
  \bibfield  {author} {\bibinfo {author} {\bibfnamefont {B.~P.}\ \bibnamefont
  {Abbott}} \emph {et~al.} (\bibinfo {collaboration} {LIGO Scientific,
  Virgo}),\ }\href {\doibase 10.1103/PhysRevLett.116.221101,
  10.1103/PhysRevLett.121.129902} {\bibfield  {journal} {\bibinfo  {journal}
  {Phys. Rev. Lett.}\ }\textbf {\bibinfo {volume} {116}},\ \bibinfo {pages}
  {221101} (\bibinfo {year} {2016})},\ \bibinfo {note} {[Erratum: Phys. Rev.
  Lett.121,no.12,129902(2018)]},\ \Eprint {http://arxiv.org/abs/1602.03841}
  {arXiv:1602.03841 [gr-qc]} \BibitemShut {NoStop}%
\bibitem [{\citenamefont {Yunes}\ \emph {et~al.}(2016)\citenamefont {Yunes},
  \citenamefont {Yagi},\ and\ \citenamefont {Pretorius}}]{Yunes:2016jcc}%
  \BibitemOpen
  \bibfield  {author} {\bibinfo {author} {\bibfnamefont {N.}~\bibnamefont
  {Yunes}}, \bibinfo {author} {\bibfnamefont {K.}~\bibnamefont {Yagi}}, \ and\
  \bibinfo {author} {\bibfnamefont {F.}~\bibnamefont {Pretorius}},\ }\href
  {\doibase 10.1103/PhysRevD.94.084002} {\bibfield  {journal} {\bibinfo
  {journal} {Phys. Rev.}\ }\textbf {\bibinfo {volume} {D94}},\ \bibinfo {pages}
  {084002} (\bibinfo {year} {2016})},\ \Eprint
  {http://arxiv.org/abs/1603.08955} {arXiv:1603.08955 [gr-qc]} \BibitemShut
  {NoStop}%
\bibitem [{\citenamefont {Berti}\ \emph
  {et~al.}(2018{\natexlab{a}})\citenamefont {Berti}, \citenamefont {Yagi},\
  and\ \citenamefont {Yunes}}]{Berti:2018cxi}%
  \BibitemOpen
  \bibfield  {author} {\bibinfo {author} {\bibfnamefont {E.}~\bibnamefont
  {Berti}}, \bibinfo {author} {\bibfnamefont {K.}~\bibnamefont {Yagi}}, \ and\
  \bibinfo {author} {\bibfnamefont {N.}~\bibnamefont {Yunes}},\ }\href
  {\doibase 10.1007/s10714-018-2362-8} {\bibfield  {journal} {\bibinfo
  {journal} {Gen. Rel. Grav.}\ }\textbf {\bibinfo {volume} {50}},\ \bibinfo
  {pages} {46} (\bibinfo {year} {2018}{\natexlab{a}})},\ \Eprint
  {http://arxiv.org/abs/1801.03208} {arXiv:1801.03208 [gr-qc]} \BibitemShut
  {NoStop}%
\bibitem [{\citenamefont {Berti}\ \emph
  {et~al.}(2018{\natexlab{b}})\citenamefont {Berti}, \citenamefont {Yagi},
  \citenamefont {Yang},\ and\ \citenamefont {Yunes}}]{Berti:2018vdi}%
  \BibitemOpen
  \bibfield  {author} {\bibinfo {author} {\bibfnamefont {E.}~\bibnamefont
  {Berti}}, \bibinfo {author} {\bibfnamefont {K.}~\bibnamefont {Yagi}},
  \bibinfo {author} {\bibfnamefont {H.}~\bibnamefont {Yang}}, \ and\ \bibinfo
  {author} {\bibfnamefont {N.}~\bibnamefont {Yunes}},\ }\href {\doibase
  10.1007/s10714-018-2372-6} {\bibfield  {journal} {\bibinfo  {journal} {Gen.
  Rel. Grav.}\ }\textbf {\bibinfo {volume} {50}},\ \bibinfo {pages} {49}
  (\bibinfo {year} {2018}{\natexlab{b}})},\ \Eprint
  {http://arxiv.org/abs/1801.03587} {arXiv:1801.03587 [gr-qc]} \BibitemShut
  {NoStop}%
\bibitem [{\citenamefont {Barack}\ \emph {et~al.}(2019)\citenamefont {Barack}
  \emph {et~al.}}]{Barack:2018yly}%
  \BibitemOpen
  \bibfield  {author} {\bibinfo {author} {\bibfnamefont {L.}~\bibnamefont
  {Barack}} \emph {et~al.},\ }\href {\doibase 10.1088/1361-6382/ab0587}
  {\bibfield  {journal} {\bibinfo  {journal} {Class. Quant. Grav.}\ }\textbf
  {\bibinfo {volume} {36}},\ \bibinfo {pages} {143001} (\bibinfo {year}
  {2019})},\ \Eprint {http://arxiv.org/abs/1806.05195} {arXiv:1806.05195
  [gr-qc]} \BibitemShut {NoStop}%
\bibitem [{\citenamefont {Abbott}\ \emph
  {et~al.}(2019{\natexlab{a}})\citenamefont {Abbott} \emph
  {et~al.}}]{Abbott:2018lct}%
  \BibitemOpen
  \bibfield  {author} {\bibinfo {author} {\bibfnamefont {B.}~\bibnamefont
  {Abbott}} \emph {et~al.} (\bibinfo {collaboration} {LIGO Scientific,
  Virgo}),\ }\href {\doibase 10.1103/PhysRevLett.123.011102} {\bibfield
  {journal} {\bibinfo  {journal} {Phys. Rev. Lett.}\ }\textbf {\bibinfo
  {volume} {123}},\ \bibinfo {pages} {011102} (\bibinfo {year}
  {2019}{\natexlab{a}})},\ \Eprint {http://arxiv.org/abs/1811.00364}
  {arXiv:1811.00364 [gr-qc]} \BibitemShut {NoStop}%
\bibitem [{\citenamefont {Abbott}\ \emph
  {et~al.}(2019{\natexlab{b}})\citenamefont {Abbott} \emph
  {et~al.}}]{LIGOScientific:2019fpa}%
  \BibitemOpen
  \bibfield  {author} {\bibinfo {author} {\bibfnamefont {B.}~\bibnamefont
  {Abbott}} \emph {et~al.} (\bibinfo {collaboration} {LIGO Scientific,
  Virgo}),\ }\href {\doibase 10.1103/PhysRevD.100.104036} {\bibfield  {journal}
  {\bibinfo  {journal} {Phys. Rev. D}\ }\textbf {\bibinfo {volume} {100}},\
  \bibinfo {pages} {104036} (\bibinfo {year} {2019}{\natexlab{b}})},\ \Eprint
  {http://arxiv.org/abs/1903.04467} {arXiv:1903.04467 [gr-qc]} \BibitemShut
  {NoStop}%
\bibitem [{\citenamefont {Cabero}\ \emph {et~al.}(2020)\citenamefont {Cabero},
  \citenamefont {Westerweck}, \citenamefont {Capano}, \citenamefont {Kumar},
  \citenamefont {Nielsen},\ and\ \citenamefont {Krishnan}}]{Cabero:2019zyt}%
  \BibitemOpen
  \bibfield  {author} {\bibinfo {author} {\bibfnamefont {M.}~\bibnamefont
  {Cabero}}, \bibinfo {author} {\bibfnamefont {J.}~\bibnamefont {Westerweck}},
  \bibinfo {author} {\bibfnamefont {C.~D.}\ \bibnamefont {Capano}}, \bibinfo
  {author} {\bibfnamefont {S.}~\bibnamefont {Kumar}}, \bibinfo {author}
  {\bibfnamefont {A.~B.}\ \bibnamefont {Nielsen}}, \ and\ \bibinfo {author}
  {\bibfnamefont {B.}~\bibnamefont {Krishnan}},\ }\href {\doibase
  10.1103/PhysRevD.101.064044} {\bibfield  {journal} {\bibinfo  {journal}
  {Phys. Rev. D}\ }\textbf {\bibinfo {volume} {101}},\ \bibinfo {pages}
  {064044} (\bibinfo {year} {2020})},\ \Eprint
  {http://arxiv.org/abs/1911.01361} {arXiv:1911.01361 [gr-qc]} \BibitemShut
  {NoStop}%
\bibitem [{\citenamefont {Abbott}\ \emph {et~al.}(2021)\citenamefont {Abbott}
  \emph {et~al.}}]{LIGOScientific:2020tif}%
  \BibitemOpen
  \bibfield  {author} {\bibinfo {author} {\bibfnamefont {R.}~\bibnamefont
  {Abbott}} \emph {et~al.} (\bibinfo {collaboration} {LIGO Scientific,
  Virgo}),\ }\href {\doibase 10.1103/PhysRevD.103.122002} {\bibfield  {journal}
  {\bibinfo  {journal} {Phys. Rev. D}\ }\textbf {\bibinfo {volume} {103}},\
  \bibinfo {pages} {122002} (\bibinfo {year} {2021})},\ \Eprint
  {http://arxiv.org/abs/2010.14529} {arXiv:2010.14529 [gr-qc]} \BibitemShut
  {NoStop}%
\bibitem [{\citenamefont {Cardoso}\ \emph {et~al.}(2016)\citenamefont
  {Cardoso}, \citenamefont {Franzin},\ and\ \citenamefont
  {Pani}}]{Cardoso:2016rao}%
  \BibitemOpen
  \bibfield  {author} {\bibinfo {author} {\bibfnamefont {V.}~\bibnamefont
  {Cardoso}}, \bibinfo {author} {\bibfnamefont {E.}~\bibnamefont {Franzin}}, \
  and\ \bibinfo {author} {\bibfnamefont {P.}~\bibnamefont {Pani}},\ }\href
  {\doibase 10.1103/PhysRevLett.117.089902, 10.1103/PhysRevLett.116.171101}
  {\bibfield  {journal} {\bibinfo  {journal} {Phys. Rev. Lett.}\ }\textbf
  {\bibinfo {volume} {116}},\ \bibinfo {pages} {171101} (\bibinfo {year}
  {2016})},\ \bibinfo {note} {[Erratum: Phys. Rev.
  Lett.117,no.8,089902(2016)]},\ \Eprint {http://arxiv.org/abs/1602.07309}
  {arXiv:1602.07309 [gr-qc]} \BibitemShut {NoStop}%
\bibitem [{\citenamefont {Okounkova}\ \emph {et~al.}(2020)\citenamefont
  {Okounkova}, \citenamefont {Stein}, \citenamefont {Moxon}, \citenamefont
  {Scheel},\ and\ \citenamefont {Teukolsky}}]{Okounkova:2019zjf}%
  \BibitemOpen
  \bibfield  {author} {\bibinfo {author} {\bibfnamefont {M.}~\bibnamefont
  {Okounkova}}, \bibinfo {author} {\bibfnamefont {L.~C.}\ \bibnamefont
  {Stein}}, \bibinfo {author} {\bibfnamefont {J.}~\bibnamefont {Moxon}},
  \bibinfo {author} {\bibfnamefont {M.~A.}\ \bibnamefont {Scheel}}, \ and\
  \bibinfo {author} {\bibfnamefont {S.~A.}\ \bibnamefont {Teukolsky}},\ }\href
  {\doibase 10.1103/PhysRevD.101.104016} {\bibfield  {journal} {\bibinfo
  {journal} {Phys. Rev. D}\ }\textbf {\bibinfo {volume} {101}},\ \bibinfo
  {pages} {104016} (\bibinfo {year} {2020})},\ \Eprint
  {http://arxiv.org/abs/1911.02588} {arXiv:1911.02588 [gr-qc]} \BibitemShut
  {NoStop}%
\bibitem [{\citenamefont {Sennett}\ \emph {et~al.}(2020)\citenamefont
  {Sennett}, \citenamefont {Brito}, \citenamefont {Buonanno}, \citenamefont
  {Gorbenko},\ and\ \citenamefont {Senatore}}]{Sennett:2019bpc}%
  \BibitemOpen
  \bibfield  {author} {\bibinfo {author} {\bibfnamefont {N.}~\bibnamefont
  {Sennett}}, \bibinfo {author} {\bibfnamefont {R.}~\bibnamefont {Brito}},
  \bibinfo {author} {\bibfnamefont {A.}~\bibnamefont {Buonanno}}, \bibinfo
  {author} {\bibfnamefont {V.}~\bibnamefont {Gorbenko}}, \ and\ \bibinfo
  {author} {\bibfnamefont {L.}~\bibnamefont {Senatore}},\ }\href {\doibase
  10.1103/PhysRevD.102.044056} {\bibfield  {journal} {\bibinfo  {journal}
  {Phys. Rev. D}\ }\textbf {\bibinfo {volume} {102}},\ \bibinfo {pages}
  {044056} (\bibinfo {year} {2020})},\ \Eprint
  {http://arxiv.org/abs/1912.09917} {arXiv:1912.09917 [gr-qc]} \BibitemShut
  {NoStop}%
\bibitem [{\citenamefont {Nair}\ \emph {et~al.}(2019)\citenamefont {Nair},
  \citenamefont {Perkins}, \citenamefont {Silva},\ and\ \citenamefont
  {Yunes}}]{Nair:2019iur}%
  \BibitemOpen
  \bibfield  {author} {\bibinfo {author} {\bibfnamefont {R.}~\bibnamefont
  {Nair}}, \bibinfo {author} {\bibfnamefont {S.}~\bibnamefont {Perkins}},
  \bibinfo {author} {\bibfnamefont {H.~O.}\ \bibnamefont {Silva}}, \ and\
  \bibinfo {author} {\bibfnamefont {N.}~\bibnamefont {Yunes}},\ }\href
  {\doibase 10.1103/PhysRevLett.123.191101} {\bibfield  {journal} {\bibinfo
  {journal} {Phys. Rev. Lett.}\ }\textbf {\bibinfo {volume} {123}},\ \bibinfo
  {pages} {191101} (\bibinfo {year} {2019})},\ \Eprint
  {http://arxiv.org/abs/1905.00870} {arXiv:1905.00870 [gr-qc]} \BibitemShut
  {NoStop}%
\bibitem [{\citenamefont {Carson}\ and\ \citenamefont
  {Yagi}(2020{\natexlab{a}})}]{Carson:2020cqb}%
  \BibitemOpen
  \bibfield  {author} {\bibinfo {author} {\bibfnamefont {Z.}~\bibnamefont
  {Carson}}\ and\ \bibinfo {author} {\bibfnamefont {K.}~\bibnamefont {Yagi}},\
  }\href {\doibase 10.1088/1361-6382/aba221} {\bibfield  {journal} {\bibinfo
  {journal} {Class. Quant. Grav.}\ }\textbf {\bibinfo {volume} {37}},\ \bibinfo
  {pages} {215007} (\bibinfo {year} {2020}{\natexlab{a}})},\ \Eprint
  {http://arxiv.org/abs/2002.08559} {arXiv:2002.08559 [gr-qc]} \BibitemShut
  {NoStop}%
\bibitem [{\citenamefont {Carson}\ and\ \citenamefont
  {Yagi}(2020{\natexlab{b}})}]{Carson:2020ter}%
  \BibitemOpen
  \bibfield  {author} {\bibinfo {author} {\bibfnamefont {Z.}~\bibnamefont
  {Carson}}\ and\ \bibinfo {author} {\bibfnamefont {K.}~\bibnamefont {Yagi}},\
  }\href {\doibase 10.1103/PhysRevD.101.104030} {\bibfield  {journal} {\bibinfo
   {journal} {Phys. Rev. D}\ }\textbf {\bibinfo {volume} {101}},\ \bibinfo
  {pages} {104030} (\bibinfo {year} {2020}{\natexlab{b}})},\ \Eprint
  {http://arxiv.org/abs/2003.00286} {arXiv:2003.00286 [gr-qc]} \BibitemShut
  {NoStop}%
\bibitem [{\citenamefont {Carson}\ and\ \citenamefont
  {Yagi}(2020{\natexlab{c}})}]{Carson:2020iik}%
  \BibitemOpen
  \bibfield  {author} {\bibinfo {author} {\bibfnamefont {Z.}~\bibnamefont
  {Carson}}\ and\ \bibinfo {author} {\bibfnamefont {K.}~\bibnamefont {Yagi}},\
  }\href {\doibase 10.1103/PhysRevD.101.084050} {\bibfield  {journal} {\bibinfo
   {journal} {Phys. Rev. D}\ }\textbf {\bibinfo {volume} {101}},\ \bibinfo
  {pages} {084050} (\bibinfo {year} {2020}{\natexlab{c}})},\ \Eprint
  {http://arxiv.org/abs/2003.02374} {arXiv:2003.02374 [gr-qc]} \BibitemShut
  {NoStop}%
\bibitem [{\citenamefont {Okounkova}(2020)}]{Okounkova:2020rqw}%
  \BibitemOpen
  \bibfield  {author} {\bibinfo {author} {\bibfnamefont {M.}~\bibnamefont
  {Okounkova}},\ }\href {\doibase 10.1103/PhysRevD.102.084046} {\bibfield
  {journal} {\bibinfo  {journal} {Phys. Rev. D}\ }\textbf {\bibinfo {volume}
  {102}},\ \bibinfo {pages} {084046} (\bibinfo {year} {2020})},\ \Eprint
  {http://arxiv.org/abs/2001.03571} {arXiv:2001.03571 [gr-qc]} \BibitemShut
  {NoStop}%
\bibitem [{\citenamefont {de~Rham}\ \emph {et~al.}(2020)\citenamefont
  {de~Rham}, \citenamefont {Francfort},\ and\ \citenamefont
  {Zhang}}]{deRham:2020ejn}%
  \BibitemOpen
  \bibfield  {author} {\bibinfo {author} {\bibfnamefont {C.}~\bibnamefont
  {de~Rham}}, \bibinfo {author} {\bibfnamefont {J.}~\bibnamefont {Francfort}},
  \ and\ \bibinfo {author} {\bibfnamefont {J.}~\bibnamefont {Zhang}},\ }\href
  {\doibase 10.1103/PhysRevD.102.024079} {\bibfield  {journal} {\bibinfo
  {journal} {Phys. Rev. D}\ }\textbf {\bibinfo {volume} {102}},\ \bibinfo
  {pages} {024079} (\bibinfo {year} {2020})},\ \Eprint
  {http://arxiv.org/abs/2005.13923} {arXiv:2005.13923 [hep-th]} \BibitemShut
  {NoStop}%
\bibitem [{\citenamefont {Perkins}\ \emph {et~al.}(2021)\citenamefont
  {Perkins}, \citenamefont {Nair}, \citenamefont {Silva},\ and\ \citenamefont
  {Yunes}}]{Perkins:2021mhb}%
  \BibitemOpen
  \bibfield  {author} {\bibinfo {author} {\bibfnamefont {S.~E.}\ \bibnamefont
  {Perkins}}, \bibinfo {author} {\bibfnamefont {R.}~\bibnamefont {Nair}},
  \bibinfo {author} {\bibfnamefont {H.~O.}\ \bibnamefont {Silva}}, \ and\
  \bibinfo {author} {\bibfnamefont {N.}~\bibnamefont {Yunes}},\ }\href
  {\doibase 10.1103/PhysRevD.104.024060} {\bibfield  {journal} {\bibinfo
  {journal} {Phys. Rev. D}\ }\textbf {\bibinfo {volume} {104}},\ \bibinfo
  {pages} {024060} (\bibinfo {year} {2021})},\ \Eprint
  {http://arxiv.org/abs/2104.11189} {arXiv:2104.11189 [gr-qc]} \BibitemShut
  {NoStop}%
\bibitem [{\citenamefont {Gross}\ and\ \citenamefont
  {Witten}(1986)}]{Gross:1986iv}%
  \BibitemOpen
  \bibfield  {author} {\bibinfo {author} {\bibfnamefont {D.~J.}\ \bibnamefont
  {Gross}}\ and\ \bibinfo {author} {\bibfnamefont {E.}~\bibnamefont {Witten}},\
  }\href {\doibase 10.1016/0550-3213(86)90429-3} {\bibfield  {journal}
  {\bibinfo  {journal} {Nucl. Phys.}\ }\textbf {\bibinfo {volume} {B277}},\
  \bibinfo {pages} {1} (\bibinfo {year} {1986})}\BibitemShut {NoStop}%
\bibitem [{\citenamefont {Gross}\ and\ \citenamefont
  {Sloan}(1987)}]{Gross:1986mw}%
  \BibitemOpen
  \bibfield  {author} {\bibinfo {author} {\bibfnamefont {D.~J.}\ \bibnamefont
  {Gross}}\ and\ \bibinfo {author} {\bibfnamefont {J.~H.}\ \bibnamefont
  {Sloan}},\ }\href {\doibase 10.1016/0550-3213(87)90465-2} {\bibfield
  {journal} {\bibinfo  {journal} {Nucl. Phys.}\ }\textbf {\bibinfo {volume}
  {B291}},\ \bibinfo {pages} {41} (\bibinfo {year} {1987})}\BibitemShut
  {NoStop}%
\bibitem [{\citenamefont {Bergshoeff}\ and\ \citenamefont
  {de~Roo}(1989)}]{Bergshoeff:1989de}%
  \BibitemOpen
  \bibfield  {author} {\bibinfo {author} {\bibfnamefont {E.~A.}\ \bibnamefont
  {Bergshoeff}}\ and\ \bibinfo {author} {\bibfnamefont {M.}~\bibnamefont
  {de~Roo}},\ }\href {\doibase 10.1016/0550-3213(89)90336-2} {\bibfield
  {journal} {\bibinfo  {journal} {Nucl. Phys.}\ }\textbf {\bibinfo {volume}
  {B328}},\ \bibinfo {pages} {439} (\bibinfo {year} {1989})}\BibitemShut
  {NoStop}%
\bibitem [{\citenamefont {Konno}\ \emph {et~al.}(2007)\citenamefont {Konno},
  \citenamefont {Matsuyama},\ and\ \citenamefont {Tanda}}]{Konno:2007ze}%
  \BibitemOpen
  \bibfield  {author} {\bibinfo {author} {\bibfnamefont {K.}~\bibnamefont
  {Konno}}, \bibinfo {author} {\bibfnamefont {T.}~\bibnamefont {Matsuyama}}, \
  and\ \bibinfo {author} {\bibfnamefont {S.}~\bibnamefont {Tanda}},\ }\href
  {\doibase 10.1103/PhysRevD.76.024009} {\bibfield  {journal} {\bibinfo
  {journal} {Phys. Rev.}\ }\textbf {\bibinfo {volume} {D76}},\ \bibinfo {pages}
  {024009} (\bibinfo {year} {2007})},\ \Eprint {http://arxiv.org/abs/0706.3080}
  {arXiv:0706.3080 [gr-qc]} \BibitemShut {NoStop}%
\bibitem [{\citenamefont {Yunes}\ and\ \citenamefont
  {Pretorius}(2009)}]{Yunes:2009hc}%
  \BibitemOpen
  \bibfield  {author} {\bibinfo {author} {\bibfnamefont {N.}~\bibnamefont
  {Yunes}}\ and\ \bibinfo {author} {\bibfnamefont {F.}~\bibnamefont
  {Pretorius}},\ }\href {\doibase 10.1103/PhysRevD.79.084043} {\bibfield
  {journal} {\bibinfo  {journal} {Phys. Rev.}\ }\textbf {\bibinfo {volume}
  {D79}},\ \bibinfo {pages} {084043} (\bibinfo {year} {2009})},\ \Eprint
  {http://arxiv.org/abs/0902.4669} {arXiv:0902.4669 [gr-qc]} \BibitemShut
  {NoStop}%
\bibitem [{\citenamefont {Pani}\ and\ \citenamefont
  {Cardoso}(2009)}]{Pani:2009wy}%
  \BibitemOpen
  \bibfield  {author} {\bibinfo {author} {\bibfnamefont {P.}~\bibnamefont
  {Pani}}\ and\ \bibinfo {author} {\bibfnamefont {V.}~\bibnamefont {Cardoso}},\
  }\href {\doibase 10.1103/PhysRevD.79.084031} {\bibfield  {journal} {\bibinfo
  {journal} {Phys. Rev.}\ }\textbf {\bibinfo {volume} {D79}},\ \bibinfo {pages}
  {084031} (\bibinfo {year} {2009})},\ \Eprint {http://arxiv.org/abs/0902.1569}
  {arXiv:0902.1569 [gr-qc]} \BibitemShut {NoStop}%
\bibitem [{\citenamefont {Kleihaus}\ \emph {et~al.}(2011)\citenamefont
  {Kleihaus}, \citenamefont {Kunz},\ and\ \citenamefont
  {Radu}}]{Kleihaus:2011tg}%
  \BibitemOpen
  \bibfield  {author} {\bibinfo {author} {\bibfnamefont {B.}~\bibnamefont
  {Kleihaus}}, \bibinfo {author} {\bibfnamefont {J.}~\bibnamefont {Kunz}}, \
  and\ \bibinfo {author} {\bibfnamefont {E.}~\bibnamefont {Radu}},\ }\href
  {\doibase 10.1103/PhysRevLett.106.151104} {\bibfield  {journal} {\bibinfo
  {journal} {Phys. Rev. Lett.}\ }\textbf {\bibinfo {volume} {106}},\ \bibinfo
  {pages} {151104} (\bibinfo {year} {2011})},\ \Eprint
  {http://arxiv.org/abs/1101.2868} {arXiv:1101.2868 [gr-qc]} \BibitemShut
  {NoStop}%
\bibitem [{\citenamefont {Pani}\ \emph {et~al.}(2011)\citenamefont {Pani},
  \citenamefont {Macedo}, \citenamefont {Crispino},\ and\ \citenamefont
  {Cardoso}}]{Pani:2011gy}%
  \BibitemOpen
  \bibfield  {author} {\bibinfo {author} {\bibfnamefont {P.}~\bibnamefont
  {Pani}}, \bibinfo {author} {\bibfnamefont {C.~F.~B.}\ \bibnamefont {Macedo}},
  \bibinfo {author} {\bibfnamefont {L.~C.~B.}\ \bibnamefont {Crispino}}, \ and\
  \bibinfo {author} {\bibfnamefont {V.}~\bibnamefont {Cardoso}},\ }\href
  {\doibase 10.1103/PhysRevD.84.087501} {\bibfield  {journal} {\bibinfo
  {journal} {Phys. Rev.}\ }\textbf {\bibinfo {volume} {D84}},\ \bibinfo {pages}
  {087501} (\bibinfo {year} {2011})},\ \Eprint {http://arxiv.org/abs/1109.3996}
  {arXiv:1109.3996 [gr-qc]} \BibitemShut {NoStop}%
\bibitem [{\citenamefont {Yagi}\ \emph {et~al.}(2012)\citenamefont {Yagi},
  \citenamefont {Yunes},\ and\ \citenamefont {Tanaka}}]{Yagi:2012ya}%
  \BibitemOpen
  \bibfield  {author} {\bibinfo {author} {\bibfnamefont {K.}~\bibnamefont
  {Yagi}}, \bibinfo {author} {\bibfnamefont {N.}~\bibnamefont {Yunes}}, \ and\
  \bibinfo {author} {\bibfnamefont {T.}~\bibnamefont {Tanaka}},\ }\href
  {\doibase 10.1103/PhysRevD.89.049902, 10.1103/PhysRevD.86.044037} {\bibfield
  {journal} {\bibinfo  {journal} {Phys. Rev.}\ }\textbf {\bibinfo {volume}
  {D86}},\ \bibinfo {pages} {044037} (\bibinfo {year} {2012})},\ \bibinfo
  {note} {[Erratum: Phys. Rev.D89,049902(2014)]},\ \Eprint
  {http://arxiv.org/abs/1206.6130} {arXiv:1206.6130 [gr-qc]} \BibitemShut
  {NoStop}%
\bibitem [{\citenamefont {Ayzenberg}\ and\ \citenamefont
  {Yunes}(2014)}]{Ayzenberg:2014aka}%
  \BibitemOpen
  \bibfield  {author} {\bibinfo {author} {\bibfnamefont {D.}~\bibnamefont
  {Ayzenberg}}\ and\ \bibinfo {author} {\bibfnamefont {N.}~\bibnamefont
  {Yunes}},\ }\href {\doibase 10.1103/PhysRevD.91.069905,
  10.1103/PhysRevD.90.044066} {\bibfield  {journal} {\bibinfo  {journal} {Phys.
  Rev.}\ }\textbf {\bibinfo {volume} {D90}},\ \bibinfo {pages} {044066}
  (\bibinfo {year} {2014})},\ \bibinfo {note} {[Erratum: Phys.
  Rev.D91,no.6,069905(2015)]},\ \Eprint {http://arxiv.org/abs/1405.2133}
  {arXiv:1405.2133 [gr-qc]} \BibitemShut {NoStop}%
\bibitem [{\citenamefont {Maselli}\ \emph {et~al.}(2015)\citenamefont
  {Maselli}, \citenamefont {Pani}, \citenamefont {Gualtieri},\ and\
  \citenamefont {Ferrari}}]{Maselli:2015tta}%
  \BibitemOpen
  \bibfield  {author} {\bibinfo {author} {\bibfnamefont {A.}~\bibnamefont
  {Maselli}}, \bibinfo {author} {\bibfnamefont {P.}~\bibnamefont {Pani}},
  \bibinfo {author} {\bibfnamefont {L.}~\bibnamefont {Gualtieri}}, \ and\
  \bibinfo {author} {\bibfnamefont {V.}~\bibnamefont {Ferrari}},\ }\href
  {\doibase 10.1103/PhysRevD.92.083014} {\bibfield  {journal} {\bibinfo
  {journal} {Phys. Rev.}\ }\textbf {\bibinfo {volume} {D92}},\ \bibinfo {pages}
  {083014} (\bibinfo {year} {2015})},\ \Eprint
  {http://arxiv.org/abs/1507.00680} {arXiv:1507.00680 [gr-qc]} \BibitemShut
  {NoStop}%
\bibitem [{\citenamefont {Kleihaus}\ \emph {et~al.}(2016)\citenamefont
  {Kleihaus}, \citenamefont {Kunz}, \citenamefont {Mojica},\ and\ \citenamefont
  {Radu}}]{Kleihaus:2015aje}%
  \BibitemOpen
  \bibfield  {author} {\bibinfo {author} {\bibfnamefont {B.}~\bibnamefont
  {Kleihaus}}, \bibinfo {author} {\bibfnamefont {J.}~\bibnamefont {Kunz}},
  \bibinfo {author} {\bibfnamefont {S.}~\bibnamefont {Mojica}}, \ and\ \bibinfo
  {author} {\bibfnamefont {E.}~\bibnamefont {Radu}},\ }\href {\doibase
  10.1103/PhysRevD.93.044047} {\bibfield  {journal} {\bibinfo  {journal} {Phys.
  Rev. D}\ }\textbf {\bibinfo {volume} {93}},\ \bibinfo {pages} {044047}
  (\bibinfo {year} {2016})},\ \Eprint {http://arxiv.org/abs/1511.05513}
  {arXiv:1511.05513 [gr-qc]} \BibitemShut {NoStop}%
\bibitem [{\citenamefont {Delsate}\ \emph {et~al.}(2018)\citenamefont
  {Delsate}, \citenamefont {Herdeiro},\ and\ \citenamefont
  {Radu}}]{Delsate:2018ome}%
  \BibitemOpen
  \bibfield  {author} {\bibinfo {author} {\bibfnamefont {T.}~\bibnamefont
  {Delsate}}, \bibinfo {author} {\bibfnamefont {C.}~\bibnamefont {Herdeiro}}, \
  and\ \bibinfo {author} {\bibfnamefont {E.}~\bibnamefont {Radu}},\ }\href
  {\doibase 10.1016/j.physletb.2018.09.060} {\bibfield  {journal} {\bibinfo
  {journal} {Phys. Lett.}\ }\textbf {\bibinfo {volume} {B787}},\ \bibinfo
  {pages} {8} (\bibinfo {year} {2018})},\ \Eprint
  {http://arxiv.org/abs/1806.06700} {arXiv:1806.06700 [gr-qc]} \BibitemShut
  {NoStop}%
\bibitem [{\citenamefont {Cardoso}\ \emph {et~al.}(2018)\citenamefont
  {Cardoso}, \citenamefont {Kimura}, \citenamefont {Maselli},\ and\
  \citenamefont {Senatore}}]{Cardoso:2018ptl}%
  \BibitemOpen
  \bibfield  {author} {\bibinfo {author} {\bibfnamefont {V.}~\bibnamefont
  {Cardoso}}, \bibinfo {author} {\bibfnamefont {M.}~\bibnamefont {Kimura}},
  \bibinfo {author} {\bibfnamefont {A.}~\bibnamefont {Maselli}}, \ and\
  \bibinfo {author} {\bibfnamefont {L.}~\bibnamefont {Senatore}},\ }\href
  {\doibase 10.1103/PhysRevLett.121.251105} {\bibfield  {journal} {\bibinfo
  {journal} {Phys. Rev. Lett.}\ }\textbf {\bibinfo {volume} {121}},\ \bibinfo
  {pages} {251105} (\bibinfo {year} {2018})},\ \Eprint
  {http://arxiv.org/abs/1808.08962} {arXiv:1808.08962 [gr-qc]} \BibitemShut
  {NoStop}%
\bibitem [{\citenamefont {Cano}\ and\ \citenamefont
  {Ruipérez}(2019)}]{Cano:2019ore}%
  \BibitemOpen
  \bibfield  {author} {\bibinfo {author} {\bibfnamefont {P.~A.}\ \bibnamefont
  {Cano}}\ and\ \bibinfo {author} {\bibfnamefont {A.}~\bibnamefont
  {Ruipérez}},\ }\href {\doibase 10.1007/JHEP05(2019)189} {\bibfield
  {journal} {\bibinfo  {journal} {JHEP}\ }\textbf {\bibinfo {volume} {05}},\
  \bibinfo {pages} {189} (\bibinfo {year} {2019})},\ \bibinfo {note} {[Erratum:
  JHEP 03, 187 (2020)]},\ \Eprint {http://arxiv.org/abs/1901.01315}
  {arXiv:1901.01315 [gr-qc]} \BibitemShut {NoStop}%
\bibitem [{\citenamefont {Adair}\ \emph {et~al.}(2020)\citenamefont {Adair},
  \citenamefont {Bueno}, \citenamefont {Cano}, \citenamefont {Hennigar},\ and\
  \citenamefont {Mann}}]{Adair:2020vso}%
  \BibitemOpen
  \bibfield  {author} {\bibinfo {author} {\bibfnamefont {C.}~\bibnamefont
  {Adair}}, \bibinfo {author} {\bibfnamefont {P.}~\bibnamefont {Bueno}},
  \bibinfo {author} {\bibfnamefont {P.~A.}\ \bibnamefont {Cano}}, \bibinfo
  {author} {\bibfnamefont {R.~A.}\ \bibnamefont {Hennigar}}, \ and\ \bibinfo
  {author} {\bibfnamefont {R.~B.}\ \bibnamefont {Mann}},\ }\href@noop {} {\
  (\bibinfo {year} {2020})},\ \Eprint {http://arxiv.org/abs/2004.09598}
  {arXiv:2004.09598 [gr-qc]} \BibitemShut {NoStop}%
\bibitem [{\citenamefont {Berti}\ \emph {et~al.}(2009)\citenamefont {Berti},
  \citenamefont {Cardoso},\ and\ \citenamefont {Starinets}}]{Berti:2009kk}%
  \BibitemOpen
  \bibfield  {author} {\bibinfo {author} {\bibfnamefont {E.}~\bibnamefont
  {Berti}}, \bibinfo {author} {\bibfnamefont {V.}~\bibnamefont {Cardoso}}, \
  and\ \bibinfo {author} {\bibfnamefont {A.~O.}\ \bibnamefont {Starinets}},\
  }\href {\doibase 10.1088/0264-9381/26/16/163001} {\bibfield  {journal}
  {\bibinfo  {journal} {Class. Quant. Grav.}\ }\textbf {\bibinfo {volume}
  {26}},\ \bibinfo {pages} {163001} (\bibinfo {year} {2009})},\ \Eprint
  {http://arxiv.org/abs/0905.2975} {arXiv:0905.2975 [gr-qc]} \BibitemShut
  {NoStop}%
\bibitem [{\citenamefont {Konoplya}\ and\ \citenamefont
  {Zhidenko}(2011)}]{Konoplya:2011qq}%
  \BibitemOpen
  \bibfield  {author} {\bibinfo {author} {\bibfnamefont {R.~A.}\ \bibnamefont
  {Konoplya}}\ and\ \bibinfo {author} {\bibfnamefont {A.}~\bibnamefont
  {Zhidenko}},\ }\href {\doibase 10.1103/RevModPhys.83.793} {\bibfield
  {journal} {\bibinfo  {journal} {Rev. Mod. Phys.}\ }\textbf {\bibinfo {volume}
  {83}},\ \bibinfo {pages} {793} (\bibinfo {year} {2011})},\ \Eprint
  {http://arxiv.org/abs/1102.4014} {arXiv:1102.4014 [gr-qc]} \BibitemShut
  {NoStop}%
\bibitem [{\citenamefont {Cardoso}\ and\ \citenamefont
  {Gualtieri}(2009)}]{Cardoso:2009pk}%
  \BibitemOpen
  \bibfield  {author} {\bibinfo {author} {\bibfnamefont {V.}~\bibnamefont
  {Cardoso}}\ and\ \bibinfo {author} {\bibfnamefont {L.}~\bibnamefont
  {Gualtieri}},\ }\href {\doibase 10.1103/PhysRevD.81.089903,
  10.1103/PhysRevD.80.064008} {\bibfield  {journal} {\bibinfo  {journal} {Phys.
  Rev.}\ }\textbf {\bibinfo {volume} {D80}},\ \bibinfo {pages} {064008}
  (\bibinfo {year} {2009})},\ \bibinfo {note} {[Erratum: Phys.
  Rev.D81,089903(2010)]},\ \Eprint {http://arxiv.org/abs/0907.5008}
  {arXiv:0907.5008 [gr-qc]} \BibitemShut {NoStop}%
\bibitem [{\citenamefont {Molina}\ \emph {et~al.}(2010)\citenamefont {Molina},
  \citenamefont {Pani}, \citenamefont {Cardoso},\ and\ \citenamefont
  {Gualtieri}}]{Molina:2010fb}%
  \BibitemOpen
  \bibfield  {author} {\bibinfo {author} {\bibfnamefont {C.}~\bibnamefont
  {Molina}}, \bibinfo {author} {\bibfnamefont {P.}~\bibnamefont {Pani}},
  \bibinfo {author} {\bibfnamefont {V.}~\bibnamefont {Cardoso}}, \ and\
  \bibinfo {author} {\bibfnamefont {L.}~\bibnamefont {Gualtieri}},\ }\href
  {\doibase 10.1103/PhysRevD.81.124021} {\bibfield  {journal} {\bibinfo
  {journal} {Phys. Rev. D}\ }\textbf {\bibinfo {volume} {81}},\ \bibinfo
  {pages} {124021} (\bibinfo {year} {2010})},\ \Eprint
  {http://arxiv.org/abs/1004.4007} {arXiv:1004.4007 [gr-qc]} \BibitemShut
  {NoStop}%
\bibitem [{\citenamefont {Blázquez-Salcedo}\ \emph {et~al.}(2016)\citenamefont
  {Blázquez-Salcedo}, \citenamefont {Macedo}, \citenamefont {Cardoso},
  \citenamefont {Ferrari}, \citenamefont {Gualtieri}, \citenamefont {Khoo},
  \citenamefont {Kunz},\ and\ \citenamefont {Pani}}]{Blazquez-Salcedo:2016enn}%
  \BibitemOpen
  \bibfield  {author} {\bibinfo {author} {\bibfnamefont {J.~L.}\ \bibnamefont
  {Blázquez-Salcedo}}, \bibinfo {author} {\bibfnamefont {C.~F.~B.}\
  \bibnamefont {Macedo}}, \bibinfo {author} {\bibfnamefont {V.}~\bibnamefont
  {Cardoso}}, \bibinfo {author} {\bibfnamefont {V.}~\bibnamefont {Ferrari}},
  \bibinfo {author} {\bibfnamefont {L.}~\bibnamefont {Gualtieri}}, \bibinfo
  {author} {\bibfnamefont {F.~S.}\ \bibnamefont {Khoo}}, \bibinfo {author}
  {\bibfnamefont {J.}~\bibnamefont {Kunz}}, \ and\ \bibinfo {author}
  {\bibfnamefont {P.}~\bibnamefont {Pani}},\ }\href {\doibase
  10.1103/PhysRevD.94.104024} {\bibfield  {journal} {\bibinfo  {journal} {Phys.
  Rev.}\ }\textbf {\bibinfo {volume} {D94}},\ \bibinfo {pages} {104024}
  (\bibinfo {year} {2016})},\ \Eprint {http://arxiv.org/abs/1609.01286}
  {arXiv:1609.01286 [gr-qc]} \BibitemShut {NoStop}%
\bibitem [{\citenamefont {Blázquez-Salcedo}\ \emph {et~al.}(2017)\citenamefont
  {Blázquez-Salcedo}, \citenamefont {Khoo},\ and\ \citenamefont
  {Kunz}}]{Blazquez-Salcedo:2017txk}%
  \BibitemOpen
  \bibfield  {author} {\bibinfo {author} {\bibfnamefont {J.~L.}\ \bibnamefont
  {Blázquez-Salcedo}}, \bibinfo {author} {\bibfnamefont {F.~S.}\ \bibnamefont
  {Khoo}}, \ and\ \bibinfo {author} {\bibfnamefont {J.}~\bibnamefont {Kunz}},\
  }\href {\doibase 10.1103/PhysRevD.96.064008} {\bibfield  {journal} {\bibinfo
  {journal} {Phys. Rev. D}\ }\textbf {\bibinfo {volume} {96}},\ \bibinfo
  {pages} {064008} (\bibinfo {year} {2017})},\ \Eprint
  {http://arxiv.org/abs/1706.03262} {arXiv:1706.03262 [gr-qc]} \BibitemShut
  {NoStop}%
\bibitem [{\citenamefont {McManus}\ \emph {et~al.}(2019)\citenamefont
  {McManus}, \citenamefont {Berti}, \citenamefont {Macedo}, \citenamefont
  {Kimura}, \citenamefont {Maselli},\ and\ \citenamefont
  {Cardoso}}]{McManus:2019ulj}%
  \BibitemOpen
  \bibfield  {author} {\bibinfo {author} {\bibfnamefont {R.}~\bibnamefont
  {McManus}}, \bibinfo {author} {\bibfnamefont {E.}~\bibnamefont {Berti}},
  \bibinfo {author} {\bibfnamefont {C.~F.}\ \bibnamefont {Macedo}}, \bibinfo
  {author} {\bibfnamefont {M.}~\bibnamefont {Kimura}}, \bibinfo {author}
  {\bibfnamefont {A.}~\bibnamefont {Maselli}}, \ and\ \bibinfo {author}
  {\bibfnamefont {V.}~\bibnamefont {Cardoso}},\ }\href {\doibase
  10.1103/PhysRevD.100.044061} {\bibfield  {journal} {\bibinfo  {journal}
  {Phys. Rev. D}\ }\textbf {\bibinfo {volume} {100}},\ \bibinfo {pages}
  {044061} (\bibinfo {year} {2019})},\ \Eprint
  {http://arxiv.org/abs/1906.05155} {arXiv:1906.05155 [gr-qc]} \BibitemShut
  {NoStop}%
\bibitem [{\citenamefont {Tattersall}\ and\ \citenamefont
  {Ferreira}(2018)}]{Tattersall:2018nve}%
  \BibitemOpen
  \bibfield  {author} {\bibinfo {author} {\bibfnamefont {O.~J.}\ \bibnamefont
  {Tattersall}}\ and\ \bibinfo {author} {\bibfnamefont {P.~G.}\ \bibnamefont
  {Ferreira}},\ }\href {\doibase 10.1103/PhysRevD.97.104047} {\bibfield
  {journal} {\bibinfo  {journal} {Phys. Rev. D}\ }\textbf {\bibinfo {volume}
  {97}},\ \bibinfo {pages} {104047} (\bibinfo {year} {2018})},\ \Eprint
  {http://arxiv.org/abs/1804.08950} {arXiv:1804.08950 [gr-qc]} \BibitemShut
  {NoStop}%
\bibitem [{\citenamefont {Konoplya}\ and\ \citenamefont
  {Zinhailo}(2020)}]{Konoplya:2020bxa}%
  \BibitemOpen
  \bibfield  {author} {\bibinfo {author} {\bibfnamefont {R.}~\bibnamefont
  {Konoplya}}\ and\ \bibinfo {author} {\bibfnamefont {A.}~\bibnamefont
  {Zinhailo}},\ }\href@noop {} {\  (\bibinfo {year} {2020})},\ \Eprint
  {http://arxiv.org/abs/2003.01188} {arXiv:2003.01188 [gr-qc]} \BibitemShut
  {NoStop}%
\bibitem [{\citenamefont {Moura}\ and\ \citenamefont
  {Rodrigues}(2021{\natexlab{a}})}]{Moura:2021eln}%
  \BibitemOpen
  \bibfield  {author} {\bibinfo {author} {\bibfnamefont {F.}~\bibnamefont
  {Moura}}\ and\ \bibinfo {author} {\bibfnamefont {J.~a.}\ \bibnamefont
  {Rodrigues}},\ }\href {\doibase 10.1016/j.physletb.2021.136407} {\bibfield
  {journal} {\bibinfo  {journal} {Phys. Lett. B}\ }\textbf {\bibinfo {volume}
  {819}},\ \bibinfo {pages} {136407} (\bibinfo {year} {2021}{\natexlab{a}})},\
  \Eprint {http://arxiv.org/abs/2103.09302} {arXiv:2103.09302 [hep-th]}
  \BibitemShut {NoStop}%
\bibitem [{\citenamefont {Moura}\ and\ \citenamefont
  {Rodrigues}(2021{\natexlab{b}})}]{Moura:2021nuh}%
  \BibitemOpen
  \bibfield  {author} {\bibinfo {author} {\bibfnamefont {F.}~\bibnamefont
  {Moura}}\ and\ \bibinfo {author} {\bibfnamefont {J.~a.}\ \bibnamefont
  {Rodrigues}},\ }\href {\doibase 10.1007/JHEP08(2021)078} {\bibfield
  {journal} {\bibinfo  {journal} {JHEP}\ }\textbf {\bibinfo {volume} {08}},\
  \bibinfo {pages} {078} (\bibinfo {year} {2021}{\natexlab{b}})},\ \Eprint
  {http://arxiv.org/abs/2105.02616} {arXiv:2105.02616 [hep-th]} \BibitemShut
  {NoStop}%
\bibitem [{\citenamefont {Tattersall}\ \emph {et~al.}(2018)\citenamefont
  {Tattersall}, \citenamefont {Ferreira},\ and\ \citenamefont
  {Lagos}}]{Tattersall:2017erk}%
  \BibitemOpen
  \bibfield  {author} {\bibinfo {author} {\bibfnamefont {O.~J.}\ \bibnamefont
  {Tattersall}}, \bibinfo {author} {\bibfnamefont {P.~G.}\ \bibnamefont
  {Ferreira}}, \ and\ \bibinfo {author} {\bibfnamefont {M.}~\bibnamefont
  {Lagos}},\ }\href {\doibase 10.1103/PhysRevD.97.044021} {\bibfield  {journal}
  {\bibinfo  {journal} {Phys. Rev. D}\ }\textbf {\bibinfo {volume} {97}},\
  \bibinfo {pages} {044021} (\bibinfo {year} {2018})},\ \Eprint
  {http://arxiv.org/abs/1711.01992} {arXiv:1711.01992 [gr-qc]} \BibitemShut
  {NoStop}%
\bibitem [{\citenamefont {Cardoso}\ \emph {et~al.}(2019)\citenamefont
  {Cardoso}, \citenamefont {Kimura}, \citenamefont {Maselli}, \citenamefont
  {Berti}, \citenamefont {Macedo},\ and\ \citenamefont
  {McManus}}]{Cardoso:2019mqo}%
  \BibitemOpen
  \bibfield  {author} {\bibinfo {author} {\bibfnamefont {V.}~\bibnamefont
  {Cardoso}}, \bibinfo {author} {\bibfnamefont {M.}~\bibnamefont {Kimura}},
  \bibinfo {author} {\bibfnamefont {A.}~\bibnamefont {Maselli}}, \bibinfo
  {author} {\bibfnamefont {E.}~\bibnamefont {Berti}}, \bibinfo {author}
  {\bibfnamefont {C.~F.}\ \bibnamefont {Macedo}}, \ and\ \bibinfo {author}
  {\bibfnamefont {R.}~\bibnamefont {McManus}},\ }\href {\doibase
  10.1103/PhysRevD.99.104077} {\bibfield  {journal} {\bibinfo  {journal} {Phys.
  Rev. D}\ }\textbf {\bibinfo {volume} {99}},\ \bibinfo {pages} {104077}
  (\bibinfo {year} {2019})},\ \Eprint {http://arxiv.org/abs/1901.01265}
  {arXiv:1901.01265 [gr-qc]} \BibitemShut {NoStop}%
\bibitem [{\citenamefont {Glampedakis}\ and\ \citenamefont
  {Silva}(2019)}]{Glampedakis:2019dqh}%
  \BibitemOpen
  \bibfield  {author} {\bibinfo {author} {\bibfnamefont {K.}~\bibnamefont
  {Glampedakis}}\ and\ \bibinfo {author} {\bibfnamefont {H.~O.}\ \bibnamefont
  {Silva}},\ }\href {\doibase 10.1103/PhysRevD.100.044040} {\bibfield
  {journal} {\bibinfo  {journal} {Phys. Rev. D}\ }\textbf {\bibinfo {volume}
  {100}},\ \bibinfo {pages} {044040} (\bibinfo {year} {2019})},\ \Eprint
  {http://arxiv.org/abs/1906.05455} {arXiv:1906.05455 [gr-qc]} \BibitemShut
  {NoStop}%
\bibitem [{\citenamefont {Silva}\ and\ \citenamefont
  {Glampedakis}(2020)}]{Silva:2019scu}%
  \BibitemOpen
  \bibfield  {author} {\bibinfo {author} {\bibfnamefont {H.~O.}\ \bibnamefont
  {Silva}}\ and\ \bibinfo {author} {\bibfnamefont {K.}~\bibnamefont
  {Glampedakis}},\ }\href {\doibase 10.1103/PhysRevD.101.044051} {\bibfield
  {journal} {\bibinfo  {journal} {Phys. Rev. D}\ }\textbf {\bibinfo {volume}
  {101}},\ \bibinfo {pages} {044051} (\bibinfo {year} {2020})},\ \Eprint
  {http://arxiv.org/abs/1912.09286} {arXiv:1912.09286 [gr-qc]} \BibitemShut
  {NoStop}%
\bibitem [{\citenamefont {Bryant}\ \emph {et~al.}(2021)\citenamefont {Bryant},
  \citenamefont {Silva}, \citenamefont {Yagi},\ and\ \citenamefont
  {Glampedakis}}]{Bryant:2021xdh}%
  \BibitemOpen
  \bibfield  {author} {\bibinfo {author} {\bibfnamefont {A.}~\bibnamefont
  {Bryant}}, \bibinfo {author} {\bibfnamefont {H.~O.}\ \bibnamefont {Silva}},
  \bibinfo {author} {\bibfnamefont {K.}~\bibnamefont {Yagi}}, \ and\ \bibinfo
  {author} {\bibfnamefont {K.}~\bibnamefont {Glampedakis}},\ }\href {\doibase
  10.1103/PhysRevD.104.044051} {\bibfield  {journal} {\bibinfo  {journal}
  {Phys. Rev. D}\ }\textbf {\bibinfo {volume} {104}},\ \bibinfo {pages}
  {044051} (\bibinfo {year} {2021})},\ \Eprint
  {http://arxiv.org/abs/2106.09657} {arXiv:2106.09657 [gr-qc]} \BibitemShut
  {NoStop}%
\bibitem [{\citenamefont {Cano}\ \emph {et~al.}(2020)\citenamefont {Cano},
  \citenamefont {Fransen},\ and\ \citenamefont {Hertog}}]{Cano:2020cao}%
  \BibitemOpen
  \bibfield  {author} {\bibinfo {author} {\bibfnamefont {P.~A.}\ \bibnamefont
  {Cano}}, \bibinfo {author} {\bibfnamefont {K.}~\bibnamefont {Fransen}}, \
  and\ \bibinfo {author} {\bibfnamefont {T.}~\bibnamefont {Hertog}},\ }\href
  {\doibase 10.1103/PhysRevD.102.044047} {\bibfield  {journal} {\bibinfo
  {journal} {Phys. Rev. D}\ }\textbf {\bibinfo {volume} {102}},\ \bibinfo
  {pages} {044047} (\bibinfo {year} {2020})},\ \Eprint
  {http://arxiv.org/abs/2005.03671} {arXiv:2005.03671 [gr-qc]} \BibitemShut
  {NoStop}%
\bibitem [{\citenamefont {Maselli}\ \emph {et~al.}(2020)\citenamefont
  {Maselli}, \citenamefont {Pani}, \citenamefont {Gualtieri},\ and\
  \citenamefont {Berti}}]{Maselli:2019mjd}%
  \BibitemOpen
  \bibfield  {author} {\bibinfo {author} {\bibfnamefont {A.}~\bibnamefont
  {Maselli}}, \bibinfo {author} {\bibfnamefont {P.}~\bibnamefont {Pani}},
  \bibinfo {author} {\bibfnamefont {L.}~\bibnamefont {Gualtieri}}, \ and\
  \bibinfo {author} {\bibfnamefont {E.}~\bibnamefont {Berti}},\ }\href
  {\doibase 10.1103/PhysRevD.101.024043} {\bibfield  {journal} {\bibinfo
  {journal} {Phys. Rev.}\ }\textbf {\bibinfo {volume} {D101}},\ \bibinfo
  {pages} {024043} (\bibinfo {year} {2020})},\ \Eprint
  {http://arxiv.org/abs/1910.12893} {arXiv:1910.12893 [gr-qc]} \BibitemShut
  {NoStop}%
\bibitem [{\citenamefont {Pani}\ \emph {et~al.}(2012)\citenamefont {Pani},
  \citenamefont {Cardoso}, \citenamefont {Gualtieri}, \citenamefont {Berti},\
  and\ \citenamefont {Ishibashi}}]{Pani:2012bp}%
  \BibitemOpen
  \bibfield  {author} {\bibinfo {author} {\bibfnamefont {P.}~\bibnamefont
  {Pani}}, \bibinfo {author} {\bibfnamefont {V.}~\bibnamefont {Cardoso}},
  \bibinfo {author} {\bibfnamefont {L.}~\bibnamefont {Gualtieri}}, \bibinfo
  {author} {\bibfnamefont {E.}~\bibnamefont {Berti}}, \ and\ \bibinfo {author}
  {\bibfnamefont {A.}~\bibnamefont {Ishibashi}},\ }\href {\doibase
  10.1103/PhysRevD.86.104017} {\bibfield  {journal} {\bibinfo  {journal} {Phys.
  Rev. D}\ }\textbf {\bibinfo {volume} {86}},\ \bibinfo {pages} {104017}
  (\bibinfo {year} {2012})},\ \Eprint {http://arxiv.org/abs/1209.0773}
  {arXiv:1209.0773 [gr-qc]} \BibitemShut {NoStop}%
\bibitem [{\citenamefont {Pani}(2013)}]{Pani:2013pma}%
  \BibitemOpen
  \bibfield  {author} {\bibinfo {author} {\bibfnamefont {P.}~\bibnamefont
  {Pani}},\ }\href {\doibase 10.1142/S0217751X13400186} {\bibfield  {journal}
  {\bibinfo  {journal} {Int. J. Mod. Phys. A}\ }\textbf {\bibinfo {volume}
  {28}},\ \bibinfo {pages} {1340018} (\bibinfo {year} {2013})},\ \Eprint
  {http://arxiv.org/abs/1305.6759} {arXiv:1305.6759 [gr-qc]} \BibitemShut
  {NoStop}%
\bibitem [{\citenamefont {Pierini}\ and\ \citenamefont
  {Gualtieri}(2021)}]{Pierini:2021jxd}%
  \BibitemOpen
  \bibfield  {author} {\bibinfo {author} {\bibfnamefont {L.}~\bibnamefont
  {Pierini}}\ and\ \bibinfo {author} {\bibfnamefont {L.}~\bibnamefont
  {Gualtieri}},\ }\href {\doibase 10.1103/PhysRevD.103.124017} {\bibfield
  {journal} {\bibinfo  {journal} {Phys. Rev. D}\ }\textbf {\bibinfo {volume}
  {103}},\ \bibinfo {pages} {124017} (\bibinfo {year} {2021})},\ \Eprint
  {http://arxiv.org/abs/2103.09870} {arXiv:2103.09870 [gr-qc]} \BibitemShut
  {NoStop}%
\bibitem [{\citenamefont {Wagle}\ \emph {et~al.}(2021)\citenamefont {Wagle},
  \citenamefont {Yunes},\ and\ \citenamefont {Silva}}]{Wagle:2021tam}%
  \BibitemOpen
  \bibfield  {author} {\bibinfo {author} {\bibfnamefont {P.}~\bibnamefont
  {Wagle}}, \bibinfo {author} {\bibfnamefont {N.}~\bibnamefont {Yunes}}, \ and\
  \bibinfo {author} {\bibfnamefont {H.~O.}\ \bibnamefont {Silva}},\ }\href@noop
  {} {\  (\bibinfo {year} {2021})},\ \Eprint {http://arxiv.org/abs/2103.09913}
  {arXiv:2103.09913 [gr-qc]} \BibitemShut {NoStop}%
\bibitem [{\citenamefont {Srivastava}\ \emph {et~al.}(2021)\citenamefont
  {Srivastava}, \citenamefont {Chen},\ and\ \citenamefont
  {Shankaranarayanan}}]{Srivastava:2021imr}%
  \BibitemOpen
  \bibfield  {author} {\bibinfo {author} {\bibfnamefont {M.}~\bibnamefont
  {Srivastava}}, \bibinfo {author} {\bibfnamefont {Y.}~\bibnamefont {Chen}}, \
  and\ \bibinfo {author} {\bibfnamefont {S.}~\bibnamefont
  {Shankaranarayanan}},\ }\href {\doibase 10.1103/PhysRevD.104.064034}
  {\bibfield  {journal} {\bibinfo  {journal} {Phys. Rev. D}\ }\textbf {\bibinfo
  {volume} {104}},\ \bibinfo {pages} {064034} (\bibinfo {year} {2021})},\
  \Eprint {http://arxiv.org/abs/2106.06209} {arXiv:2106.06209 [gr-qc]}
  \BibitemShut {NoStop}%
\bibitem [{\citenamefont {Endlich}\ \emph {et~al.}(2017)\citenamefont
  {Endlich}, \citenamefont {Gorbenko}, \citenamefont {Huang},\ and\
  \citenamefont {Senatore}}]{Endlich:2017tqa}%
  \BibitemOpen
  \bibfield  {author} {\bibinfo {author} {\bibfnamefont {S.}~\bibnamefont
  {Endlich}}, \bibinfo {author} {\bibfnamefont {V.}~\bibnamefont {Gorbenko}},
  \bibinfo {author} {\bibfnamefont {J.}~\bibnamefont {Huang}}, \ and\ \bibinfo
  {author} {\bibfnamefont {L.}~\bibnamefont {Senatore}},\ }\href {\doibase
  10.1007/JHEP09(2017)122} {\bibfield  {journal} {\bibinfo  {journal} {JHEP}\
  }\textbf {\bibinfo {volume} {09}},\ \bibinfo {pages} {122} (\bibinfo {year}
  {2017})},\ \Eprint {http://arxiv.org/abs/1704.01590} {arXiv:1704.01590
  [gr-qc]} \BibitemShut {NoStop}%
\bibitem [{\citenamefont {Gruzinov}\ and\ \citenamefont
  {Kleban}(2007)}]{Gruzinov:2006ie}%
  \BibitemOpen
  \bibfield  {author} {\bibinfo {author} {\bibfnamefont {A.}~\bibnamefont
  {Gruzinov}}\ and\ \bibinfo {author} {\bibfnamefont {M.}~\bibnamefont
  {Kleban}},\ }\href {\doibase 10.1088/0264-9381/24/13/N02} {\bibfield
  {journal} {\bibinfo  {journal} {Class. Quant. Grav.}\ }\textbf {\bibinfo
  {volume} {24}},\ \bibinfo {pages} {3521} (\bibinfo {year} {2007})},\ \Eprint
  {http://arxiv.org/abs/hep-th/0612015} {arXiv:hep-th/0612015 [hep-th]}
  \BibitemShut {NoStop}%
\bibitem [{\citenamefont {Camanho}\ \emph {et~al.}(2016)\citenamefont
  {Camanho}, \citenamefont {Edelstein}, \citenamefont {Maldacena},\ and\
  \citenamefont {Zhiboedov}}]{Camanho:2014apa}%
  \BibitemOpen
  \bibfield  {author} {\bibinfo {author} {\bibfnamefont {X.~O.}\ \bibnamefont
  {Camanho}}, \bibinfo {author} {\bibfnamefont {J.~D.}\ \bibnamefont
  {Edelstein}}, \bibinfo {author} {\bibfnamefont {J.}~\bibnamefont
  {Maldacena}}, \ and\ \bibinfo {author} {\bibfnamefont {A.}~\bibnamefont
  {Zhiboedov}},\ }\href {\doibase 10.1007/JHEP02(2016)020} {\bibfield
  {journal} {\bibinfo  {journal} {JHEP}\ }\textbf {\bibinfo {volume} {02}},\
  \bibinfo {pages} {020} (\bibinfo {year} {2016})},\ \Eprint
  {http://arxiv.org/abs/1407.5597} {arXiv:1407.5597 [hep-th]} \BibitemShut
  {NoStop}%
\bibitem [{\citenamefont {Bern}\ \emph {et~al.}(2021)\citenamefont {Bern},
  \citenamefont {Kosmopoulos},\ and\ \citenamefont {Zhiboedov}}]{Bern:2021ppb}%
  \BibitemOpen
  \bibfield  {author} {\bibinfo {author} {\bibfnamefont {Z.}~\bibnamefont
  {Bern}}, \bibinfo {author} {\bibfnamefont {D.}~\bibnamefont {Kosmopoulos}}, \
  and\ \bibinfo {author} {\bibfnamefont {A.}~\bibnamefont {Zhiboedov}},\ }\href
  {\doibase 10.1088/1751-8121/ac0e51} {\bibfield  {journal} {\bibinfo
  {journal} {J. Phys. A}\ }\textbf {\bibinfo {volume} {54}},\ \bibinfo {pages}
  {344002} (\bibinfo {year} {2021})},\ \Eprint
  {http://arxiv.org/abs/2103.12728} {arXiv:2103.12728 [hep-th]} \BibitemShut
  {NoStop}%
\bibitem [{\citenamefont {Martel}\ and\ \citenamefont
  {Poisson}(2005)}]{Martel:2005ir}%
  \BibitemOpen
  \bibfield  {author} {\bibinfo {author} {\bibfnamefont {K.}~\bibnamefont
  {Martel}}\ and\ \bibinfo {author} {\bibfnamefont {E.}~\bibnamefont
  {Poisson}},\ }\href {\doibase 10.1103/PhysRevD.71.104003} {\bibfield
  {journal} {\bibinfo  {journal} {Phys. Rev. D}\ }\textbf {\bibinfo {volume}
  {71}},\ \bibinfo {pages} {104003} (\bibinfo {year} {2005})},\ \Eprint
  {http://arxiv.org/abs/gr-qc/0502028} {arXiv:gr-qc/0502028} \BibitemShut
  {NoStop}%
\bibitem [{\citenamefont {Regge}\ and\ \citenamefont
  {Wheeler}(1957)}]{Regge:1957td}%
  \BibitemOpen
  \bibfield  {author} {\bibinfo {author} {\bibfnamefont {T.}~\bibnamefont
  {Regge}}\ and\ \bibinfo {author} {\bibfnamefont {J.~A.}\ \bibnamefont
  {Wheeler}},\ }\href {\doibase 10.1103/PhysRev.108.1063} {\bibfield  {journal}
  {\bibinfo  {journal} {Phys. Rev.}\ }\textbf {\bibinfo {volume} {108}},\
  \bibinfo {pages} {1063} (\bibinfo {year} {1957})}\BibitemShut {NoStop}%
\bibitem [{\citenamefont {Zerilli}(1970{\natexlab{a}})}]{Zerilli:1970wzz}%
  \BibitemOpen
  \bibfield  {author} {\bibinfo {author} {\bibfnamefont {F.~J.}\ \bibnamefont
  {Zerilli}},\ }\href {\doibase 10.1103/PhysRevD.2.2141} {\bibfield  {journal}
  {\bibinfo  {journal} {Phys. Rev. D}\ }\textbf {\bibinfo {volume} {2}},\
  \bibinfo {pages} {2141} (\bibinfo {year} {1970}{\natexlab{a}})}\BibitemShut
  {NoStop}%
\bibitem [{\citenamefont {Zerilli}(1970{\natexlab{b}})}]{Zerilli:1970se}%
  \BibitemOpen
  \bibfield  {author} {\bibinfo {author} {\bibfnamefont {F.~J.}\ \bibnamefont
  {Zerilli}},\ }\href {\doibase 10.1103/PhysRevLett.24.737} {\bibfield
  {journal} {\bibinfo  {journal} {Phys. Rev. Lett.}\ }\textbf {\bibinfo
  {volume} {24}},\ \bibinfo {pages} {737} (\bibinfo {year}
  {1970}{\natexlab{b}})}\BibitemShut {NoStop}%
\bibitem [{\citenamefont {Leaver}(1986)}]{Leaver:1986gd}%
  \BibitemOpen
  \bibfield  {author} {\bibinfo {author} {\bibfnamefont {E.~W.}\ \bibnamefont
  {Leaver}},\ }\href {\doibase 10.1103/PhysRevD.34.384} {\bibfield  {journal}
  {\bibinfo  {journal} {Phys. Rev. D}\ }\textbf {\bibinfo {volume} {34}},\
  \bibinfo {pages} {384} (\bibinfo {year} {1986})}\BibitemShut {NoStop}%
\bibitem [{\citenamefont {Zimmerman}\ \emph {et~al.}(2015)\citenamefont
  {Zimmerman}, \citenamefont {Yang}, \citenamefont {Mark}, \citenamefont
  {Chen},\ and\ \citenamefont {Lehner}}]{Zimmerman:2014aha}%
  \BibitemOpen
  \bibfield  {author} {\bibinfo {author} {\bibfnamefont {A.}~\bibnamefont
  {Zimmerman}}, \bibinfo {author} {\bibfnamefont {H.}~\bibnamefont {Yang}},
  \bibinfo {author} {\bibfnamefont {Z.}~\bibnamefont {Mark}}, \bibinfo {author}
  {\bibfnamefont {Y.}~\bibnamefont {Chen}}, \ and\ \bibinfo {author}
  {\bibfnamefont {L.}~\bibnamefont {Lehner}},\ }\bibfield  {booktitle} {\emph
  {\bibinfo {booktitle} {{Proceedings, 3rd Session of the Sant Cugat Forum on
  Astrophysics : Gravitational Waves Astrophysics: Sant Cugat, Catalonia,
  Spain, April 22-25, 2014}}},\ }\href {\doibase 10.1007/978-3-319-10488-1_19}
  {\bibfield  {journal} {\bibinfo  {journal} {Astrophys. Space Sci. Proc.}\
  }\textbf {\bibinfo {volume} {40}},\ \bibinfo {pages} {217} (\bibinfo {year}
  {2015})},\ \Eprint {http://arxiv.org/abs/1406.4206} {arXiv:1406.4206 [gr-qc]}
  \BibitemShut {NoStop}%
\bibitem [{\citenamefont {Mark}\ \emph {et~al.}(2015)\citenamefont {Mark},
  \citenamefont {Yang}, \citenamefont {Zimmerman},\ and\ \citenamefont
  {Chen}}]{Mark:2014aja}%
  \BibitemOpen
  \bibfield  {author} {\bibinfo {author} {\bibfnamefont {Z.}~\bibnamefont
  {Mark}}, \bibinfo {author} {\bibfnamefont {H.}~\bibnamefont {Yang}}, \bibinfo
  {author} {\bibfnamefont {A.}~\bibnamefont {Zimmerman}}, \ and\ \bibinfo
  {author} {\bibfnamefont {Y.}~\bibnamefont {Chen}},\ }\href {\doibase
  10.1103/PhysRevD.91.044025} {\bibfield  {journal} {\bibinfo  {journal} {Phys.
  Rev.}\ }\textbf {\bibinfo {volume} {D91}},\ \bibinfo {pages} {044025}
  (\bibinfo {year} {2015})},\ \Eprint {http://arxiv.org/abs/1409.5800}
  {arXiv:1409.5800 [gr-qc]} \BibitemShut {NoStop}%
\bibitem [{\citenamefont {Cardoso}\ \emph {et~al.}(2009)\citenamefont
  {Cardoso}, \citenamefont {Miranda}, \citenamefont {Berti}, \citenamefont
  {Witek},\ and\ \citenamefont {Zanchin}}]{Cardoso:2008bp}%
  \BibitemOpen
  \bibfield  {author} {\bibinfo {author} {\bibfnamefont {V.}~\bibnamefont
  {Cardoso}}, \bibinfo {author} {\bibfnamefont {A.~S.}\ \bibnamefont
  {Miranda}}, \bibinfo {author} {\bibfnamefont {E.}~\bibnamefont {Berti}},
  \bibinfo {author} {\bibfnamefont {H.}~\bibnamefont {Witek}}, \ and\ \bibinfo
  {author} {\bibfnamefont {V.~T.}\ \bibnamefont {Zanchin}},\ }\href {\doibase
  10.1103/PhysRevD.79.064016} {\bibfield  {journal} {\bibinfo  {journal} {Phys.
  Rev.}\ }\textbf {\bibinfo {volume} {D79}},\ \bibinfo {pages} {064016}
  (\bibinfo {year} {2009})},\ \Eprint {http://arxiv.org/abs/0812.1806}
  {arXiv:0812.1806 [hep-th]} \BibitemShut {NoStop}%
\bibitem [{\citenamefont {Sago}\ \emph {et~al.}(2021)\citenamefont {Sago},
  \citenamefont {Isoyama},\ and\ \citenamefont {Nakano}}]{Sago:2021gbq}%
  \BibitemOpen
  \bibfield  {author} {\bibinfo {author} {\bibfnamefont {N.}~\bibnamefont
  {Sago}}, \bibinfo {author} {\bibfnamefont {S.}~\bibnamefont {Isoyama}}, \
  and\ \bibinfo {author} {\bibfnamefont {H.}~\bibnamefont {Nakano}},\ }\href
  {\doibase 10.3390/universe7100357} {\bibfield  {journal} {\bibinfo  {journal}
  {Universe}\ }\textbf {\bibinfo {volume} {7}},\ \bibinfo {pages} {357}
  (\bibinfo {year} {2021})},\ \Eprint {http://arxiv.org/abs/2108.13017}
  {arXiv:2108.13017 [gr-qc]} \BibitemShut {NoStop}%
\bibitem [{\citenamefont {Oshita}(2021)}]{Oshita:2021iyn}%
  \BibitemOpen
  \bibfield  {author} {\bibinfo {author} {\bibfnamefont {N.}~\bibnamefont
  {Oshita}},\ }\href@noop {} {\  (\bibinfo {year} {2021})},\ \Eprint
  {http://arxiv.org/abs/2109.09757} {arXiv:2109.09757 [gr-qc]} \BibitemShut
  {NoStop}%
\bibitem [{\citenamefont {Glampedakis}\ \emph {et~al.}(2017)\citenamefont
  {Glampedakis}, \citenamefont {Johnson},\ and\ \citenamefont
  {Kennefick}}]{Glampedakis:2017rar}%
  \BibitemOpen
  \bibfield  {author} {\bibinfo {author} {\bibfnamefont {K.}~\bibnamefont
  {Glampedakis}}, \bibinfo {author} {\bibfnamefont {A.~D.}\ \bibnamefont
  {Johnson}}, \ and\ \bibinfo {author} {\bibfnamefont {D.}~\bibnamefont
  {Kennefick}},\ }\href {\doibase 10.1103/PhysRevD.96.024036} {\bibfield
  {journal} {\bibinfo  {journal} {Phys. Rev. D}\ }\textbf {\bibinfo {volume}
  {96}},\ \bibinfo {pages} {024036} (\bibinfo {year} {2017})},\ \Eprint
  {http://arxiv.org/abs/1702.06459} {arXiv:1702.06459 [gr-qc]} \BibitemShut
  {NoStop}%
\end{thebibliography}%

\end{document}